\newcommand{\sbb}{mag/$\sq\arcsec$}

\def\rr{$R^{\star}$}
\newfont{\vs}{cmssdc10 scaled 1050}

\documentclass[traditabstract]{aa}
\usepackage{longtable,lscape}
\usepackage{graphicx}
\usepackage{txfonts}
\usepackage{natbib}
\usepackage[figuresright]{rotating}
\usepackage{subfigure}
\usepackage{multirow}
\usepackage{rotating}
\usepackage{lscape}
\usepackage{color}
\usepackage{afterpage}
\usepackage{natbib}

\begin{document} 
   \title{The ionized gas in the CALIFA early-type galaxies I. Mapping two representative cases: NGC~6762 and NGC~5966 \thanks{Based on observations collected at the Centro Astron\'omico Hispano Alem\'an (CAHA) at Calar Alto, operated jointly by the Max-Planck-Institut f\"ur Astronomie and the Instituto de Astrof\'isica de Andaluc\'ia (CSIC).}}

   
   \author{C. Kehrig\inst{1,2}, A. Monreal-Ibero\inst{2}, P. Papaderos\inst{3},
            J.M. V\'{\i}lchez\inst{2}, J.M. Gomes\inst{3}, J. Masegosa\inst{2}, 
            S.F. S\'anchez\inst{4}, M.D. Lehnert\inst{5}, R. Cid Fernandes\inst{6,2}, J. Bland-Hawthorn\inst{7}, D.J. Bomans\inst{8}, I. Marquez\inst{2}, D. Mast\inst{2,4}, J.A.L. Aguerri\inst{9}, \'A.R. L\'opez-S\'anchez\inst{10}, R.A. Marino\inst{11}, A. Pasquali\inst{12}, I. Perez\inst{13}, M.M. Roth\inst{1}, P. S\'anchez-Bl\'azquez\inst{14}, \and B. Ziegler\inst{15}}

\offprints{C. Kehrig}

\institute{Leibniz-Institut f\"{u}r Astrophysik Potsdam, innoFSPEC Potsdam, An der Sternwarte 16, 14482 Potsdam, Germany 
         \and
           Instituto de Astrof\'{\i}sica de Andaluc\'{\i}a (CSIC),
           Apartado 3004, 18080 Granada, Spain\\
        \email{kehrig@iaa.es}  
         \and 
            Centro de Astrof{\'\i}sica and Faculdade de Ci\^encias, Rua das Estrelas 4150-762 Porto, Portugal 
         \and 
             Centro Astron\'omico Hispano Alem\'an, Calar Alto (CSIC-MPG), C/ Jes\'us Durb\'an Rem\'on 2, E-04004 Almer\'{\i}a, Spain 
         \and 
            GEPI, Observatoire de Paris, UMR 8111, CNRS, Université Paris Diderot, 5 place Jules Janssen, 92190 Meudon, France 
         \and 
           Departamento de F\'{\i}sica-CFM, Universidade Federal de Santa Catarina, C.P. 476, 88040-900, Florian\'opolis, SC, Brazil 
          \and          
Sydney Institute for Astronomy, School of Physics, University of Sydney, NSW 2006, Australia 
         \and 
           Astronomical Institute of the Ruhr-University Bochum, Universitätsstr. 150, 44780 Bochum, Germany  
\and
Astrof\'{\i}sica de Can\'arias (IAC), v\'{\i}a L\'actea s/n, 38200, La Laguna, Spain 
\and
Australian Astronomical Observatory, PO Box 296, Epping, NSW, 1710, Australia, Department of Physics and Astronomy, Macquarie University, NSW 2109, Australia
\and
Departamento de Astrof\'{\i}sica y CC$.$ de la Atm\'{o}sfera, Facultad de CC$.$ F\'{i}sicas, Universidad Complutense de Madrid, Avda.\,Complutense s/n, 28040 Madrid, Spain 
\and
Astronomisches Rechen Institut, Zentrum fuer Astronomie der Universitaet Heidelberg, Moenchhofstrasse 12 - 14, 69120 Heidelberg, Germany 
\and
Dep. F\'{\i}sica Te\'orica y del Cosmos, Campus de Fuentenueva, Universidad de Granada, 18071, Granada, Spain 
\and
Departamento de F\'{\i}sica Te\'orica, Universidad Autonoma de Madrid, Cantoblanco, 28049, Madrid, Spain 
\and
University of Vienna, T\"{u}rkenschanzstrasse 17, 1180 Vienna, Austria.
}
 
\date{} 

\titlerunning{The ionized gas in the CALIFA ETGs I: NGC~6762 and NGC~5966}
\authorrunning{C. Kehrig et al.}

 
 
\abstract{As part of the ongoing CALIFA survey, we have conducted a thorough bidimensional analysis of the ionized gas in two E/S0 galaxies, NGC~6762 and NGC~5966,
aiming to shed light on the nature of their warm ionized ISM. Specifically, we
present optical (3745-7300 \AA) integral field spectroscopy obtained with the
PMAS/PPAK integral field spectrophotometer. Its wide field-of-view
(1$\arcmin$ x 1$\arcmin$) covers the entire optical extent of each galaxy down to faint continuum surface brightnesses. To recover the nebular lines, we modeled and subtracted the underlying
stellar continuum from the observed spectra using the STARLIGHT
spectral synthesis code. The pure emission-line spectra were used to
investigate the gas properties and determine the possible sources of ionization. We show the
advantages of IFU data in interpreting the complex nature of the
ionized gas in NGC~6762 and NGC~5966. In NGC~6762, the ionized gas and
stellar emission display similar morphologies, while the emission line morphology is elongated in NGC~5966, spanning $\sim$ 6 kpc, and is oriented roughly orthogonal to
the major axis of the stellar continuum ellipsoid. Whereas gas and stars are kinematically aligned in NGC~6762, the gas is kinematically
decoupled from the stars in NGC~5966. A decoupled rotating disk or an ``ionization cone'' are
two possible interpretations of the elongated ionized gas structure in
NGC~5966. The latter would be the first ``ionization cone'' of
such a dimension detected within a weak emission-line galaxy.  
Both galaxies have weak emission-lines relative
to the continuum [EW(H$\alpha$) $\lesssim$ 3 \AA] and have very low excitation, log([O{\sc iii}]$\lambda$5007/H$\beta$) $\lesssim$ 0.5. 
Based on optical diagnostic ratios ([O{\sc iii}]$\lambda$5007/H$\beta$,
[N{\sc ii}]$\lambda$6584/H$\alpha$, [S{\sc ii}]$\lambda$6717,6731/H$\alpha$, [O{\sc i}]$\lambda$6300/H$\alpha$),
both objects contain a LINER nucleus and an extended LINER-like gas
emission. The emission line ratios do not vary significantly with radius or
aperture, which indicates that the nebular properties are spatially
homogeneous. The gas emission in 
NGC~6762 can be best explained by photoionization by pAGB stars without
the need of invoking any other excitation mechanism.
In the case of NGC~5966, the presence of a
nuclear ionizing source seems to be required to shape the elongated
gas emission feature in the ``ionization cone'' scenario, although
ionization by pAGB stars cannot be ruled out. Further study of this
object is needed to clarify the nature of its elongated gas
structure.}

\maketitle 
 
\section{Introduction}\label{intro} 

Decades ago, early-type galaxies (ETGs) were thought to contain very
little, if any, gas \citep[e.g.][]{mat71,bre78,whi83}. Subsequently,
there have been many multiwavelength studies of ETGs that reveal a
substantial multiphase interstellar medium \citep[ISM;
e.g.][]{tri91,gou94b,mac96,cao00}. The dominant component of their ISM
is a hot (T $\sim10^{6}-10^{7}$~K) gaseous component that emits in the
X-rays \citep[e.g.][]{for79,fab92,osu01}. Moreover, a warm
(T$\sim$10$^4$ K), less significant phase of the ISM has been generally
detected with masses that are an order of magnitude lower than
observed in spiral galaxies \citep[][]{mac96}.  The frequency of ETGs
with a detectable warm ionized component in their ISM is significant,
ranging from 60$\%$ to 80$\%$, despite the differences in sample
selection criteria and sample sizes. Narrowband H$\alpha$
images of ETGs often reveal extended line emission, with radii of 5-10 kpc,
which mostly have morphologies similar to the underlying
stellar population \citep[][]{dem84,kim89,tri91}.

The nebular emission lines provide information about the physical
properties and the ionization source(s) of the warm ISM. Understanding
the sources required to ionize the gas is needed to investigate
fundamental questions of the origin and the nature of the ionized gas
in ETGs, which are still largely unsolved despite many studies.  Most
of the ETGs are optically classified as Seyfert nuclei or
low-ionization nuclear emission-line regions (LINERs) based on their
spectroscopic properties \citep[see][and references therein]{ani10}.
Several studies have discussed possible excitation mechanisms in ETGs,
for example post-AGB (pAGB) stars, shocks, active galactic nuclei
(AGNs), and OB
stars \citep[e.g.][]{bin94,sta08,sar10,ani10,fin10}. \cite{bin94}
claimed that white dwarfs and hot post-AGB stars provide sufficient
ionizing photons to explain the observed generally low H$\alpha$
equivalent widths (EW) and the LINER-like emission-line ratios
observed in such galaxies \citep[see
also][]{sod99,sta08}. \citet{sar10} investigate the ionizing sources
for the gas in ETGs based on SAURON integral-field spectroscopy data
whose spectra are limited to a relatively narrow wavelength range. The
authors conclude that pAGB stars are the main source of ionizing
photons in ETGs, and not fast shocks. In contrast, \citet{ani10}, by
analyzing optical long-slit spectra of 65 ETGs, claim that their
nuclear line-emission can be explained by excitation from the hard
ionizing continuum from an AGN and/or fast shocks.  However, they do
not completely rule out a contribution from pAGB stars at large radii,
even if, their study of spatial variations in the warm ISM was limited
by the area covered by their slits. Furthermore, it seems that ongoing
star formation might be ocurring in some ETGs, and the photoionization
by hot young stars in some of these galaxies cannot be
dismissed \citep[e.g.][]{vil98,sch07,sha10}. This leaves us in the
puzzling situation where all processes (photoionization from the old
stellar population, young stellar population, and an AGN, or heating
owing to fast shocks) may or may not all contribute to the excitation
of the warm ionized medium in ETGs.

To address questions like "What are the sources of ionization that
contribute to exciting line emission in ETGs?", we initiated a program
to analyze the warm ISM in ETGs within the context of the {\it Calar
Alto Legacy Integral Field Area (CALIFA)
survey} \citep[][]{Sanchez2011}. CALIFA, through the use of
wide-field, optical integral-field spectroscopy (IFS), it offers a unique
observing capability to study the detailed properties of the extended
optical emission-line gas in galaxies.

In this paper, we describe the results of a pilot study of two ETGs,
NGC~6762 and NGC~5966.  Our goal is to probe the properties of their
warm ISM via spatially resolved emission-line diagnostics. These
objects represent the two morphological types (S0 and E) among those
ETGs first observed within CALIFA showing extended gas emission. As
far as we know, this is the first investigation of the properties of
the ionized gas in NGC~6762 and NGC~5966. The results for the
remaining CALIFA ETGs hosting ionized gas will be presented in
forthcoming papers.

The paper is organized as follows. In Sect.~\ref{overview} we
provide an overview of our two galaxies.  Sect.~\ref{obs_datared}
describes the observations and data reduction. The methodology used to
subtract the stellar population from our spectra is presented in
Sect.~\ref{2D}. In Sects. 5 and 6 we present a detailed analysis of the
properties of the warm ionized medium in both galaxies. In
Sect.~\ref{discussion} we discuss our results. Finally, in
Sect.~\ref{fim} we summarize our main conclusions from this
study.

\section{The sample: NGC~6762 and NGC~5966}\label{overview}

Both NGC~6762 and NGC~5966 belong to the Uppsala General Catalogue
(UGC) of Galaxies (Nilson 1973) and are among the most luminous
objects (in $z$-band absolute magnitude, M$_{z}$) in the CALIFA survey
sample \citep[see Fig. 3 in][]{Sanchez2011}. The basic data of both galaxies are summarized in Table~\ref{sample}.

\begin{figure*}
\begin{minipage}{0.5\linewidth}
\includegraphics[width=0.9\textwidth,height=0.8\textwidth,angle=0.0]{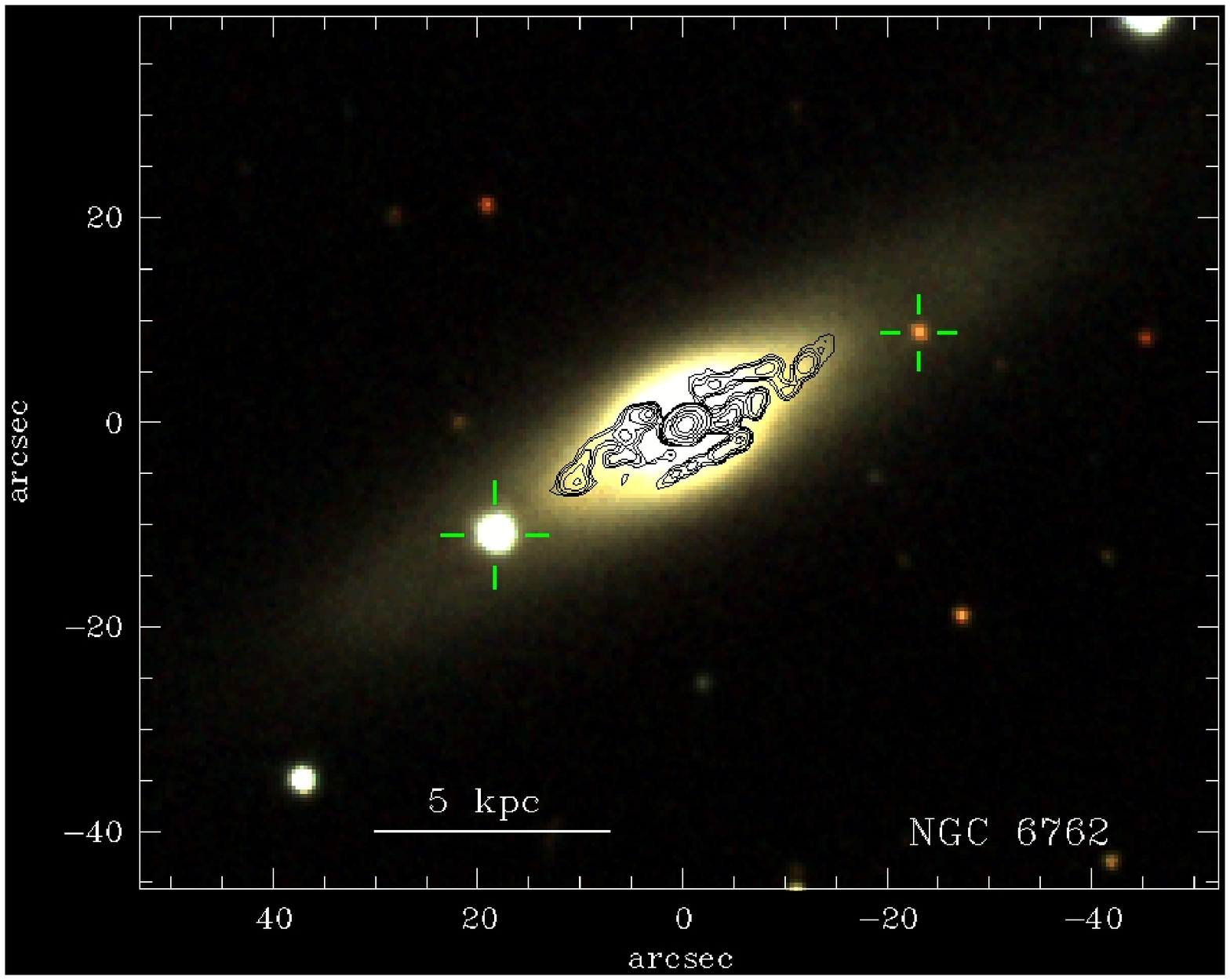}
\end{minipage}
\begin{minipage}{0.5\linewidth}
\includegraphics[width=0.9\textwidth,height=0.8\textwidth,angle=0.0]{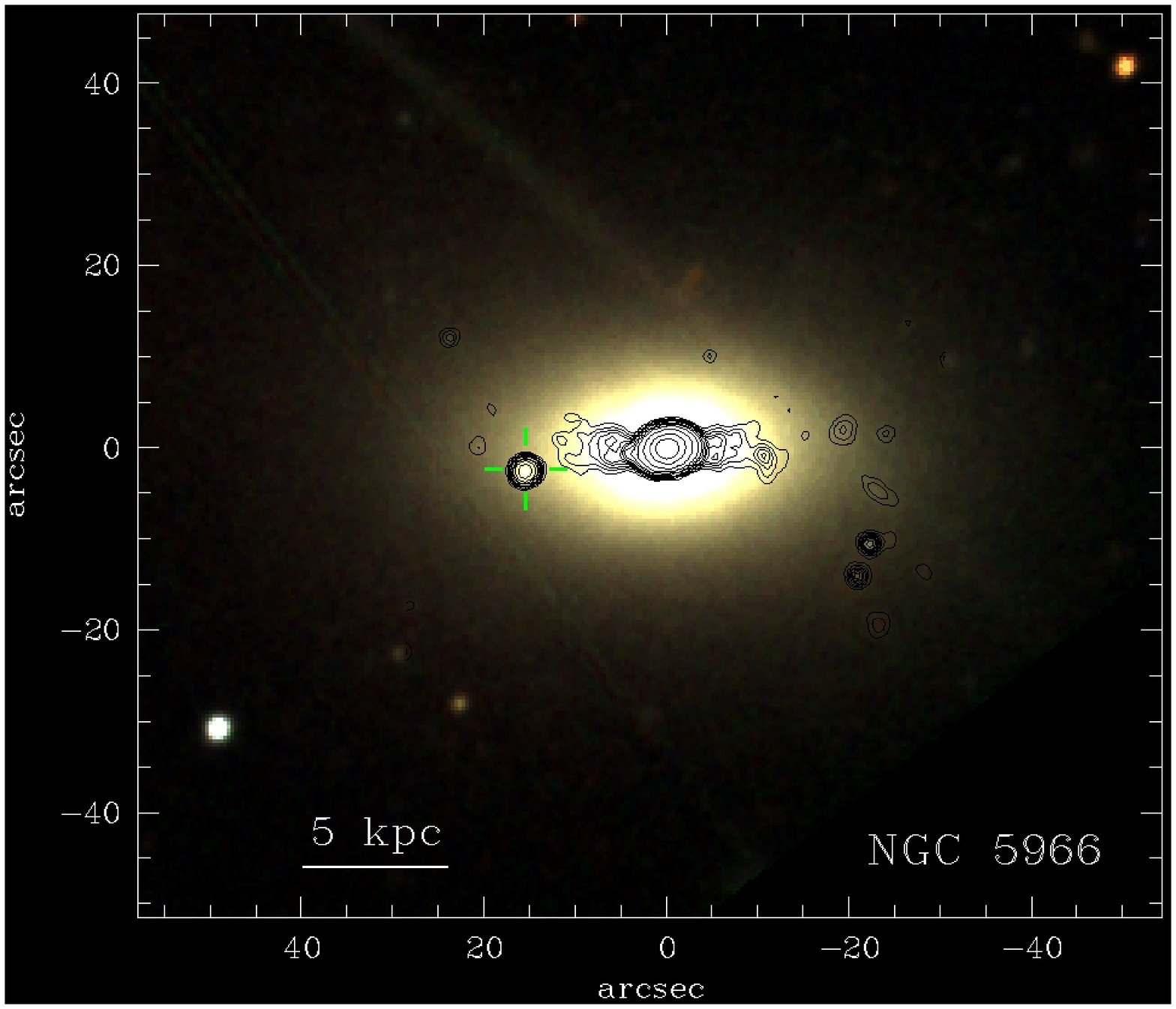}
\end{minipage}
\caption[]{{\it Left panel}: Three-color, $g$, $r$, and $i$, composite SDSS image of NGC~6762. 
Foreground sources that have been disregarded in the subsequent spectral
modeling and analysis are marked by crosses.
Contours delineate an extremely faint spiral feature with a projected size 
of 31\arcsec$\times$9\arcsec\ disclosed by unsharp masking in the bulge of the 
galaxy. {\it Right panel}: SDSS color composite of NGC~5966 with contours from an unsharp masked
image delineating a faint bar-like feature centered at the nucleus. North is up and east to the left in both images.}
\label{SDSS} 
\end{figure*}


In Fig.~\ref{SDSS} we show three-color broad band image from the Sloan
Digital Sky Survey \citep[SDSS,][]{yor00} of NGC~6762 (left panel) and
NGC~5966 (right panel).  For the lenticular galaxy NGC~6762, the SDSS
images reveal a featureless disk-like morphology with no spiral arms,
as expected from its morphological type. However, applying a
flux-conserving unsharp masking technique
\citep{Papaderos1998} on the SDSS images reveals an extended (31\arcsec$\times$9\arcsec) inclined spiral-like pattern that is traceable at extremely faint ($\leq$1\%) levels. Although not detected in the color maps owing to its extreme faintness, this feature indicates a complex formation history for the bulge component of NGC~6762 than what a cursory inspection of SDSS images may suggest. For this galaxy, we found overall red colors ($g$--$i$=1.13$\pm$0.03; $g$--$r$=0.76$\pm$0.04), characteristic of an old dominant stellar component ($>$ 6 Gyr) in both
the disk and the bulge. In NGC~5966 (right panel of Fig.~\ref{SDSS}) unsharp masking reveals a compact bar-like structure 24\arcsec\ across centered on the compact nucleus. This structure has been associated with a radio source that is very likely powered by an AGN, instead of star formation, due to its low far-infrared flux upper limit \citep[see][and references therein]{con02}. 
NGC~5966 shows approximately
constant colors across its optical continuum emission with colors of
$g$--$i$=1.20$\pm$0.02 and $g$--$r$=0.80$\pm$0.01. These red and uniform
colors imply that NGC~5966 has an old dominant stellar population. As is
the case for NGC~6762, no X-ray detection has been reported for NGC~5966.

\begin{table}
\caption{Basic data\label{sample}}
\centering
\renewcommand{\footnoterule}{}
\begin{minipage}{17.9cm}
\begin{tabular}{lcc} \hline
Parameter      &NGC~6762  &NGC~5966 \\ \hline
Other designations  & UGC~11405   & UGC~09923   \\
Morphological type & S0  & E \\
R.A. (J2000.0)  & 19h 05m 37.1s  &  15h 35m 52.1s       \\
DEC. (J2000.0)      & +63d 56' 03''   & +39d 46' 08''      \\
redshift   & 0.0098       &  0.015       \\
$r_{\rm eff}$\footnote{$g$ band effective radius from this work} & $7.2^{\prime\prime}$ & $10.6^{\prime\prime}$ \\
D\footnote{Distance to the galaxy from NED}(Mpc) & 45 & 69  \\
Scale (pc/$\prime\prime$)     & 217 &  334\\
M$_{u}$\footnote{Absolute magnitude in the SDSS corrected for Galactic extinction}(mag)   & -17.56  & -19.08         \\
M$_{g}$$^c$(mag)    & -19.36 & -20.95          \\
M$_{r}$$^c$(mag)    &  -20.15 & -21.73         \\
M$_{i}$$^c$(mag)    &  -20.58  & -22.14          \\
M$_{z}$$^c$(mag)    &   -20.85 & -22.43         \\
{\it u}\footnote{Apparent magnitude in the SDSS corrected for Galactic extinction} (mag) & 15.70 & 15.11  \\
{\it g}$^d$ (mag) & 13.90  & 13.24  \\
{\it r}$^d$ (mag) & 13.11 & 12.46 \\
{\it i}$^d$ (mag) & 12.68 & 12.05 \\
{\it z}$^d$ (mag) & 12.41 & 11.76 \\
A$_{V}$\footnote{Galactic extinction from Schlegel et al. (1998)} (mag) & 0.18 & 0.076 \\
\hline
\end{tabular}
\end{minipage}
\end{table}

 
\section{Observations and data reduction}\label{obs_datared} 
 
The observations of NGC~6762 and NGC~5966 were performed within the
CALIFA survey which aims to carry out a statistically complete IFS
survey of over the full range Hubble types present in
the local universe \citep[][]{Sanchez2011}. The survey is being
conducted at the 3.5 m telescope of the Calar Alto observatory using
the \emph{Potsdam MultiAperture Spectrograph} (PMAS) in its PPAK mode
\citep{kel06}. A new CCD was installed in PMAS in 2009 \citep{rot10}
which is being used for the entire survey. Fibers in the PPAK bundle
have a projected diameter on the sky of $2.7^{\prime\prime}$ and 331 out
of 382 of the fibers form a hexagonal area covering a field of view of
$\sim72^{\prime\prime}\times64^{\prime\prime}$ with a filling factor of
$\sim$65$\%$. The remaining 51 fibers are dedicated to sky background
(36 fibers) or used to obtain exposures of calibration lamps (15 fibers). 

NGC~6762 and NGC~5966 were observed on 12 July 2010 and 1 April 2011,
respectively, under photometric conditions and with a seeing of about
0\farcs8-1\farcs2. For this study, we used the V500 grating, which covers
from 3745 to 7300 \AA\ and has an effective spectral resolution of $\sim$
6.5 \AA\ full width at half-maximum (FWHM) at $\sim$ 5000 \AA\ and a
resolving power of $R\sim850$. A dithering scheme with three pointings was used to
sample the whole optical extent of each galaxy \citep[see][for details]{Sanchez2011}. Each
pointing was observed for a total of 900s, divided to three individual exposures
to facilitate the removal of cosmic rays. The
estimated limiting surface brightness in the V band for NGC~6762 and NGC~5966 are $\sim$23.9 and 23.4
mag arcsec$^{-2}$, respectively.


Data was reduced using the CALIFA pipeline (version
1.2). A detailed description of the steps followed during the
reduction can be found in \citet[][]{Sanchez2011} and the references
therein. Briefly, first a master bias was created by averaging all the
bias frames observed during the night and then subtracted from the
science frames. Second, the location of the spectra in the CCD was
determined using a continuum illuminated exposure taken before the
science exposures.  Then each spectrum was extracted from the science
frames.  Next, wavelength calibration and distortion correction were
performed using arc lamps. Differences in the fiber-to-fiber
transmission throughput were corrected by comparing the
wavelength-calibrated extracted science frames with the corresponding
continuum illuminated ones. The night-sky background spectrum,
obtained by combining the spectra from the 36 dedicated sky fibers, is
subtracted from the science-fiber spectra of the corresponding frame.
Flux calibration was performed by comparing the extracted spectra of
spectrophotometric standards stars from the Oke Catalogue
\citep{oke90}. The data were also corrected for the atmospheric extinction using the airmass and the extinction of the observations
as measured by the Calar Alto Extinction monitor. Both galaxies were observed with airmass $<$ 1.14. To construct the
final data cube, all three pointings are combined. Using their relative
positions and the PPAK position table, they are reformated into a single
datacube with a spatial sampling of $1^{\prime\prime}$. Finally, the
present version of the pipeline corrects for the effect of the Galactic
extinction as reported by \citet{sch98}: $E(B-V)$ = 0.055 and 0.023 for
NGC~6762 and NGC~5966, respectively.



\section{2D modeling of the underlying stellar population}\label{2D} 

The emission lines in ETGs are generally extremely faint and often
have EWs that are less than a few \AA.  This is a particular problem for the
Balmer lines since the underlying absorption features from the stellar
population can have EWs of the same order or
more \citep[e.g.][]{ani10}. Determining emission-line
intensities and intensity ratios is therefore a challenging
task. To measure precisely emission-line fluxes and EWs, it
is critical to model and remove the underlying stellar continuum that
dilutes emission-line features. For this purpose, we used the
STARLIGHT\footnote{The STARLIGHT project is supported by the Brazilian
agencies CNPq, CAPES and FAPESP and by the France-Brazil CAPES/Cofecub
program.} spectral synthesis code \citep[][]{cid04a} to model
the stellar spectral energy distribution (SED) at each spaxel of the
PPAK data cube.  

The best-fitting stellar SED was then subtracted from
the observed spectrum in order to isolate the pure emission line
spectrum, which was then used to study the nebular component in ETGs.
This way, in many cases, even weak emission lines (such as, e.g. the
[O{\sc i}]$\lambda$6300 line) that seem to be absent in the observed
spectra could be detected and measured accurately enough to yield
astrophysically constraining results on the sources ionizing the gas.
STARLIGHT uses various techniques for combining synthetic stellar
populations to compute the best-fitting stellar SED. The best-fitting
linear combination of $N_\star$ single stellar populations (SSPs), is
obtained by using a nonuniform sampling of the parameter space based
on the Markov Chain Monte Carlo algorithm, plus an approach called
simulated annealing, and a convergence criterion similar to that
proposed by \cite{GelmanRubin1992}, to approximately determine a global $\chi^2$ minimum.

We chose for our analysis SSPs from \citet[][hereafter, BC03]{bru03}, which are based on the Padova 2000 evolutionary
tracks \citep{gir00} and the Salpeter initial mass function (IMF)
between 0.1 and 100 $M_\odot$. The SSP library used here consists of
three metallicities (0.5, 1 and 1.5~Z$_{\odot}$) for 34 ages between
5~Myr and 13.6~Gyr.  The intrinsic extinction was modeled as an
uniform dust screen, adopting the extinction law by \cite{car89}.
Line broadening effects, due to line-of-sight stellar motions, are
accounted for in STARLIGHT by a convoluted Gaussian function.

The spectral synthesis models were computed spaxel by spaxel in
an automatic manner using a pipeline written in the
MIDAS\footnote{Munich Image Data Analysis System, provided by the
European Southern Observatory (ESO).} script language and in Fortran,
with additional modules that make use of PGplot and CFITSIO routines.
Prior to fitting, the spectrum at each spaxel of the binned data cube
was extracted, shifted to the rest frame, and resampled to a 1 \AA\
resolution.  Emission lines and spurious spectral features were
flagged using either a predefined spectral mask or, additionally, a
3$\sigma$ clipping routine applied on the net emission line component
of each spectrum, after a coarse prefitting with a reduced SSP
library. Spectral fits were carried out in the spectral region between
4000 \AA\ and 6900 \AA, because the signal-to-noise (S/N) bluewards of
4000 \AA\ is generally too low for a reliable modeling of the stellar
component. We disregarded the red end of the spectra (6900--7300 \AA)
because of its generally lower S/N, as well as spectral artifacts
induced by vignetting in the outer zones of the PPAK
FOV \citep[see][]{Sanchez2011}

\begin{figure*}
\center
\includegraphics[width=0.28\textwidth,clip]{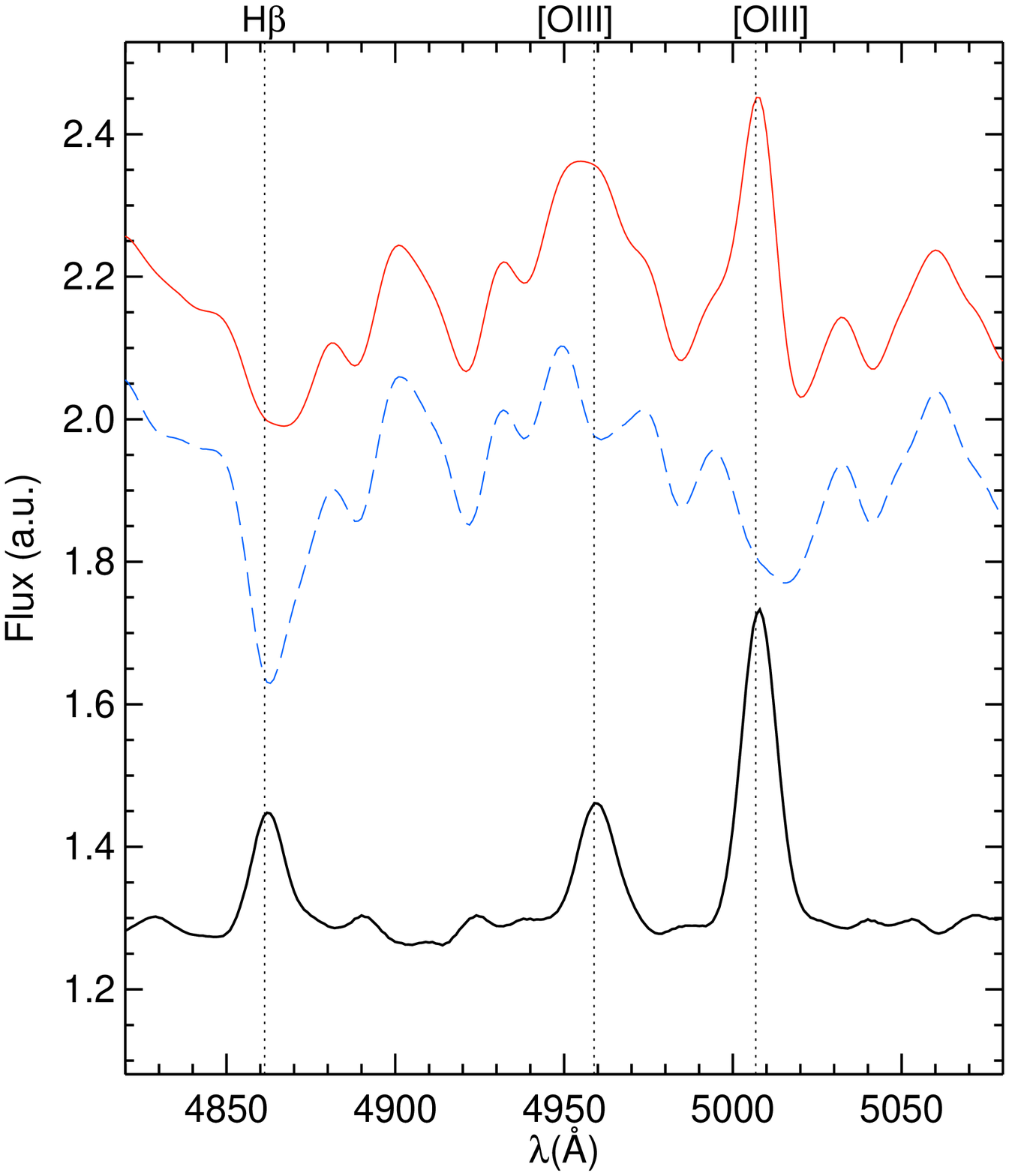}
\includegraphics[width=0.28\textwidth,clip]{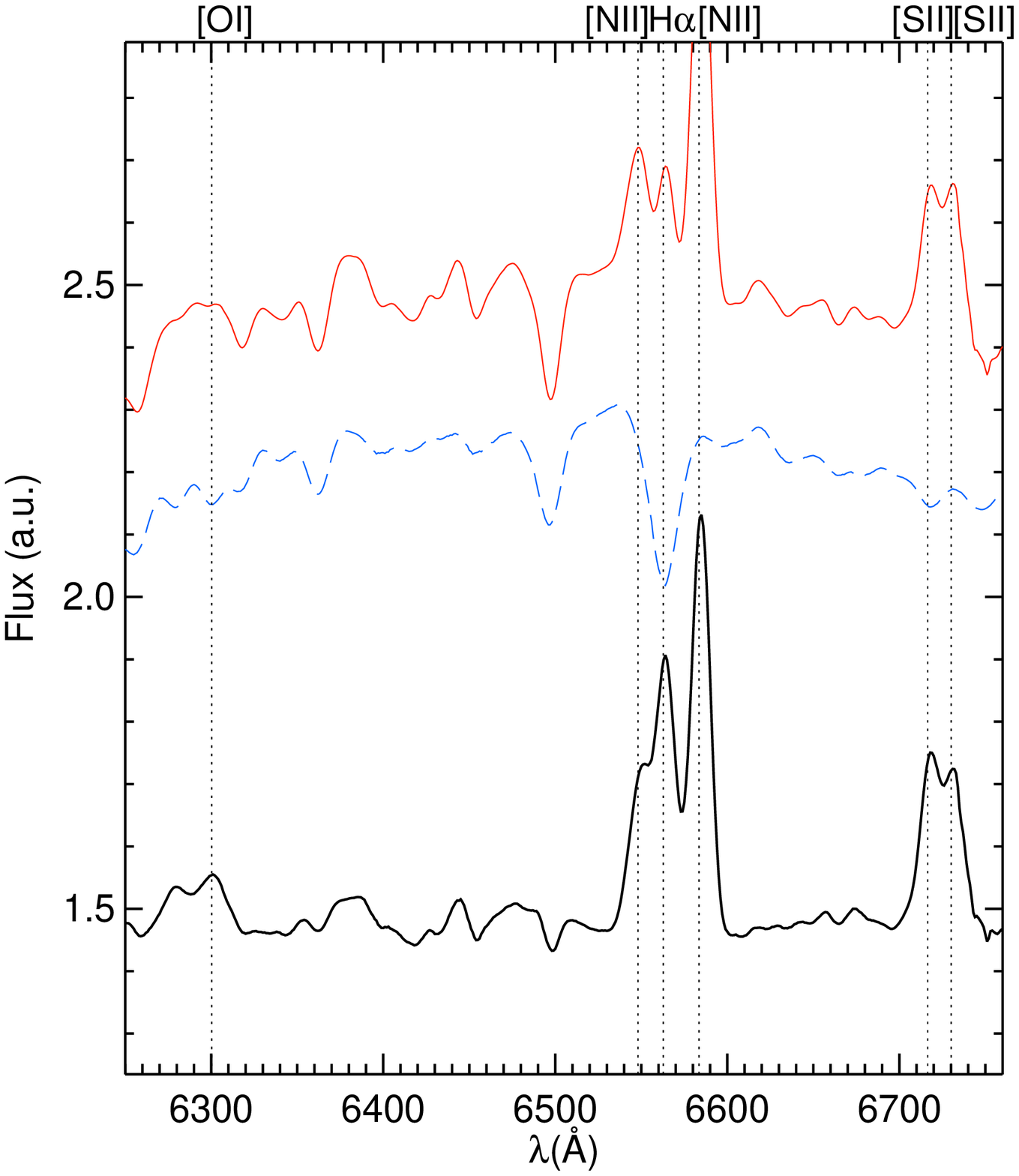}\\
\includegraphics[width=0.28\textwidth,clip]{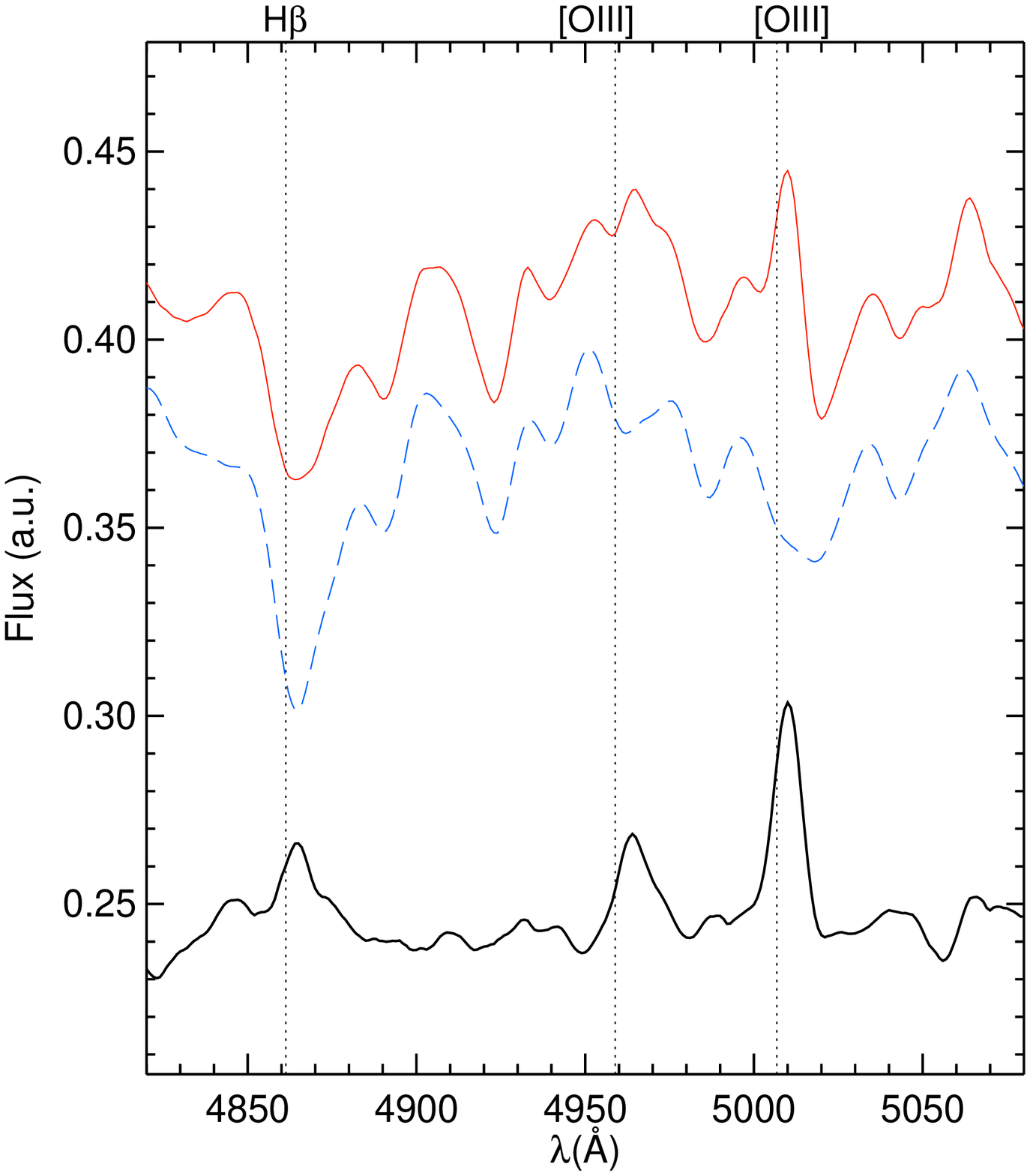}
\includegraphics[width=0.28\textwidth,clip]{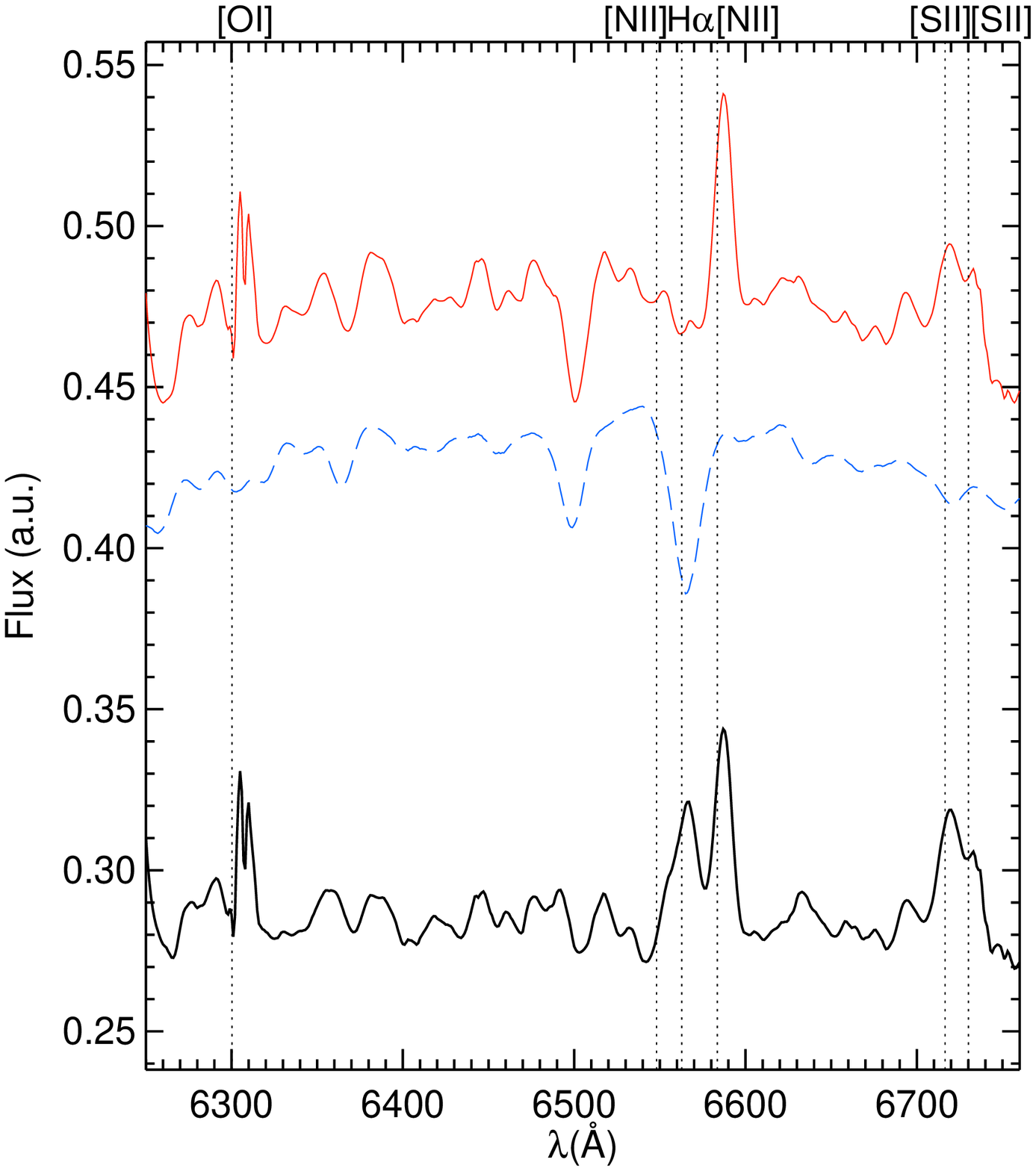}
\caption{NGC~6762: the panels to the left and right illustrate the main
output from spectral synthesis in the blue and red spectral range,
respectively, for a single spaxel in the brightest nuclear zone (top panels)
and in the fainter periphery of the galaxy ($\sim$ $10^{\prime\prime}$ from the nucleus;
bottom panels). The red and blue lines correspond to the observed and modeled stellar
spectrum, respectively. By subtracting the latter from the former we obtain
the pure nebular emission line spectrum (black). The spectra are offset in arbitrary units (a.u.) for the sake of clarity.}
\label{stellar_sub_n6762} 
\end{figure*}
\begin{figure*}
\center
\includegraphics[width=0.28\textwidth,clip]{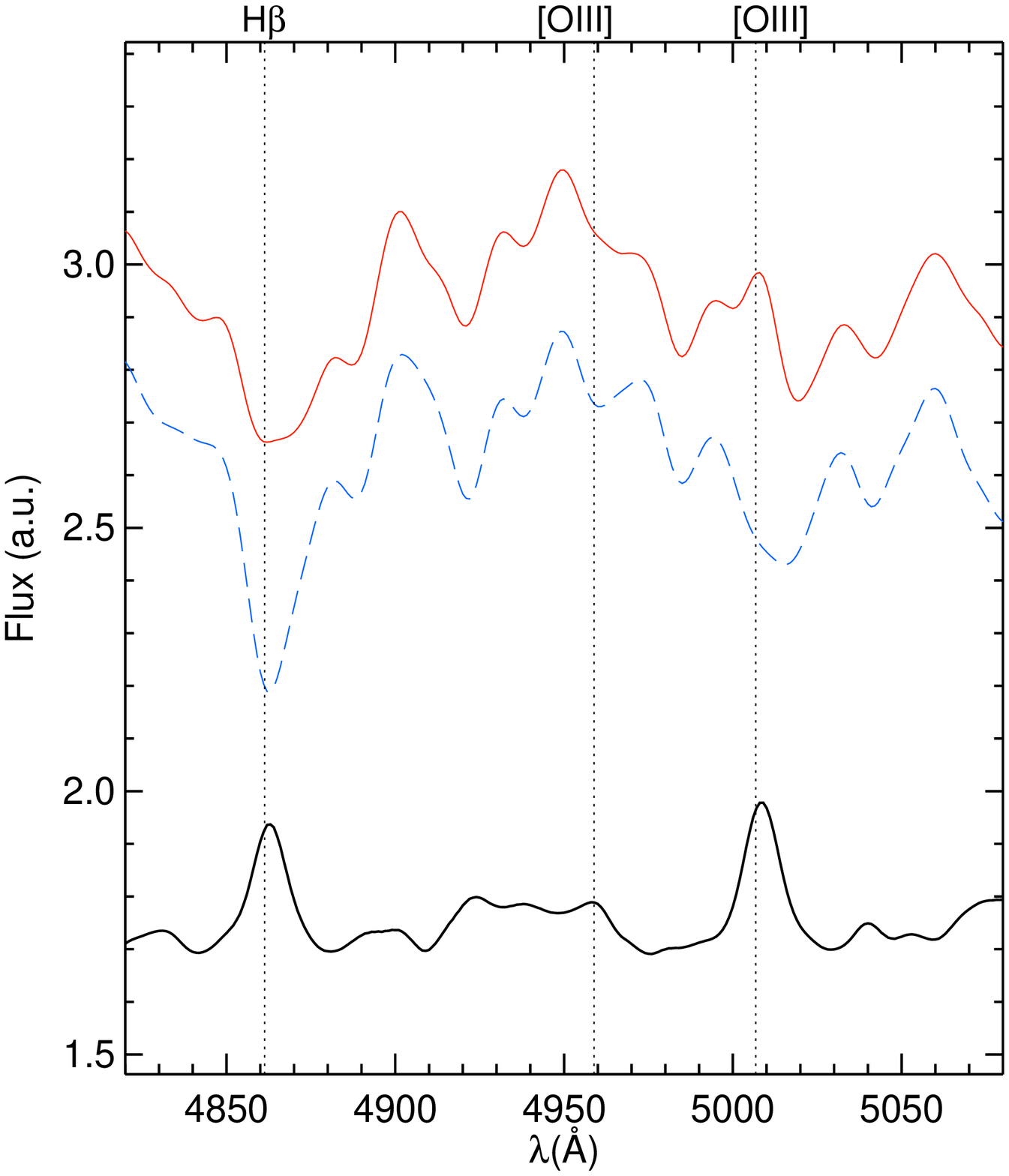}
\includegraphics[width=0.28\textwidth,clip]{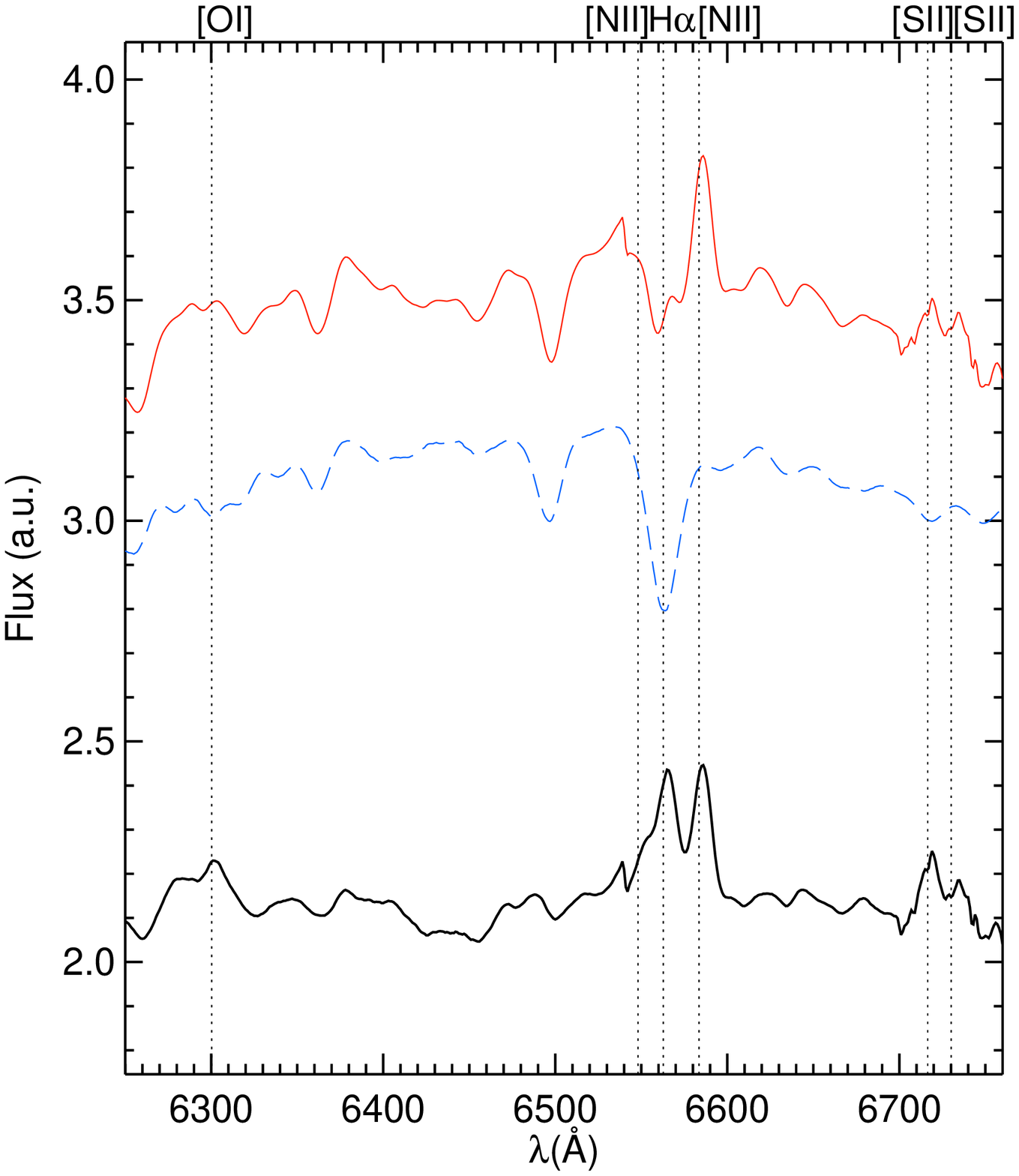}\\
\includegraphics[width=0.28\textwidth,clip]{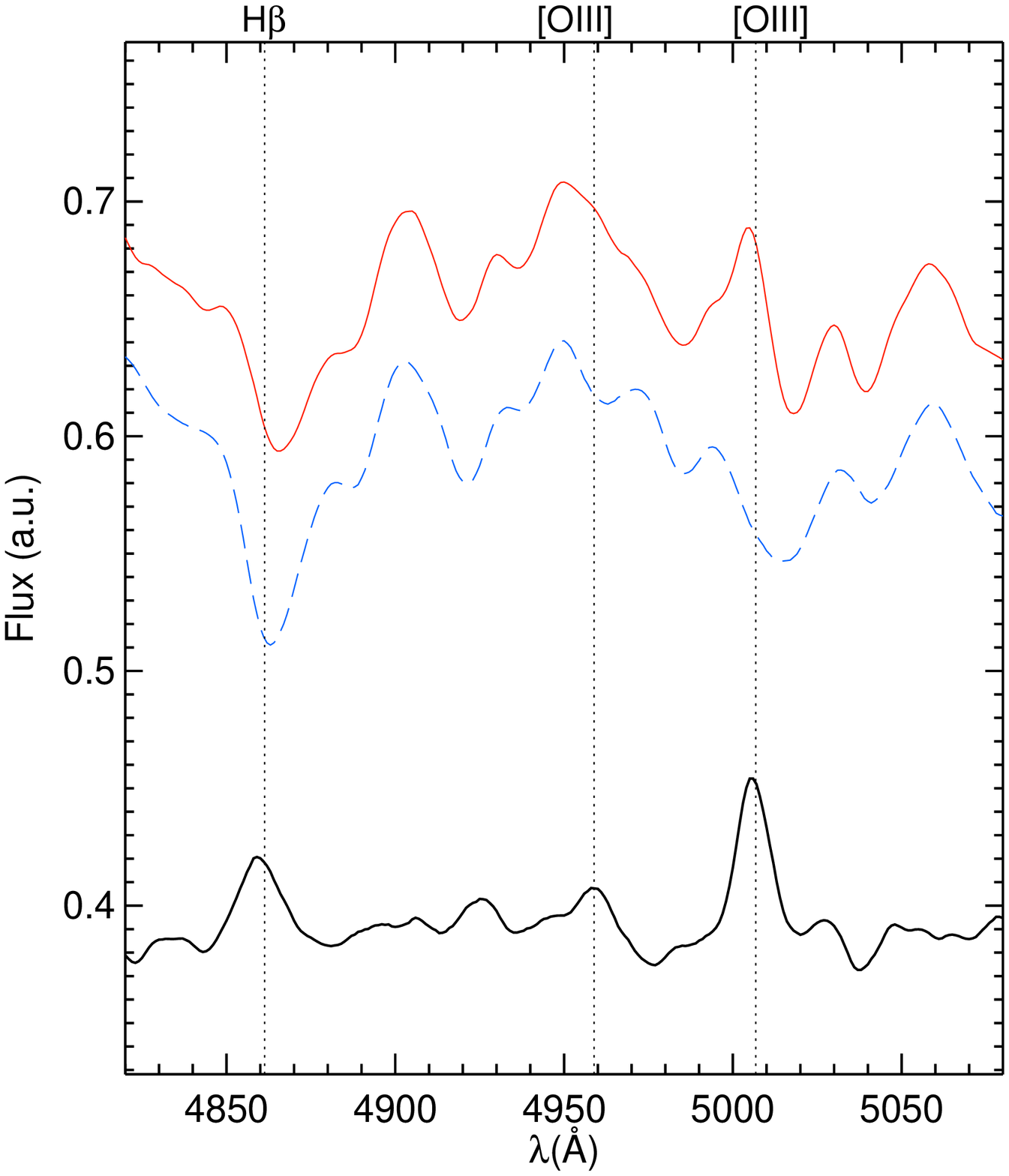}
\includegraphics[width=0.28\textwidth,clip]{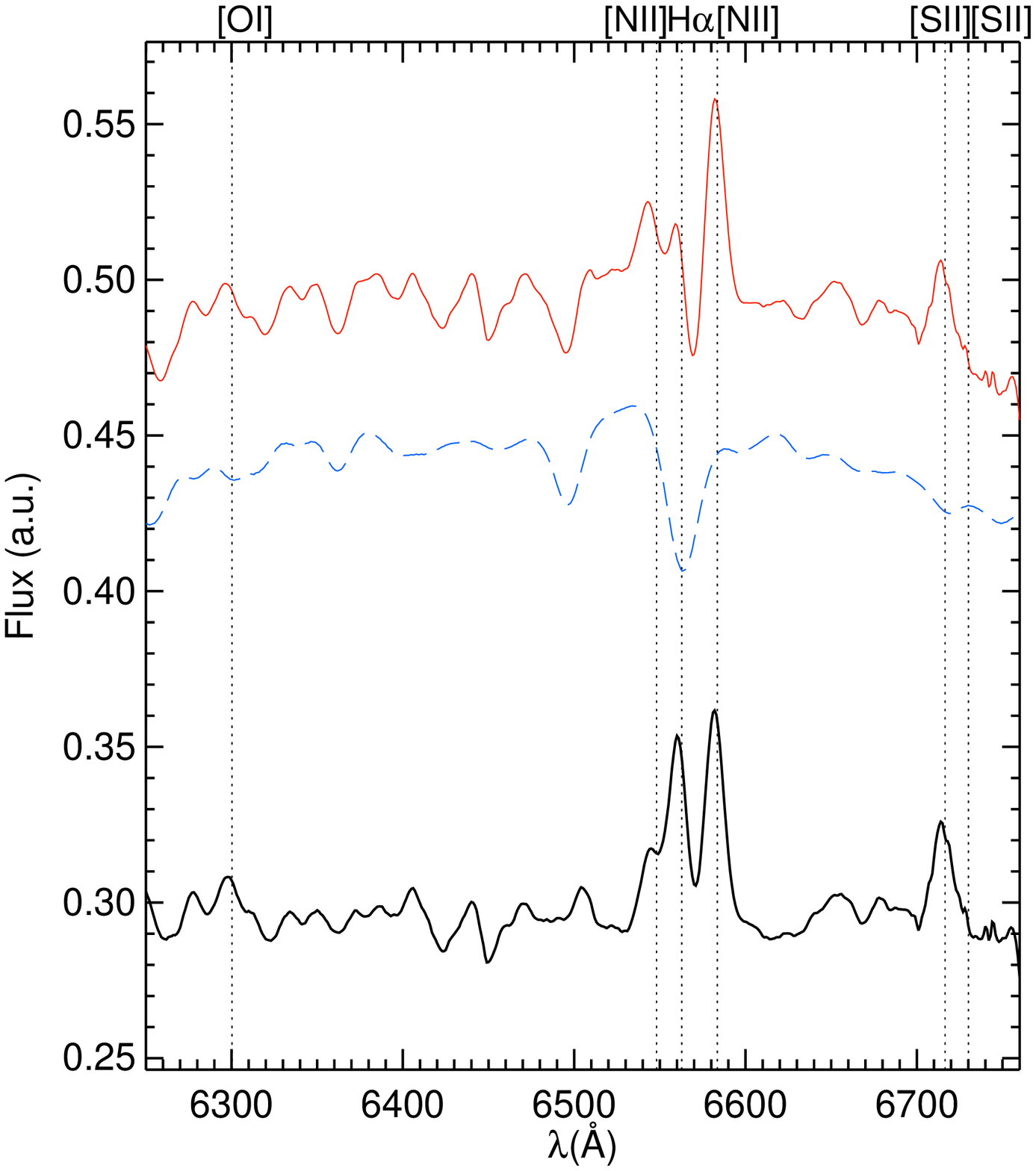}
\caption{NGC~5966: the same as Fig.~\ref{stellar_sub_n6762}.}
\label{stellar_sub_n5966} 
\end{figure*}

The output from the pipeline was then put into data cubes of
dimension $ 80 \times 75 \times 2800$ where 
each spaxel contains the post-processed observed spectrum (3D$_{\rm obs}$),
best-fitting stellar SED (3D$_{\rm mod}$), and pure emission line spectrum
(3D$_{\rm neb}$). Additionally, the pipeline stores in a
fourth data cube (3D$_{\rm res}$) the spatially resolved distribution
of various relevant quantities returned by STARLIGHT or computed
subsequently from its output.  These include, among others, the
reduced $\chi^2$ and the absolute deviation (ADEV) between the input
spectrum and its best-fitting SED, the derived $V$ band extinction
($A_V$) for the stellar component, the stellar velocity, the
luminosity-weighted and mass-weighted stellar age and metallicity, and
the surface density, as well as mass and luminosity fraction of stellar
population older than 10$^8$ yr. Illustrative examples of the
observed, modeled, and emission-line spectrum for single spaxels in
central and peripheral zones of NGC~6762 and NGC~5966 are displayed in
Figs. \ref{stellar_sub_n6762} and \ref{stellar_sub_n5966}.

As an approximate check of systematic uncertainties in the derived
emission line fluxes (Sect. 3), we additionally ran STARLIGHT models
based on MILES SSPs \citep{san06,vaz10} for similar range of stellar
population parameters. We find that the typical variation of emission
line fluxes for the two analyses is lower than 10\%, or about the
same order of magnitude as those due to uncertainties in emission line measurements
themselves.

\section{A 2D view of the warm ISM}\label{ISM} 

\subsection{Line fitting and map creation}\label{emlin} 

After subtracting the underlying stellar population from the data cubes, we
performed a Gaussian fit to the emission lines using the IDL-based routine
{\it mpfitexpr} \citep{mar09} and derived the quantities of interest for each
individual spaxel following the methodology described in \citet{mon11} and references therein. Then, we used these with the position of
the spaxels within the data cube to create an image suitable to manipulation
with standard astronomical software.

Line fitting in these galaxies is challenging since emission lines are
very weak even after stellar continuum subtraction. Therefore, the
solutions in velocity for H$\alpha$, and [N{\sc ii}]$\lambda$6584 were
used as initial guesses for the other remaining significant
emission lines in order to guarantee robust flux estimates. In the
emission-line maps presented in the following section, we show only
the line fluxes with relative error $<$ 0.10 and spaxels with ADEV $<$
4\%. This empirical threshold was found to provide a reasonable
compromise between the galaxy area studied and the goodness of the
fits to individual spaxels. Furthermore, such a conservative cut-off
ensures that Balmer absorption features are adequately reproduced by
synthetic stellar SEDs.  The chosen ADEV cutoff selects spaxels within
approximately the 22 and 22.5 $g$ \sbb\ isophote for NGC~6762 and
NGC~5966, respectively, or equivalently the inner 8 and 13 zones of
the respective plots in Fig.~\ref{annuli}. These isophotal levels
correspond to photometric radii of $\sim$1.5\,$r_{\rm eff}$ for NGC~6762  and
$\sim$1.7\,$r_{\rm eff}$ for NGC~5966, and
thus the area studied contains most of the galaxy's
optical luminosity for both galaxies.

\subsection{Continuum and Emission Line Intensity Maps}\label{maps}

\begin{figure*}
\includegraphics[width=4.5cm]{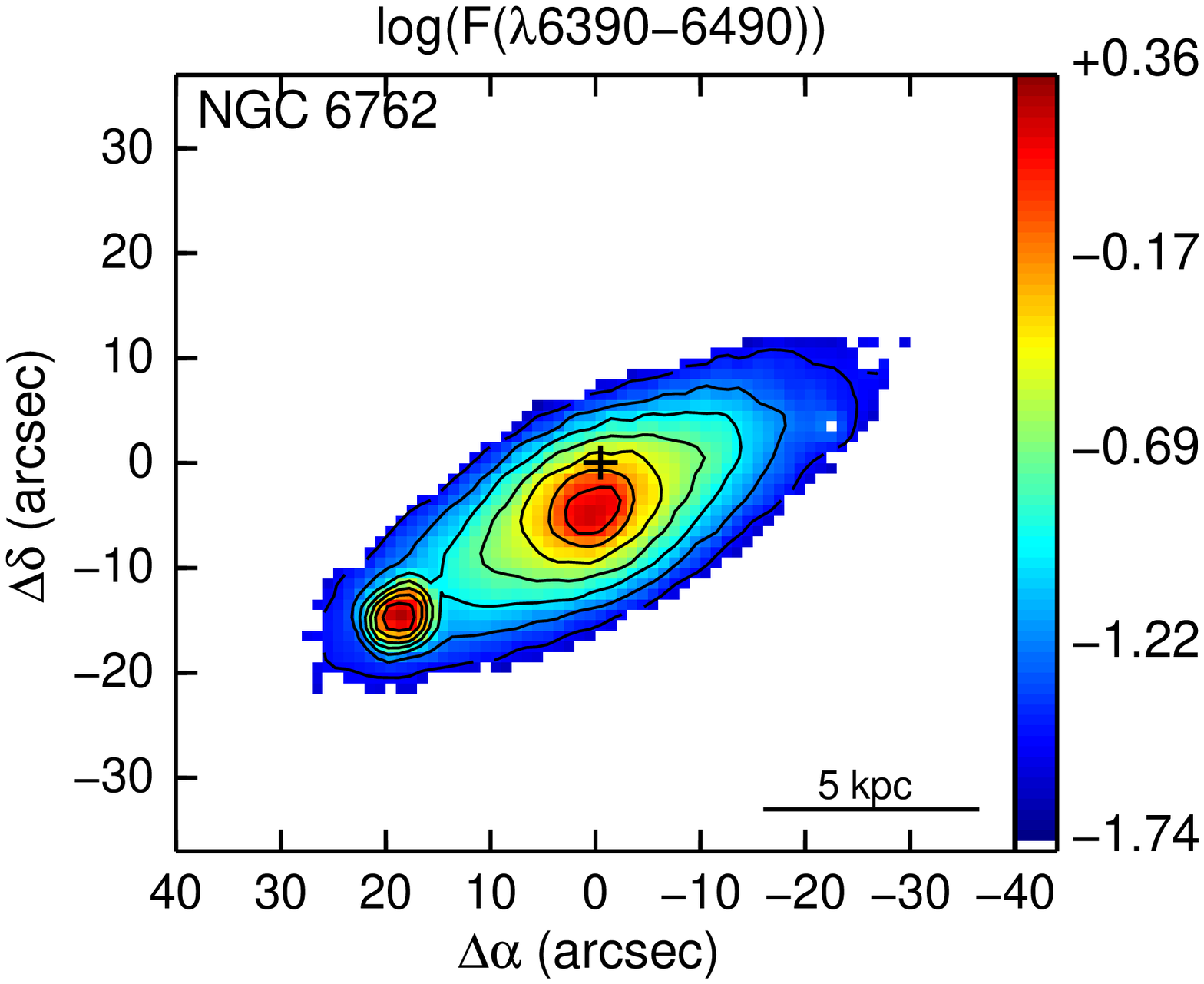} 
\includegraphics[width=4.5cm]{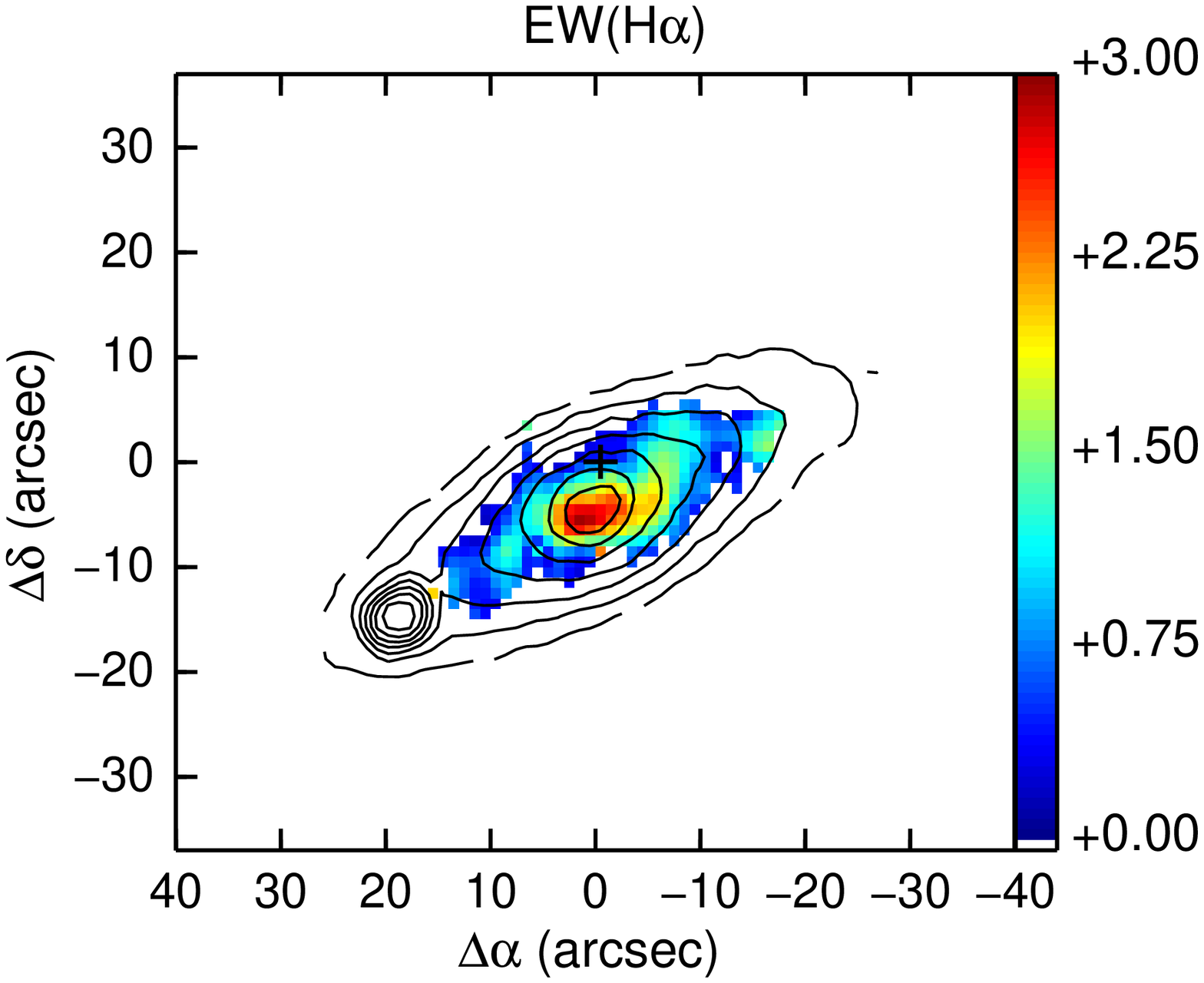}
\includegraphics[width=4.5cm]{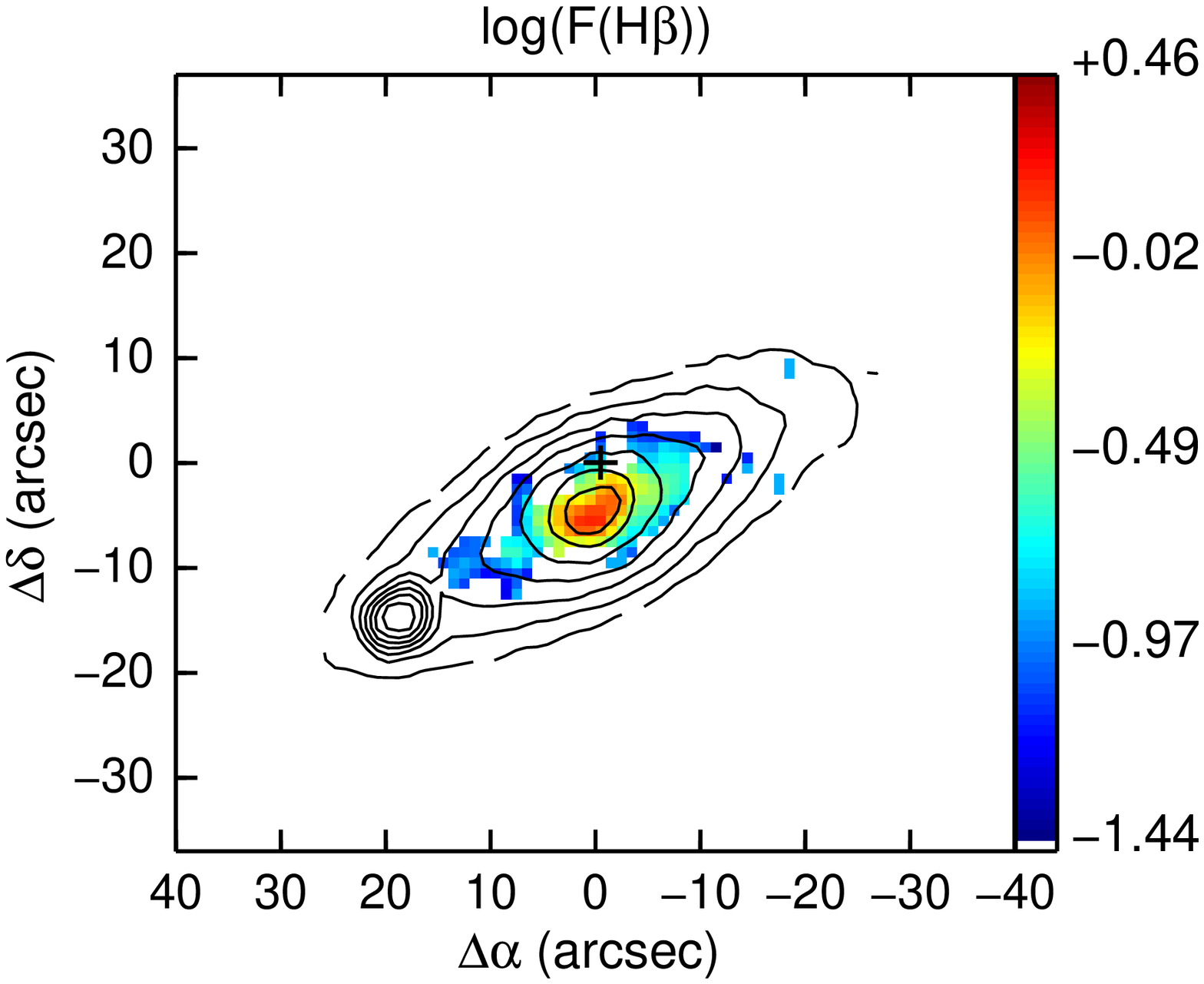}
\includegraphics[width=4.5cm]{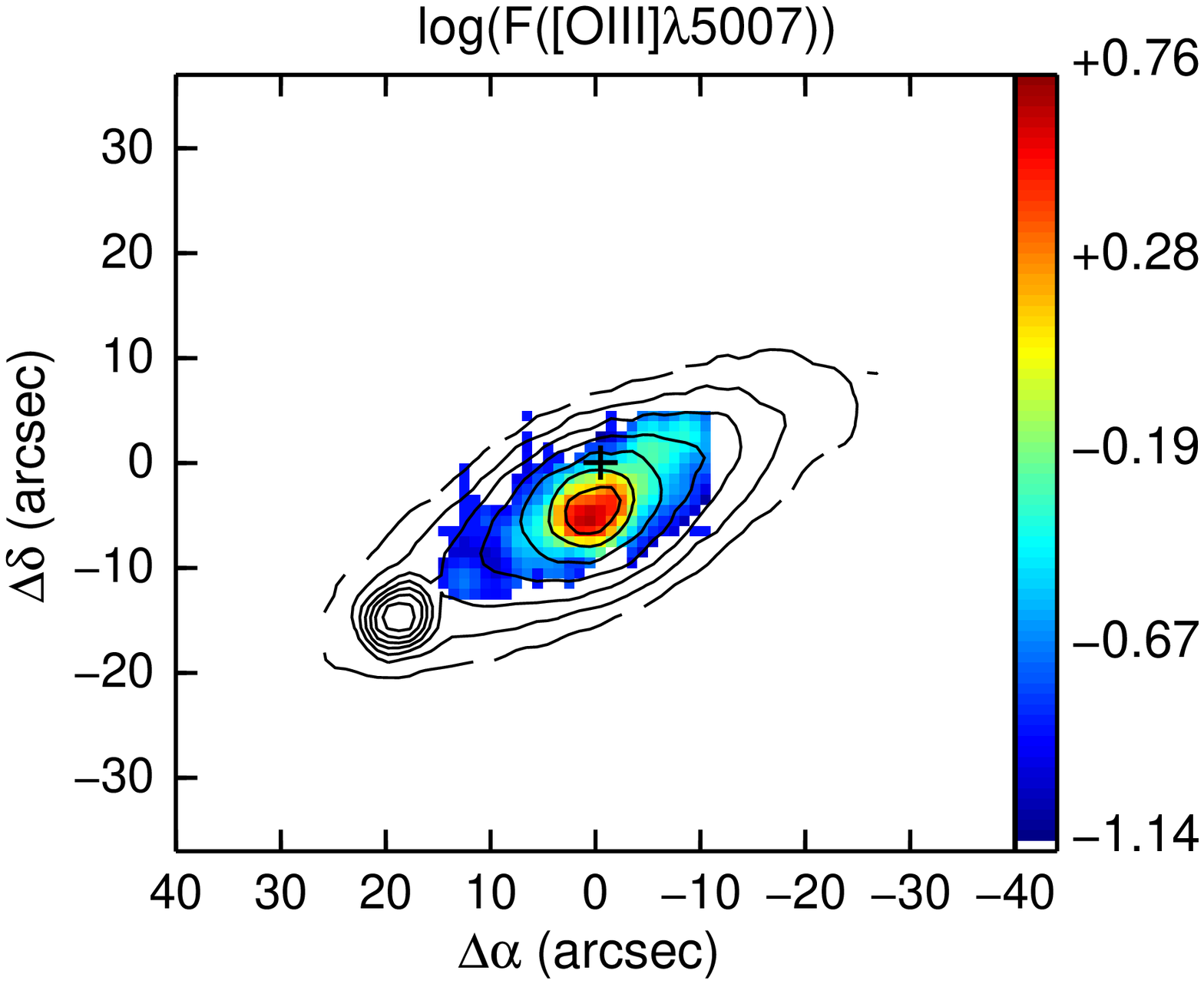}\\
\includegraphics[width=4.5cm]{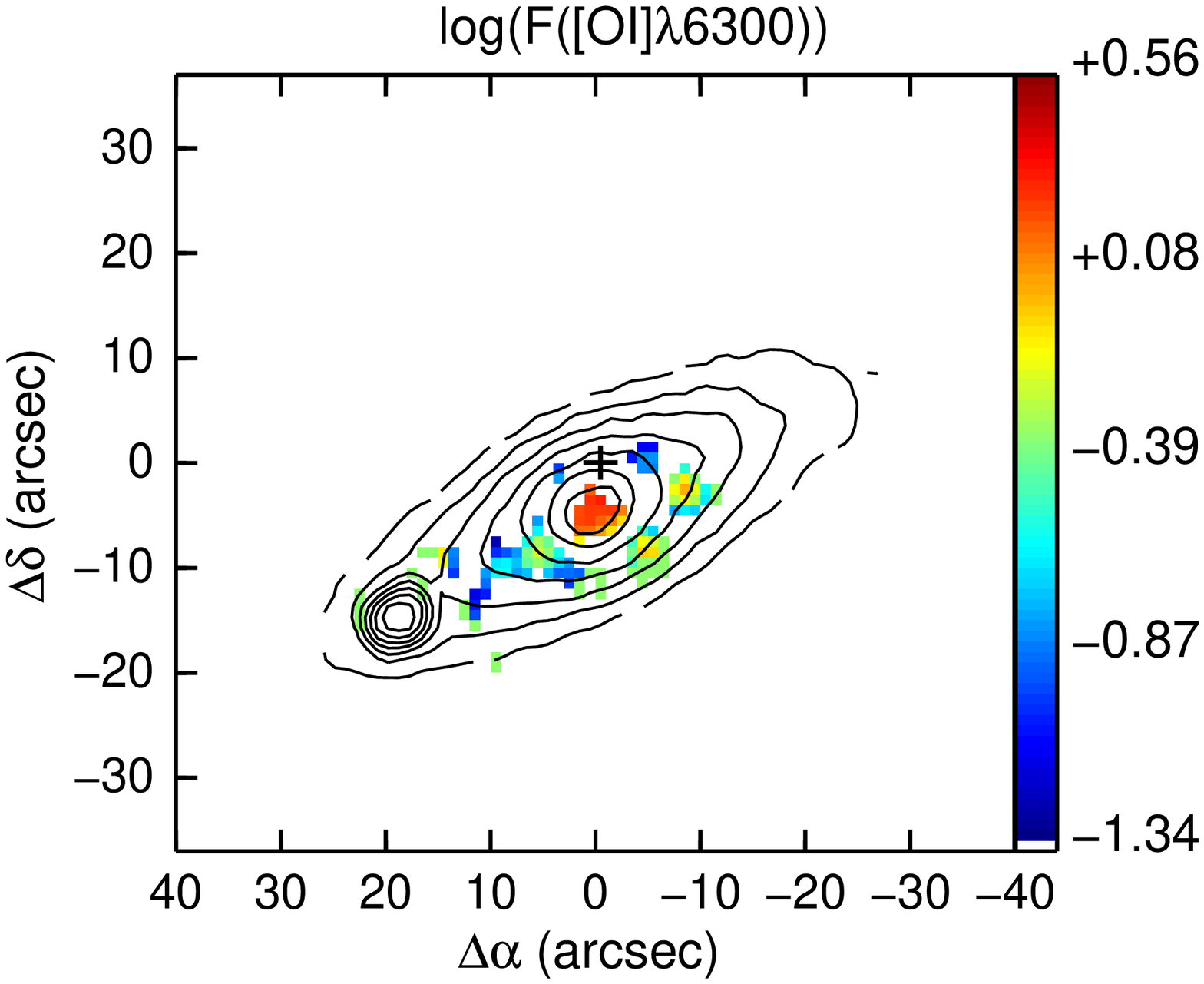}
\includegraphics[width=4.5cm]{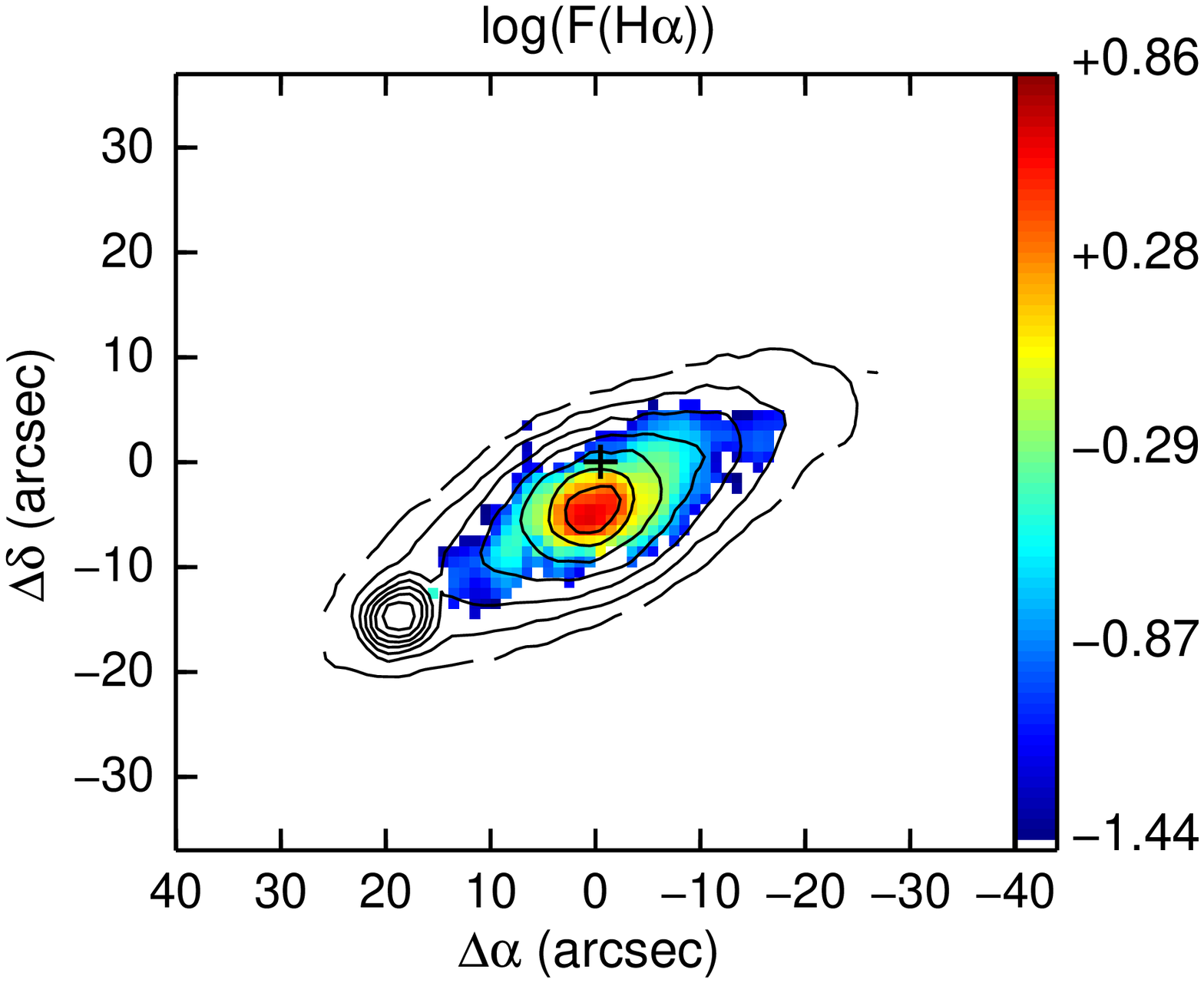}
\includegraphics[width=4.5cm]{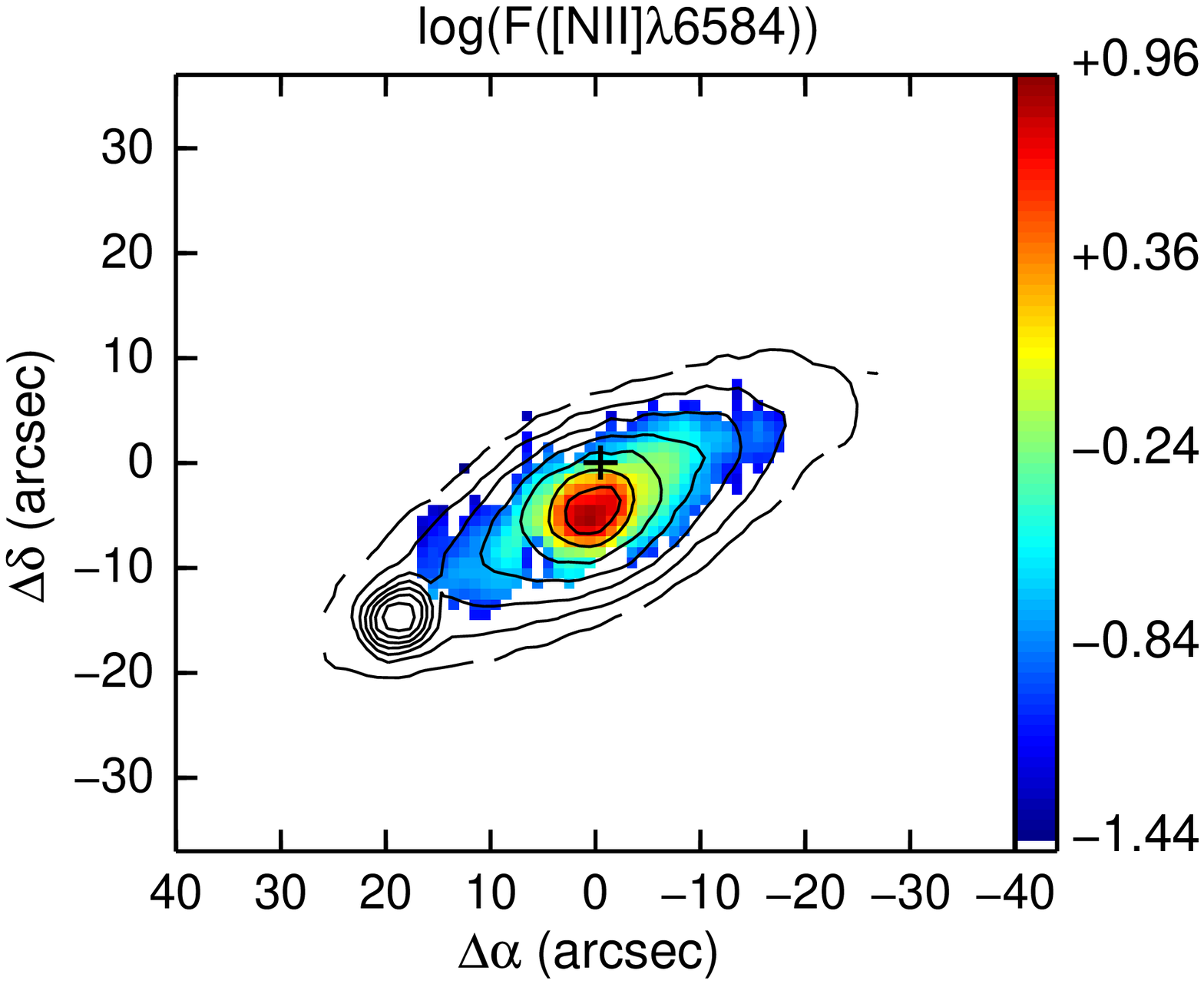}
\includegraphics[width=4.5cm]{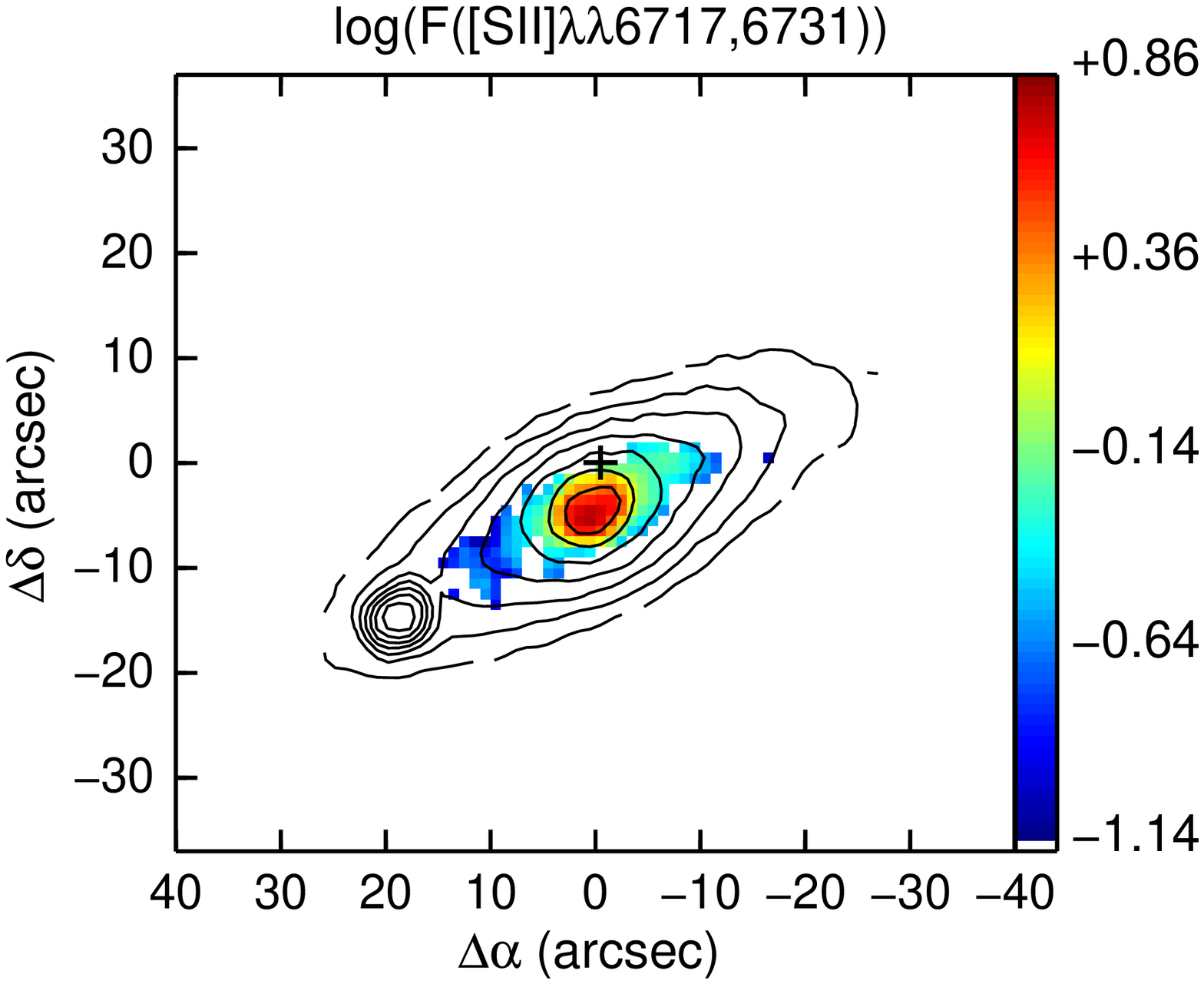} 
\caption{Maps of the emission from NGC~6762. Upper row from left to right: the stellar
component distribution as traced by a continuum map made from the median flux
between 6390-6490~\AA, a spectral region free of line emission ("pure
continuum"); the EW(H$\alpha$); H$\beta$; [O{\sc iii}]$\lambda$5007.    
Bottom row from left to right: [O{\sc i}]$\lambda$6300; H$\alpha$; [N{\sc ii}]$\lambda$6584; 
[S{\sc ii}]$\lambda$6717,6731. All maps are presented in logarithmic scale to emphasize the relevant morphological features. Flux units are 
10$^{-16}$ erg cm$^{-2}$ s$^{-1}$. The axis origin  (i.e. the center of the PPak bundle) is marked with a cross.  Contours corresponding to the stellar continuum map are overplotted on all maps
for reference. The contour corresponding to the minimun level is -1.70 dex
and the interval between contours is 0.25 dex. North
is up and east to the left. The pixel size is 1$\arcsec$ ($\sim$ 217 pc at our assumed distance to NGC~6762 of 45 Mpc). A bar showing the physical
scale of 5 kpc is plotted at the lower righthand corner in the continuum map.}
\label{ngc_fluxes}
\end{figure*}

\begin{figure*}
\includegraphics[width=4.5cm]{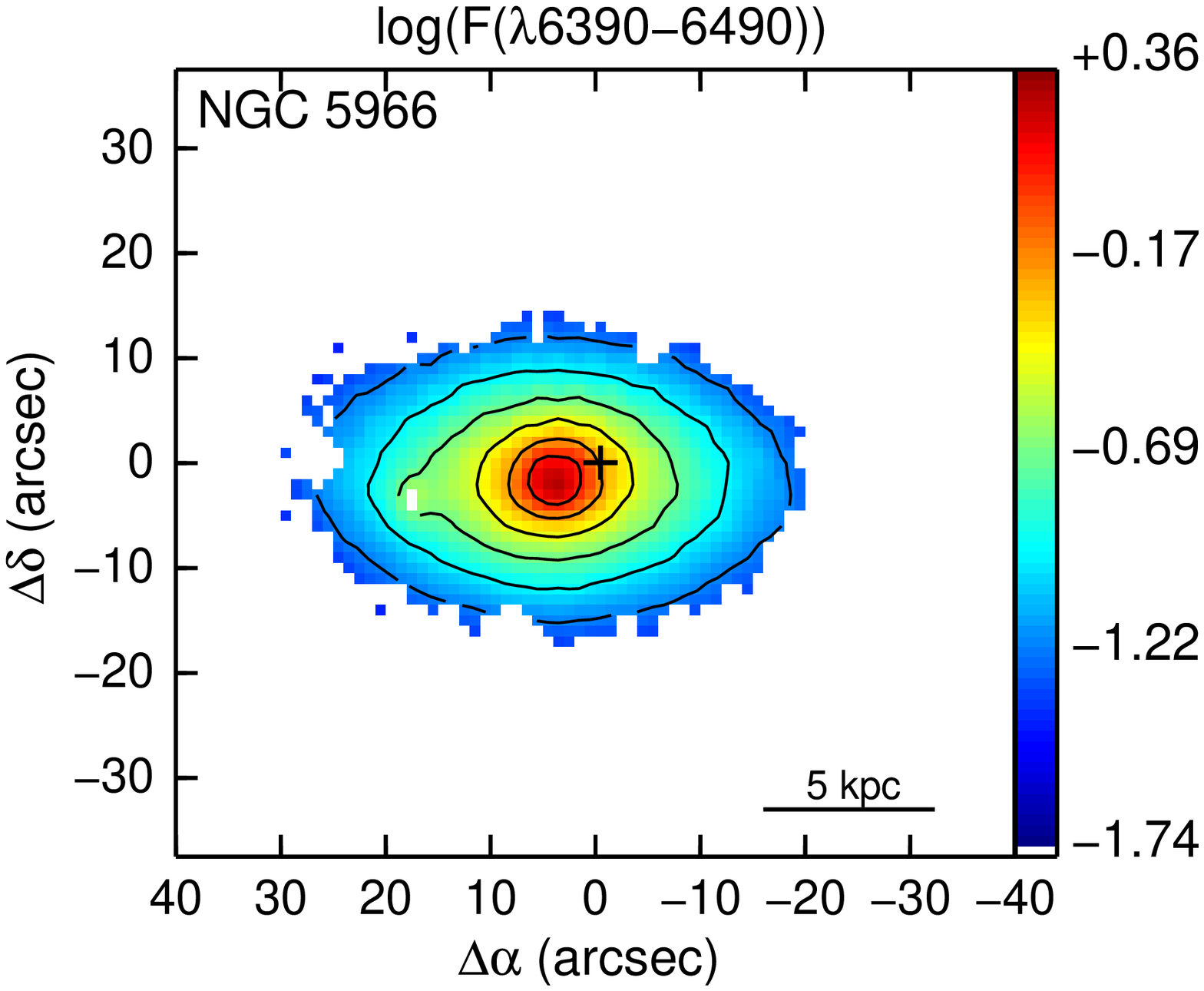}
\includegraphics[width=4.5cm]{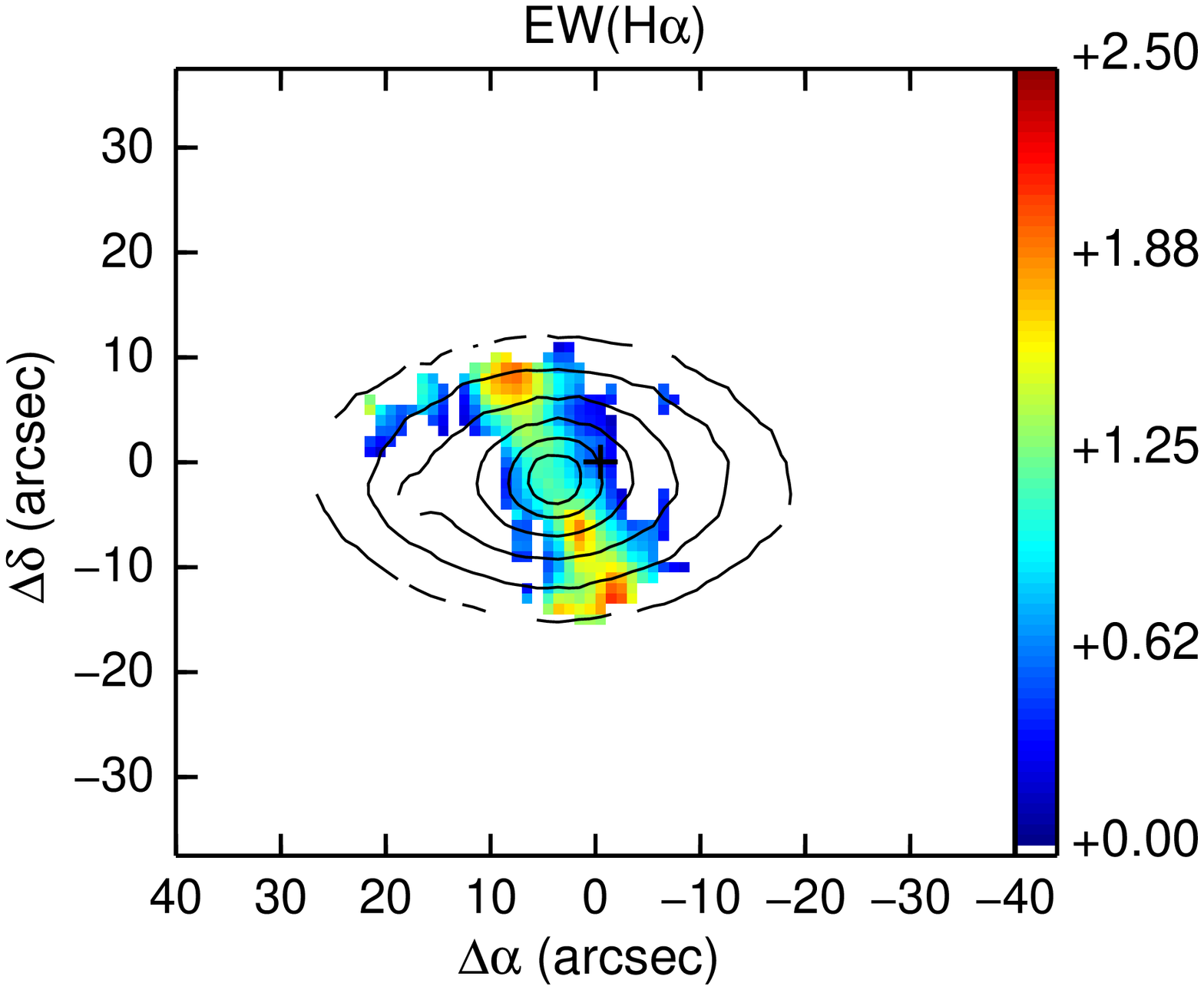}
\includegraphics[width=4.5cm]{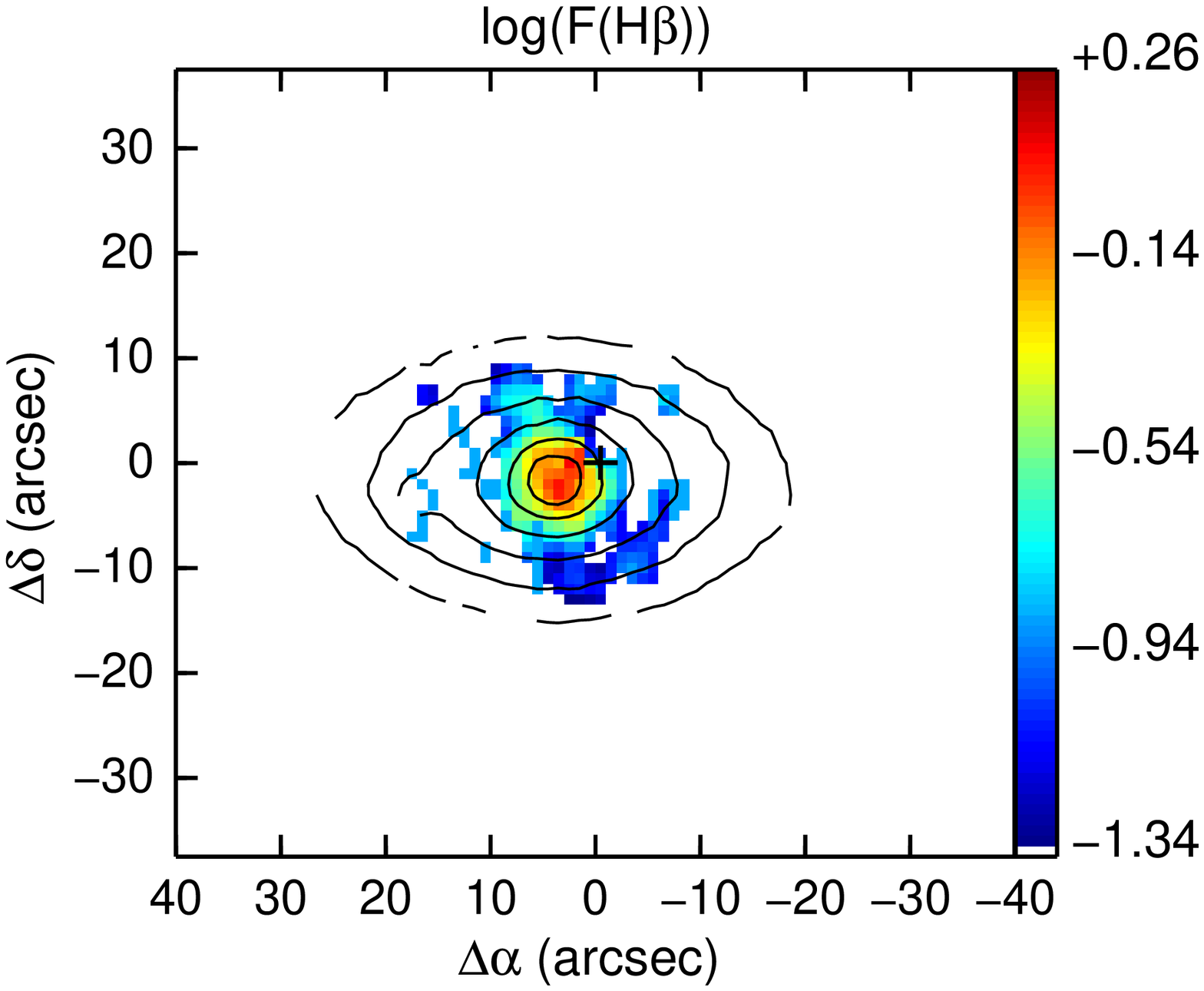}
\includegraphics[width=4.5cm]{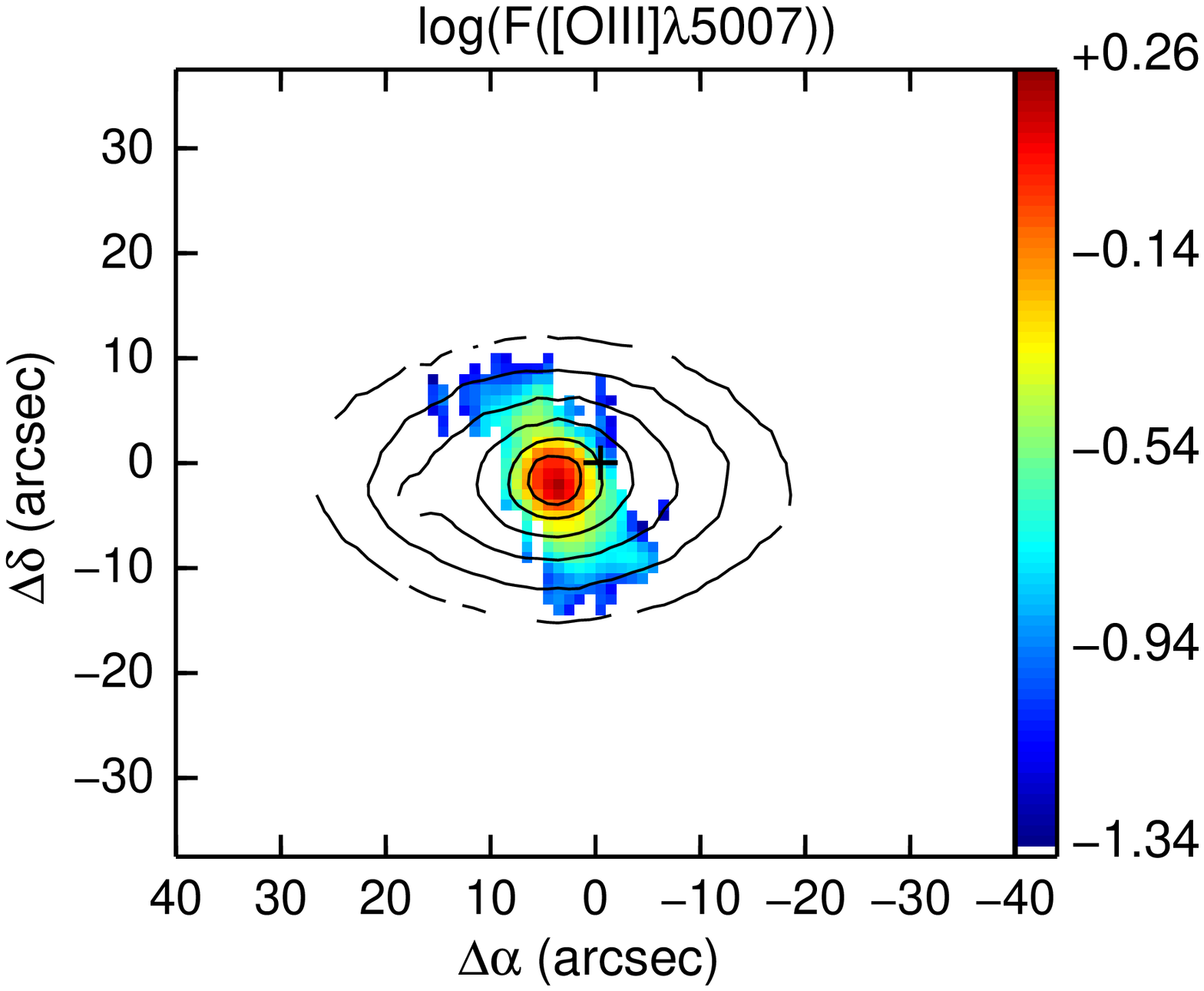}\\
\includegraphics[width=4.5cm]{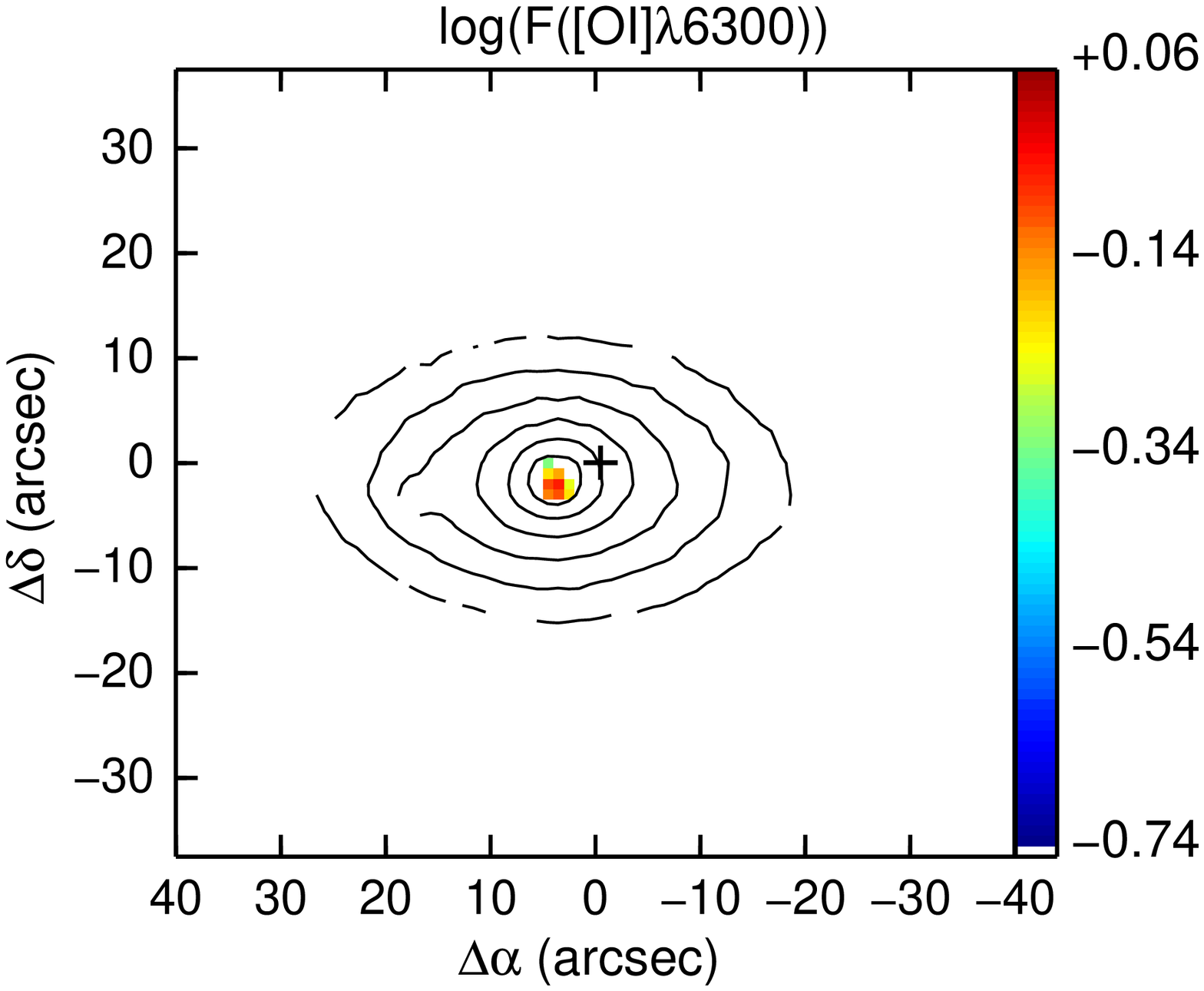}
\includegraphics[width=4.5cm]{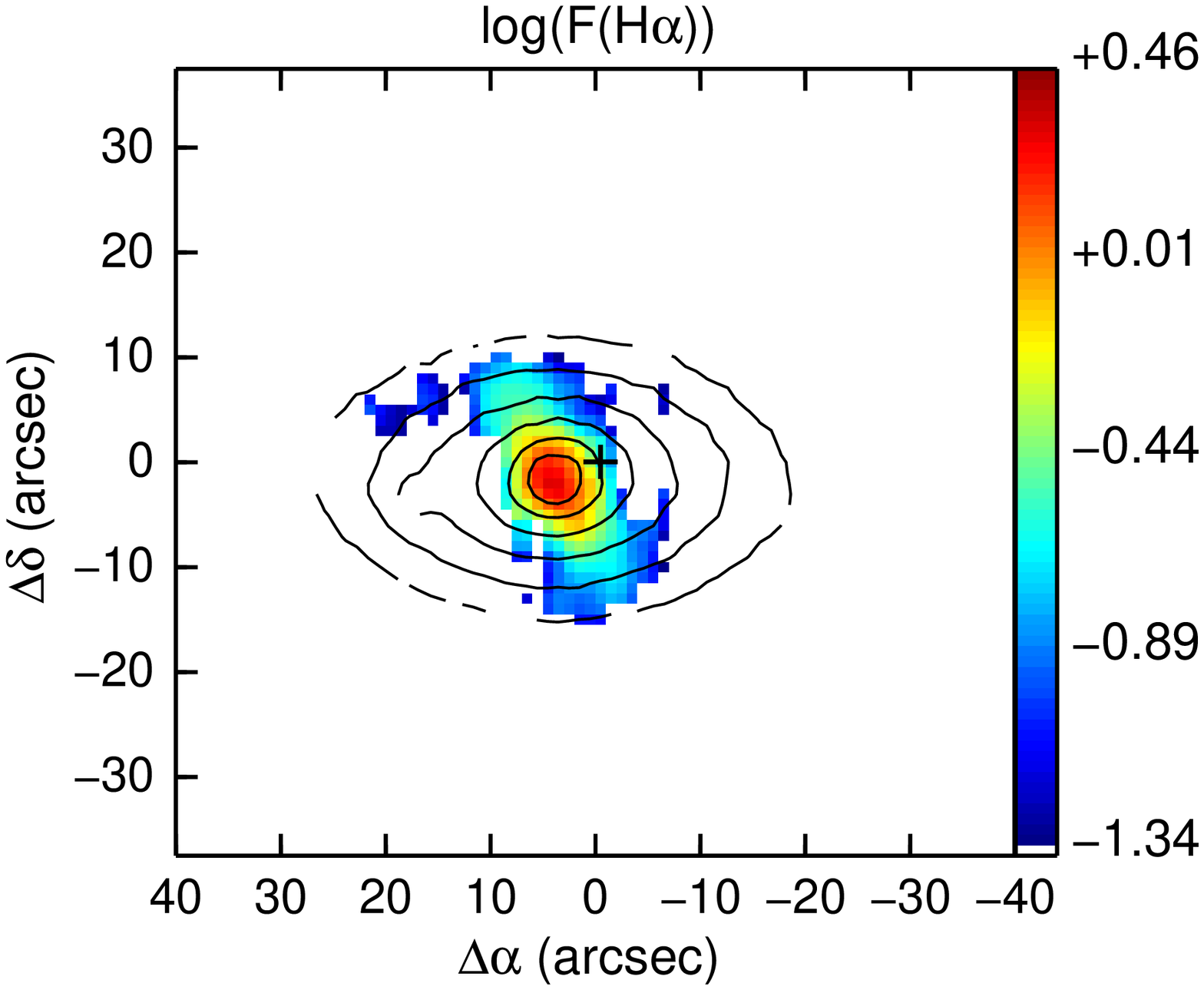}
\includegraphics[width=4.5cm]{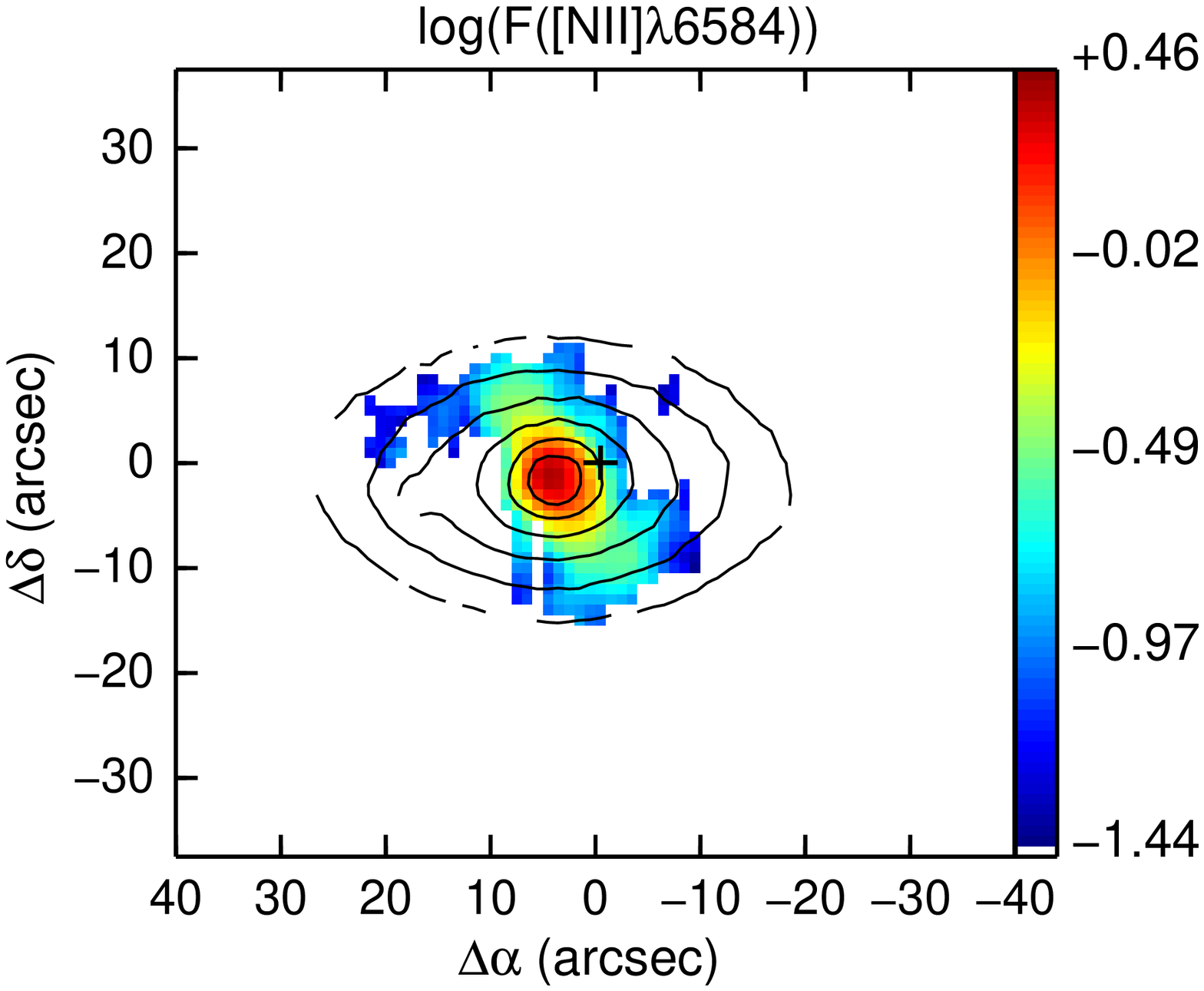}
\includegraphics[width=4.5cm]{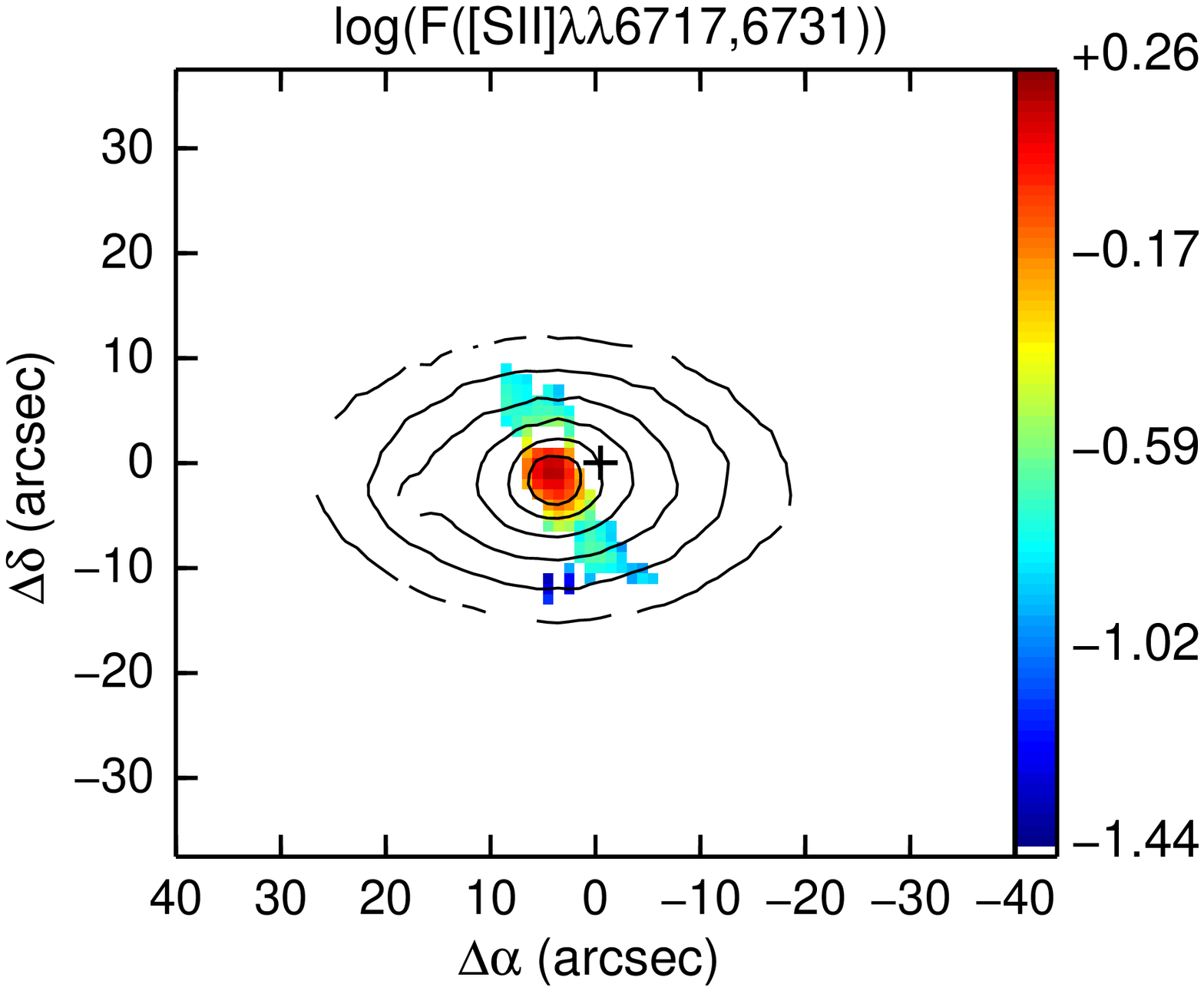}
\caption{Maps of the emission from NGC~5966. The panels show the same
maps as outlined in Fig.~\ref{ngc_fluxes}. See the caption to that
figure for details. The pixel size (1$\arcsec$) corresponds to $\sim$ 334 
pc at our assumed distance to NGC~5966 of 69 Mpc.}
\label{ngc_5966_fluxes}
\end{figure*}

We constructed continuum and flux maps of the relevant emission lines
in NGC~6762 and NGC~5966 (Figs.~\ref{ngc_fluxes}
and \ref{ngc_5966_fluxes}). Not all emission-line maps display the
same area. The H$\alpha$ and [N{\sc ii}]$\lambda$6584 maps show a
larger area than the [O{\sc i}]$\lambda$6300 maps, for example,
because those lines are the brightest optical emission lines in our
data sets.  It is important to note that the quality criteria outlined
in Sec.~\ref{emlin} are relatively conservative. There are regions of
emission at greater distances from the nucleus than what is shown in
Figs.~\ref{ngc_fluxes} and \ref{ngc_5966_fluxes}, particularly in
H$\alpha$, [N{\sc ii}]$\lambda$6584, and [S{\sc
ii}]$\lambda\lambda$6717,6731. However, in such regions, the residuals
due to the subtraction of the continuum are about the same order as
the flux of the lines, so the line flux estimates are unreliable.

Figures~\ref{ngc_fluxes} and \ref{ngc_5966_fluxes} also present the
distribution of the H$\alpha$ equivalent width, EW(H$\alpha$), defined
as the ratio between the H$\alpha$ line flux and the neighboring,
line-free continuum flux ($\sim$ 6390 - 6490 \AA). This was
measured in the observed spectra. The values of EW(H$\alpha$) are
very low ($\leq$ 3 \AA) in comparison to values measured in
star-forming galaxies, indicating that the stellar continuum dominates
the overall emission in our galaxies. Our EW(H$\alpha$) measurements
are within the range of values derived by \citet{ani10} for their
sample of ETGs.


The emission line morphology of NGC~6762 is disk-like and similar to
the continuum emission.  However, this is not the case for NGC~5966.
The emission line maps of NGC~5966 show an elongated structure (along
an axis at P.A. $\sim$ 30$^{\circ}$) extending out to at least
$10^{\prime\prime}$ ($\sim$ 3 kpc) on either side of the nucleus,
while the stellar continuum map displays an elliptical shape. The
difference or similarity of the line and continuum emission is
interesting and provides important clues to the nature of the gas
that we discuss in Sect.~\ref{discussion}.

\subsection{The physical conditions in the warm ISM}

The integral field spectra allow us to {\it spatially probe the relative role of the
various sources of ionization that could be responsible for the nebular
emission observed in early-type galaxies}. In this section we present the
radial and 2D spatial distribution of diagnostic emission-line ratios used to distinguish between different excitation mechanisms, and compare our
measurements with those predicted by ionization models available in the
literature.
 
\subsubsection{Spatial distribution of diagnostic line ratios}\label{spatial_line_ratios}

\begin{figure*}
\center
\includegraphics[width=4.5cm]{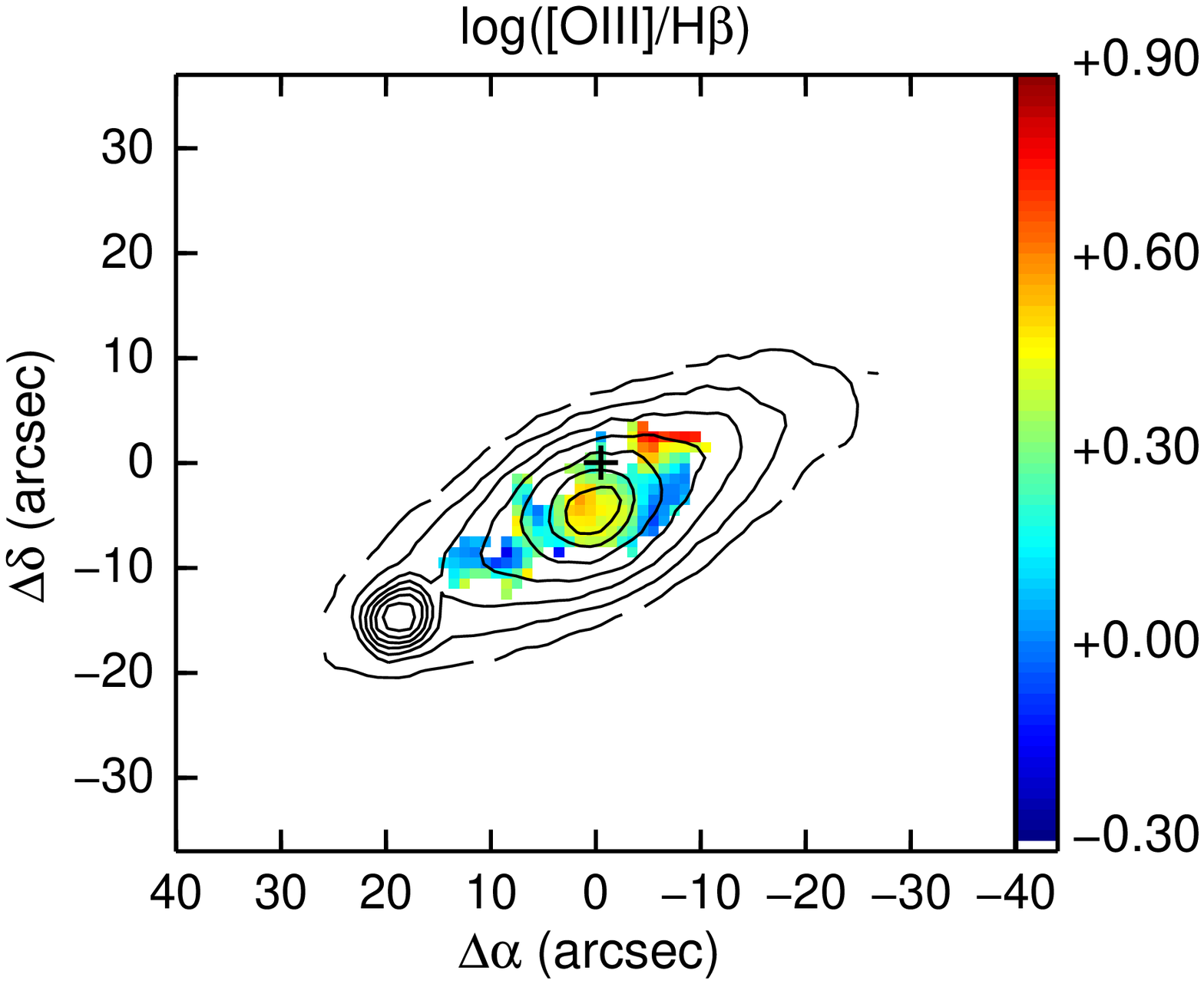}
\includegraphics[width=4.5cm]{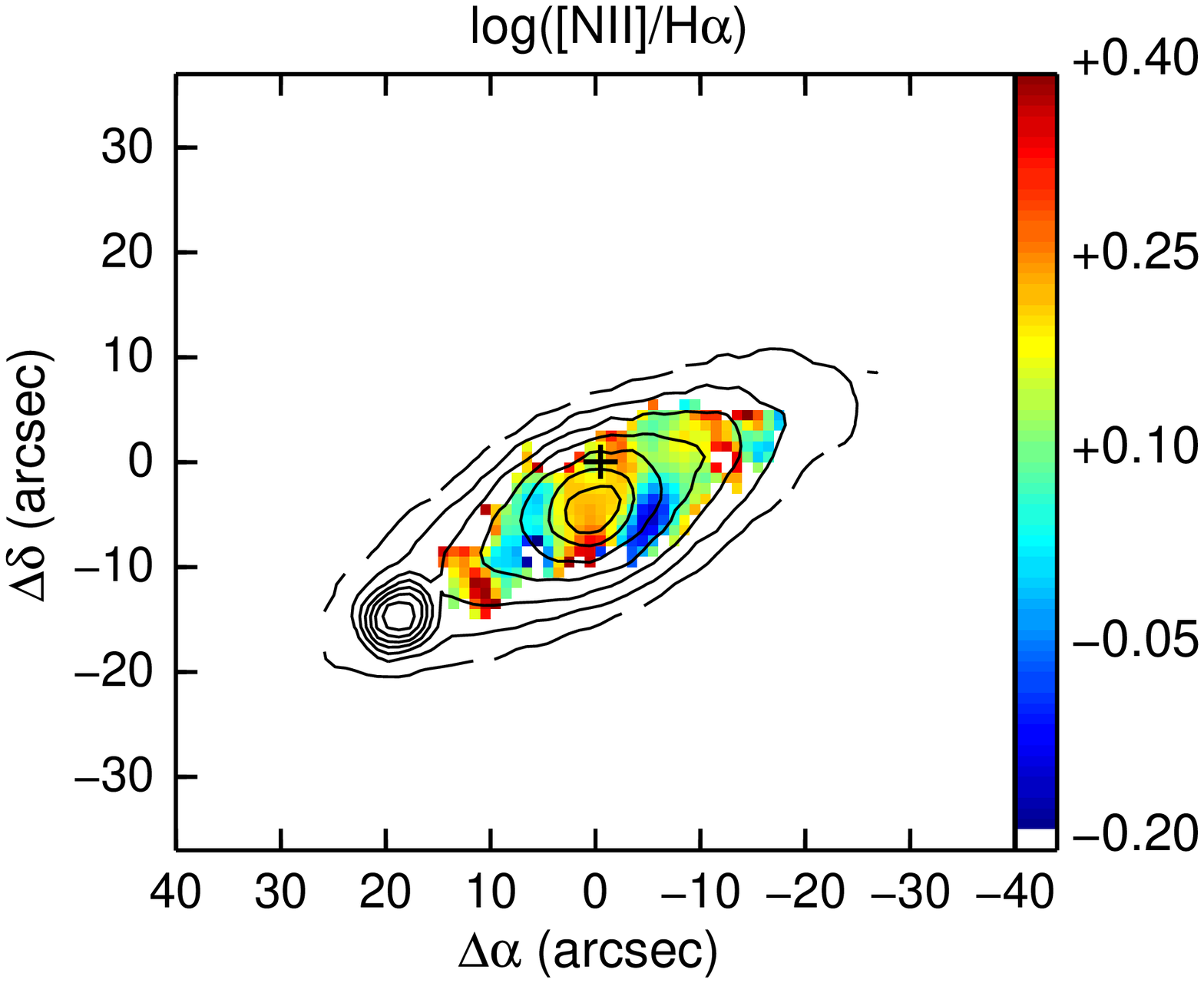}
\includegraphics[width=4.5cm]{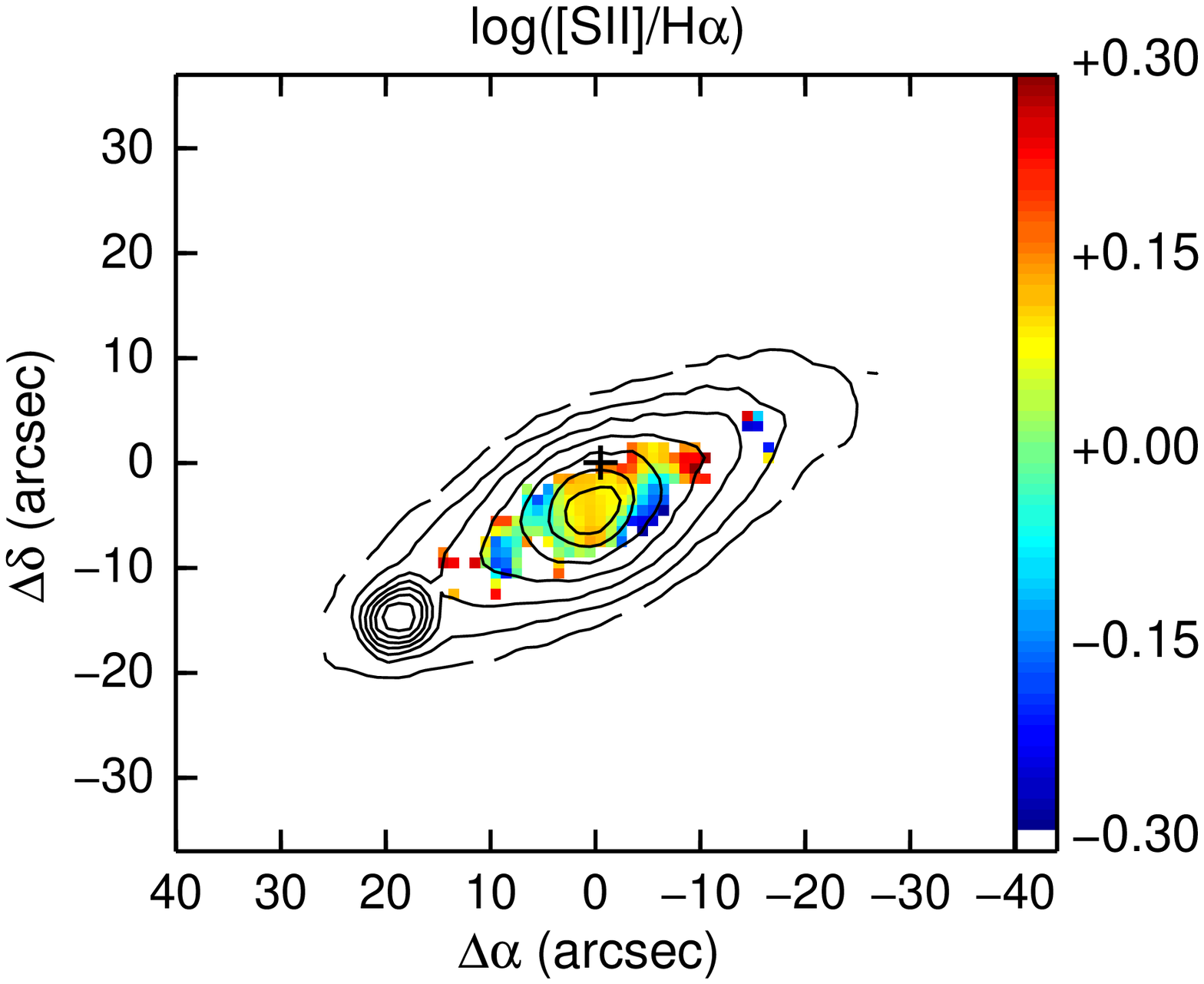}
\includegraphics[width=4.5cm]{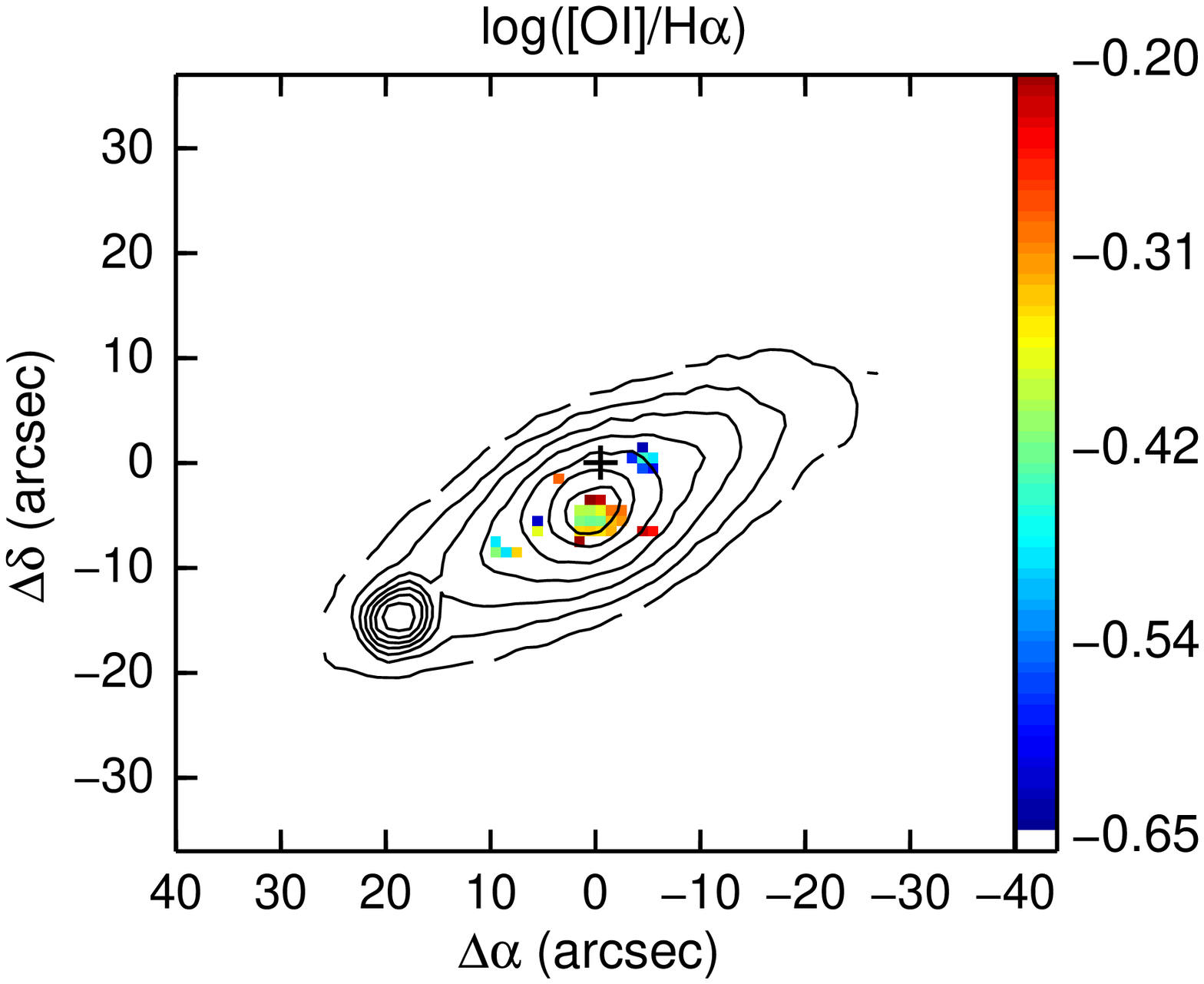}\\
\includegraphics[width=4.5cm]{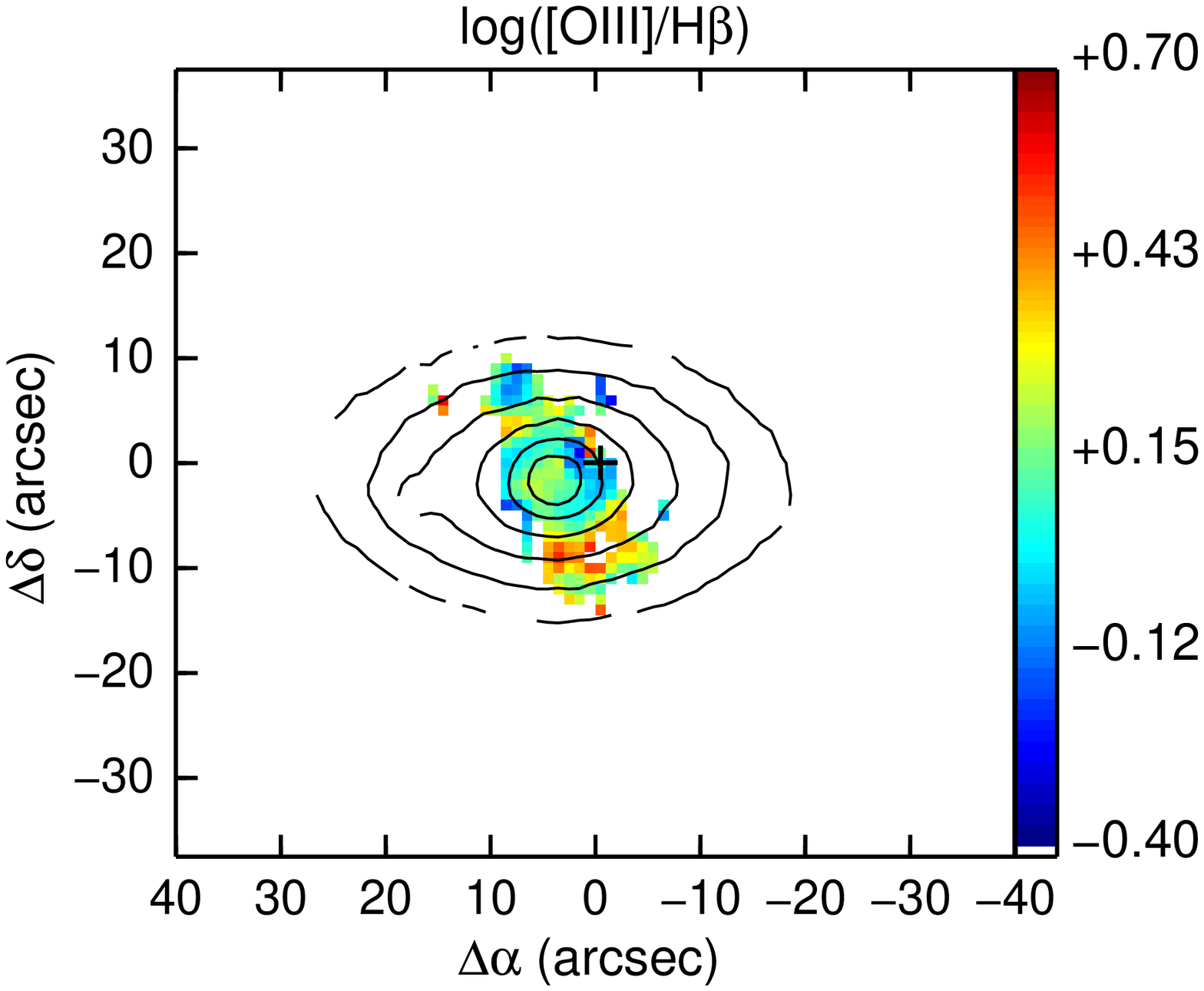}
\includegraphics[width=4.5cm]{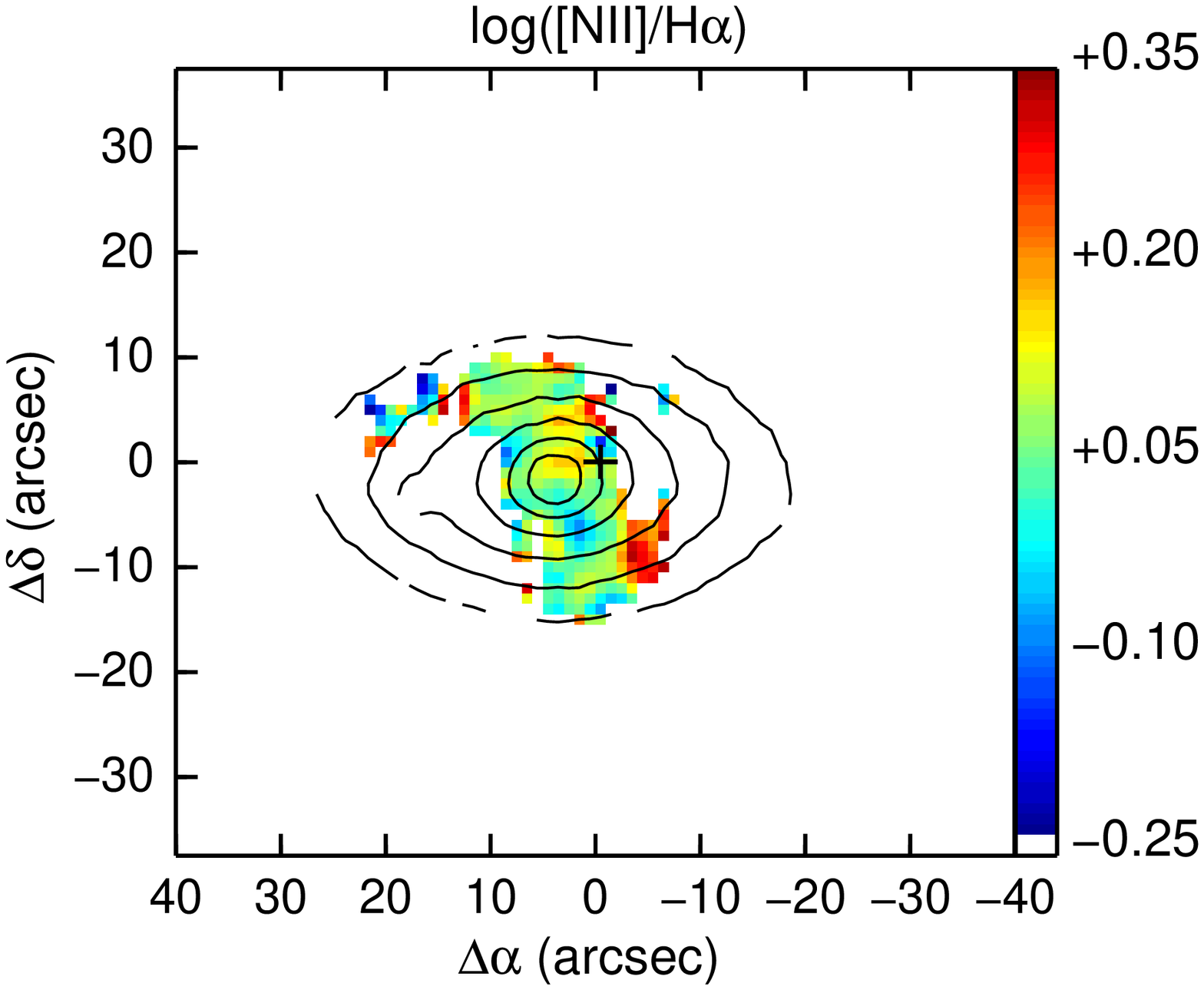}
\includegraphics[width=4.5cm]{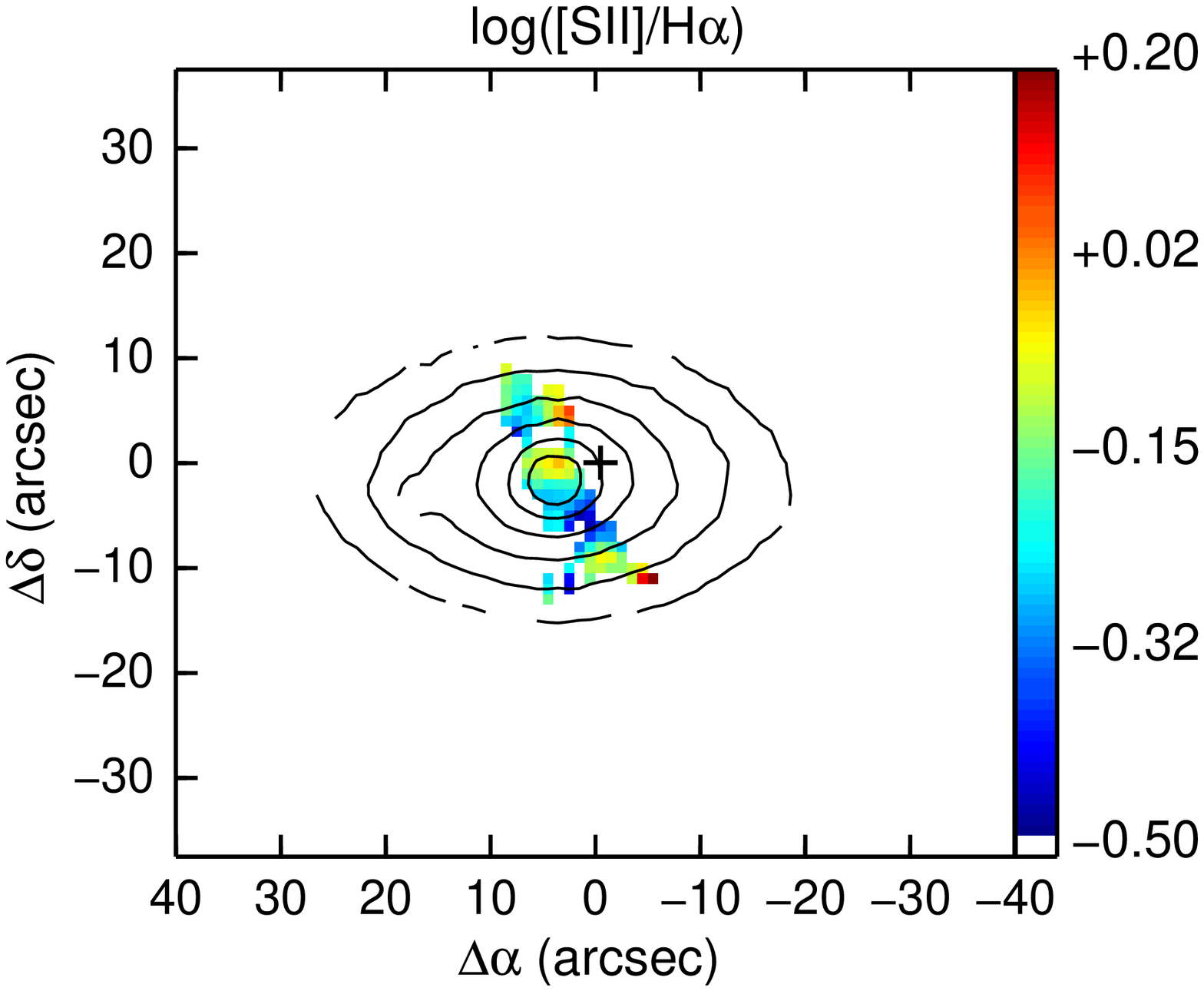}
\includegraphics[width=4.5cm]{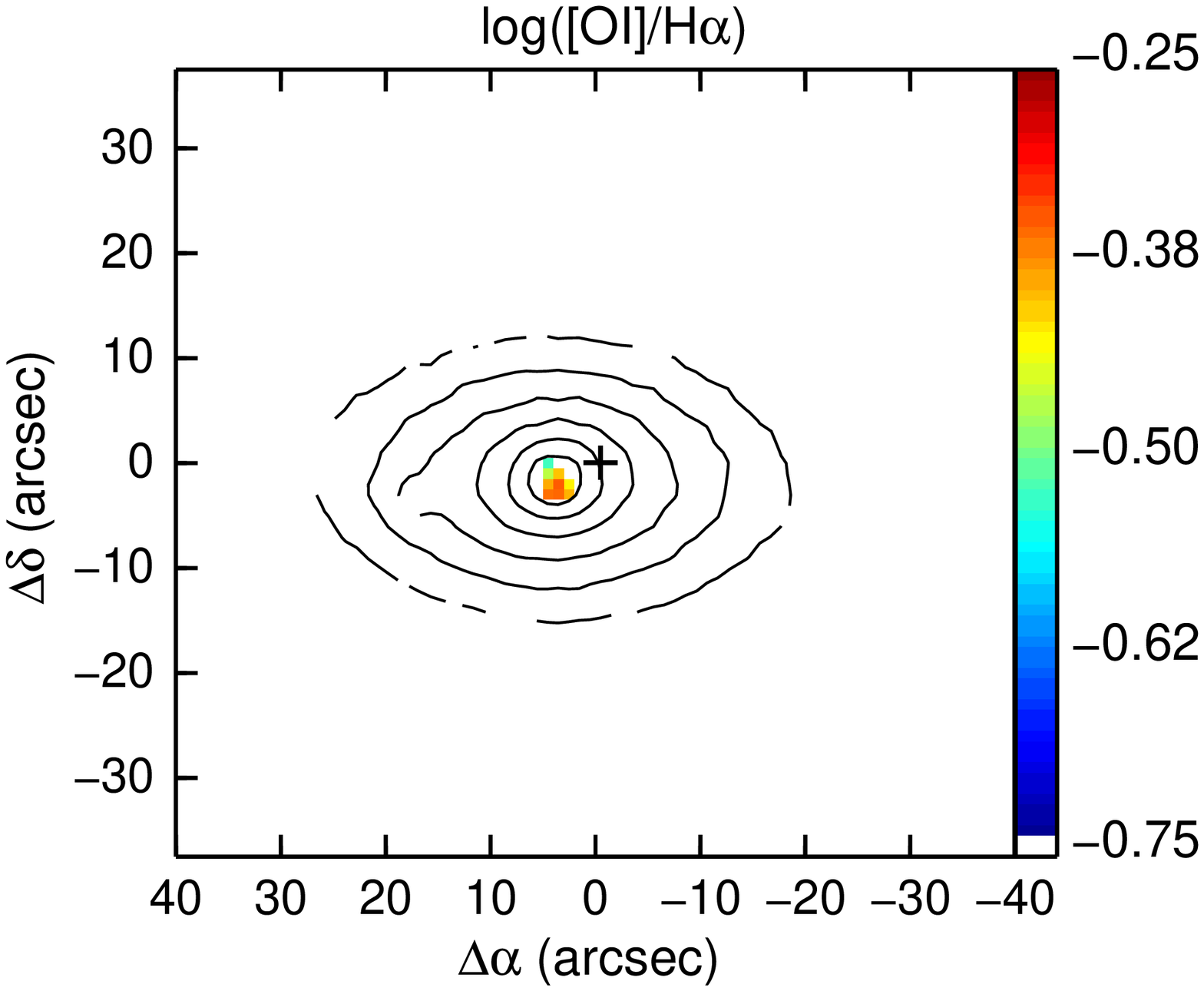}
\caption{Emission line ratio maps for NGC~6762 (top row) and NGC~5966 (bottom row). All maps are in logarithmic
scale. The label at the top of each map indicates what emission line
ratios are being displayed. The contours, linear scale, and orientation are the same as in Figs.~\ref{ngc_fluxes} and \ref{ngc_5966_fluxes}.}
\label{ngc_elr}
\end{figure*}

The [O{\sc iii}]$\lambda$5007/H$\beta$, [N{\sc ii}]$\lambda$6584/H$\alpha$,
[S{\sc ii}]$\lambda$$\lambda$6717,6731/H$\alpha$, and [O{\sc i}]$\lambda$6300/H$\alpha$ line ratio maps for NGC~6762 and NGC~5966 are displayed in
Fig.~\ref{ngc_elr}.  For each galaxy, all the excitation maps display similar morphology, with relatively small spaxel-to-spaxel
variations. 

\begin{figure*}
\includegraphics[width=9cm,clip]{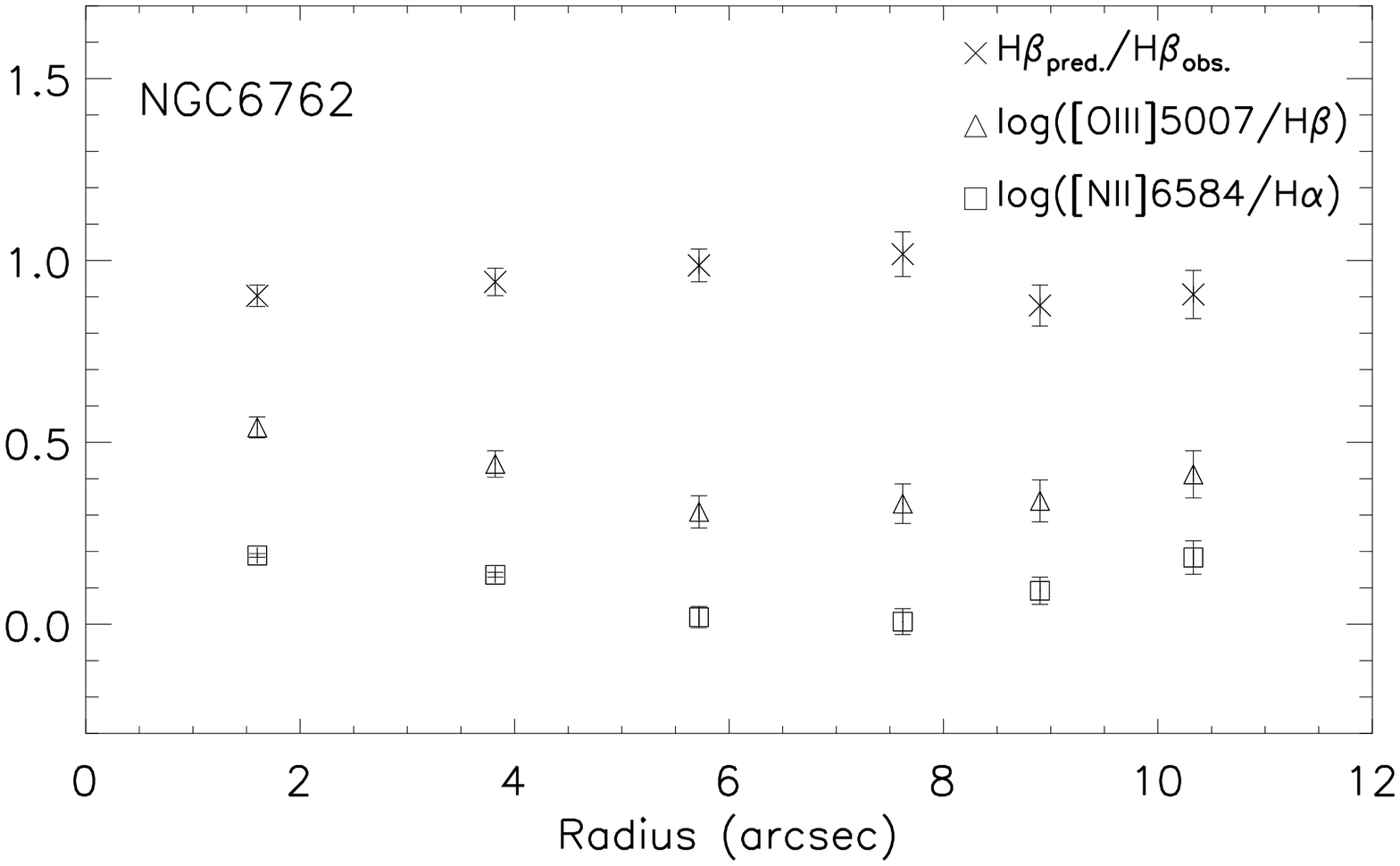}
\includegraphics[width=9cm,clip]{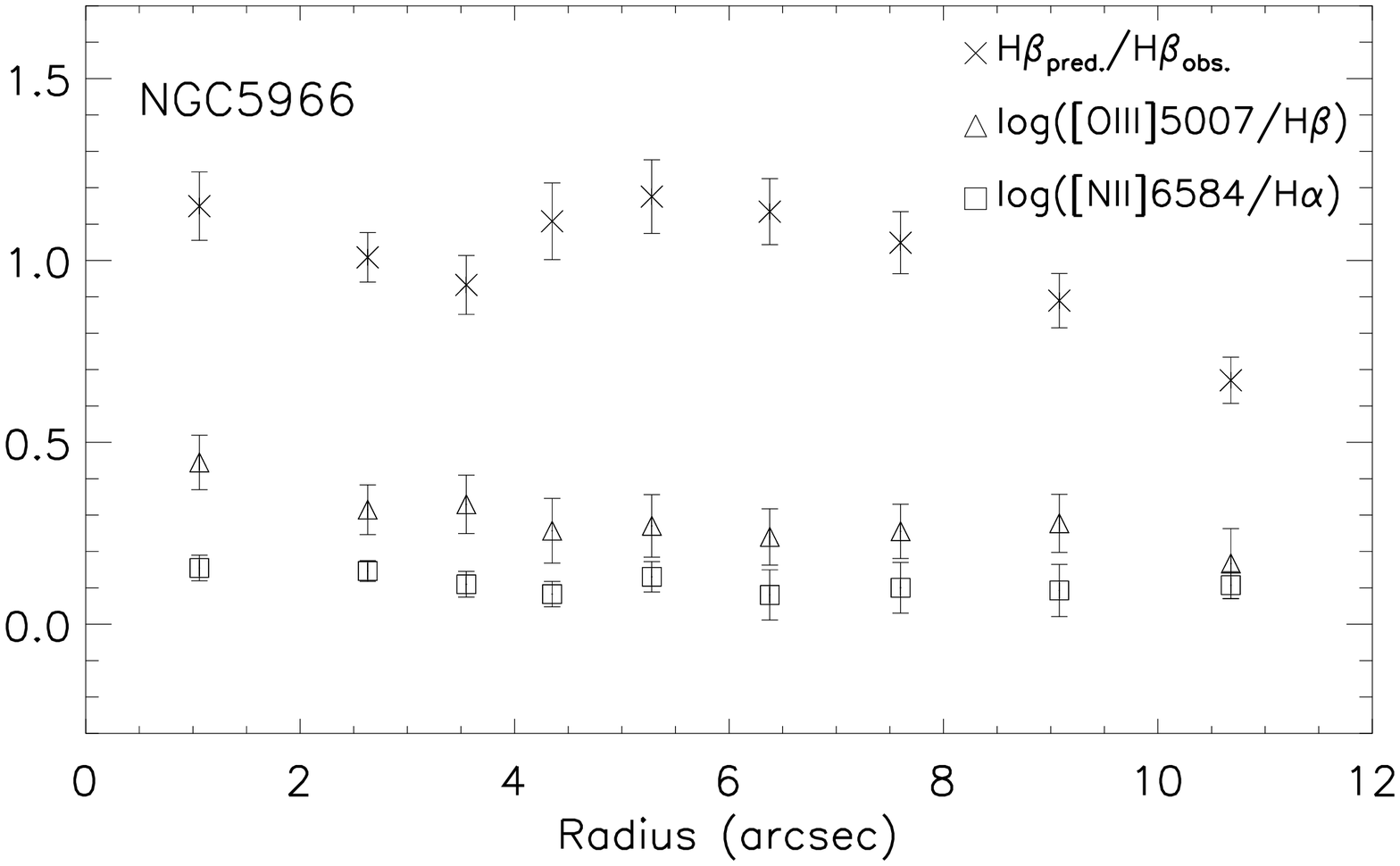}
\caption{Radial distribution of the [O{\sc iii}]$\lambda$5007/H$\beta$ (triangle),
[N{\sc ii}]$\lambda$6584/H$\alpha$ (square), and H$\beta_{pred.}$/H$\beta_{obs.}$ (cross; this ratio is discussed in Sect. 7)  for
NGC~6762 (left panel) and NGC~5966 (right panel).
}
\label{zones} 
\end{figure*}

The radial profiles of the diagnostic line ratios provide constraints on the nature of the warm ionized medium in ETGs. We calculated the values of [O{\sc iii}]$\lambda$5007/H$\beta$ and
[N{\sc ii}]$\lambda$6584/H$\alpha$ within different annuli (computed as described in Appendix A) and plotted them as a
function of the photometric radius \rr, for both galaxies (Fig.~\ref{zones}). Emission-line fluxes are measured in the integrated pure emission line spectrum in each annulus using the
IRAF\footnote{IRAF is distributed by the National Optical Astronomical
Observatories, which are operated by the Association of Universities for
Research in Astronomy, Inc., under cooperative agreement with the National
Science Foundation.} task splot. The derived line fluxes were computed by fitting a Gaussian
to each line. The line-flux errors are calculated using the expression by
\citet{cas02}, 
\begin{equation}  
\sigma_{line}=\sigma_{cont}N^{1/2}\left(1 + \frac{\rm EW}{N\Delta\lambda}\right)^{1/2} 
\end{equation} 
where $\sigma_{cont}$ is the standard deviation of the continuum near the 
emission line, $N$ the width of the region used to measure the lines in 
pixels, $\Delta\lambda$ the spectral dispersion in \AA~pix$^{-1}$, and EW 
represents the EW of the line in \AA. The EW values
are obtained from the ratio between the line flux
(measured in the pure emission line spectra) and the corresponding adjacent continuum flux (measured in the observed spectra). Since we are dealing with very
faint emission with EW values $<$ 3 \AA~(EW/N$\Delta$$\lambda$ $\ll$  1; see Figs.~\ref{ngc_ew_bpt} and ~\ref{ngc_5966_ew_bpt}), we
can neglect the addendum that involves the EW in the equation above. 

Figure~\ref{zones} shows that no significant radial trend is apparent in
both [O{\sc iii}]$\lambda$5007/H$\beta$ and [N{\sc
ii}]$\lambda$6584/H$\alpha$, well outside the nucleus with merely a
weak tendency for decreasing values in NGC~6762. This suggests that
there is no significant variation in nebular properties within the two
objects and that the dominant ionization source is not confined to the
nucleus since a decrease in excitation is expected for central source
photoionization \citep[e.g.][]{rob94,whi05}.

The majority of the spaxels in the cubes of NGC~6762 and NGC~5966 are
characterized by log([O{\sc iii}]$\lambda$5007/H$\beta$) $\leq$ 0.5, indicating relatively low excitation (Figs.~\ref{ngc_elr}
and \ref{zones}). \citet{sar10} find that 75$\%$ of their sample of
ETGs show log([O{\sc iii}]$\lambda$5007/H$\beta$) in the range
0.0-0.5. High values of [N{\sc ii}]$\lambda$6584/H$\alpha$, [S{\sc ii}]$\lambda$$\lambda$6717,6731/H$\alpha$, and [O{\sc i}]$\lambda$6300/H$\alpha$ are found in both galaxies, unlike in
star-forming galaxies \citep[e.g.][]{alo10}. In our galaxies,
generally [N{\sc ii}]$\lambda$6584 is brighter than H$\alpha$ ([N{\sc
ii}]$\lambda$6584/H$\alpha$ $>$ 1.00 for most of spaxels), and [O{\sc
i}]$\lambda$6300/H$\alpha$ can be as high as $\sim$ 0.40.

\subsubsection{Diagnostic diagrams}\label{BPT}

\begin{figure*}
\center
\includegraphics[width=0.95\textwidth,clip]{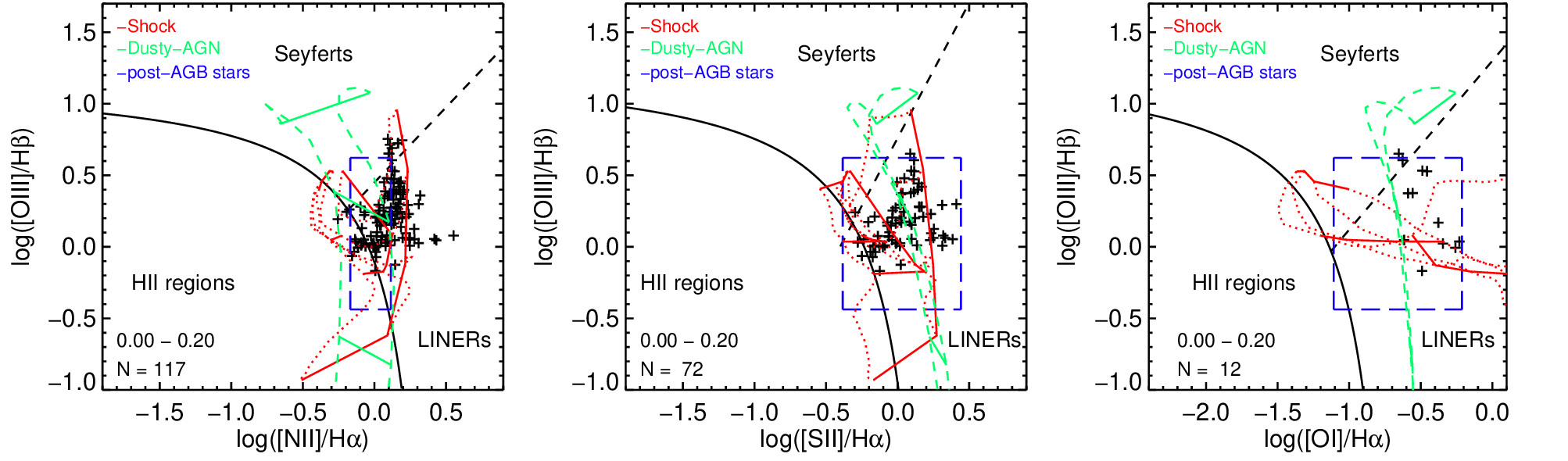}\\
\includegraphics[width=0.95\textwidth,clip]{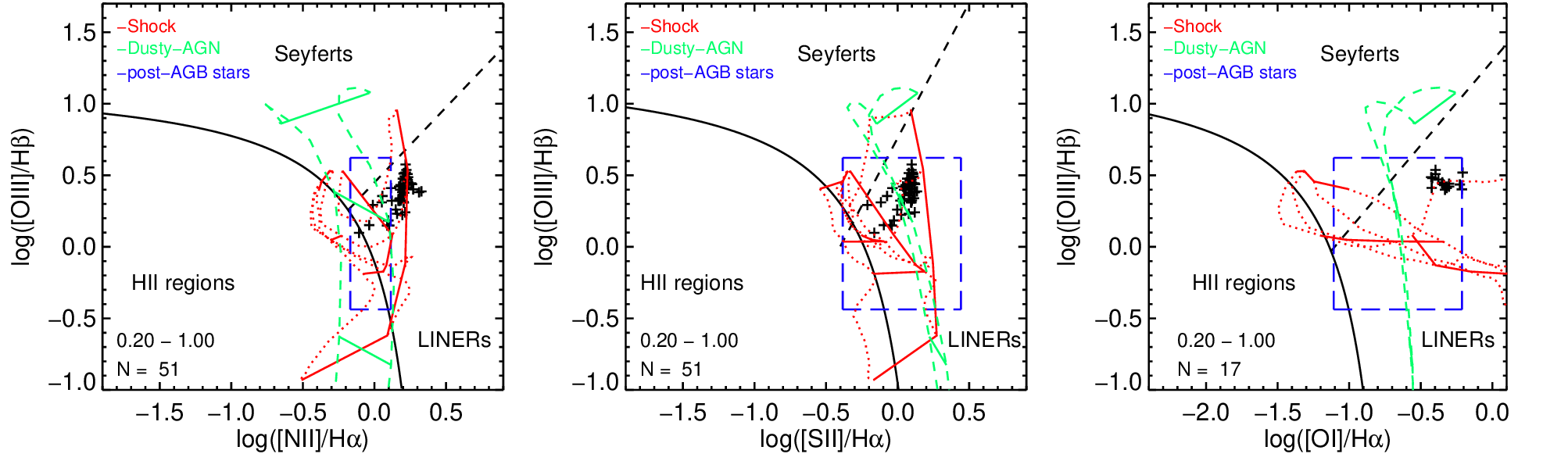}
\caption{Diagnostic diagrams for NGC~6762. From left to right: log ([O{\sc iii}]$\lambda$5007/H$\beta$) vs. log ([N{\sc ii}]$\lambda$6584/H$\alpha$), log ([O{\sc iii}]$\lambda$5007/H$\beta$) vs. log ([S{\sc ii}]$\lambda$6731,6717/H$\alpha$) and log ([O{\sc iii}]$\lambda$5007/H$\beta$) vs. log ([O{\sc i}]$\lambda$6300/H$\alpha$). The upper and bottom rows show the spaxels with F(H$\alpha$)/F(H$\alpha$)$_{peak}$ $\leq$ 0.20 and 0.20 $\lesssim$ F(H$\alpha$)/F(H$\alpha$)$_{peak}$ $\lesssim$ 1.00, respectively. 
The black solid curve (in all three panels) is the theoretical maximum
starburst model from \citet{kew01a}, devised to isolate objects whose emission line
ratios can be accounted for by the photoionization by massive stars (below
and to the left of the curve) from those where some other source of
ionization is required.. The black-dashed curves in the
[S{\sc ii}]$\lambda$6731,6717/H$\alpha$ and [O{\sc
i}]$\lambda$6300/H$\alpha$ diagrams represent the Seyfert-LINER dividing line
from \citet{kew06a} and transposed to the [N{\sc ii}]$\lambda$6584/H$\alpha$ by \citet{sch07}. The predictions of different ionization models for ionizing the gas are overplotted in each diagram.
The red lines represent the shock grids of \citet{all08} with solar
metallicity and preshock magnetic field {\it B}=1.00 $\mu$G. For the grid of shock models,
the solid lines show models with increasing shock velocity V$_{s}$ = 100, 200,
300, 1000 km s$^{-1}$, and dotted lines the grids with densities $n_{e}$ from
0.1 cm$^{-3}$ to 100 cm$^{-3}$. Grids of photoionization by an AGN \citep{gro04} are indicated by green curves, with $n_{e}$ = 100 cm$^{-3}$ and a
power-law spectral index of $-$1.4. The corresponding dashed-lines show models
for $Z=Z_{\odot}$ and $Z=2Z_{\odot}$ (from left to right), and solid lines
trace the ionization parameter log {\it U}, which increases with the
[O{\sc iii}]$\lambda$5007/H$\beta$ ratio from log {\it U} =  -3.6, -3.0, 0.0. We downloaded the
shock and AGN grids from the web page http://www.strw.leidenuniv.nl/$\sim$brent/itera.html. The boxes show the predictions of photoionization models by pAGB stars for $Z=Z_{\odot}$ and a burst age of 13 Gyr \citep{bin94}.}
\label{ngc_bpt}
\end{figure*}

\begin{figure*}
\center
\includegraphics[width=0.95\textwidth,clip]{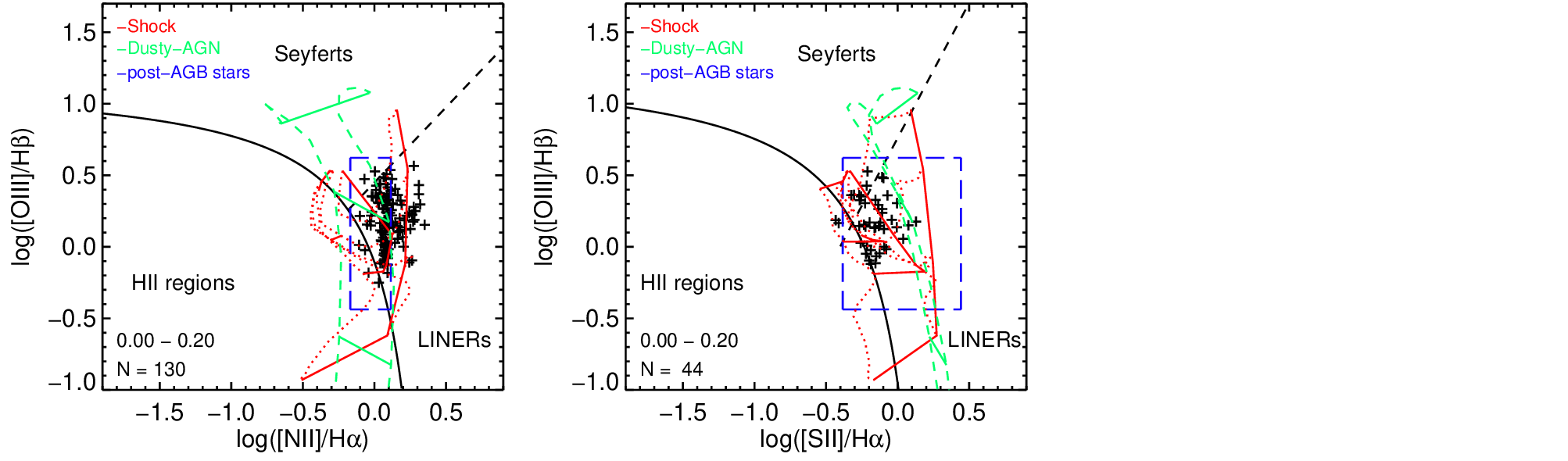}\\
\includegraphics[width=0.95\textwidth,clip]{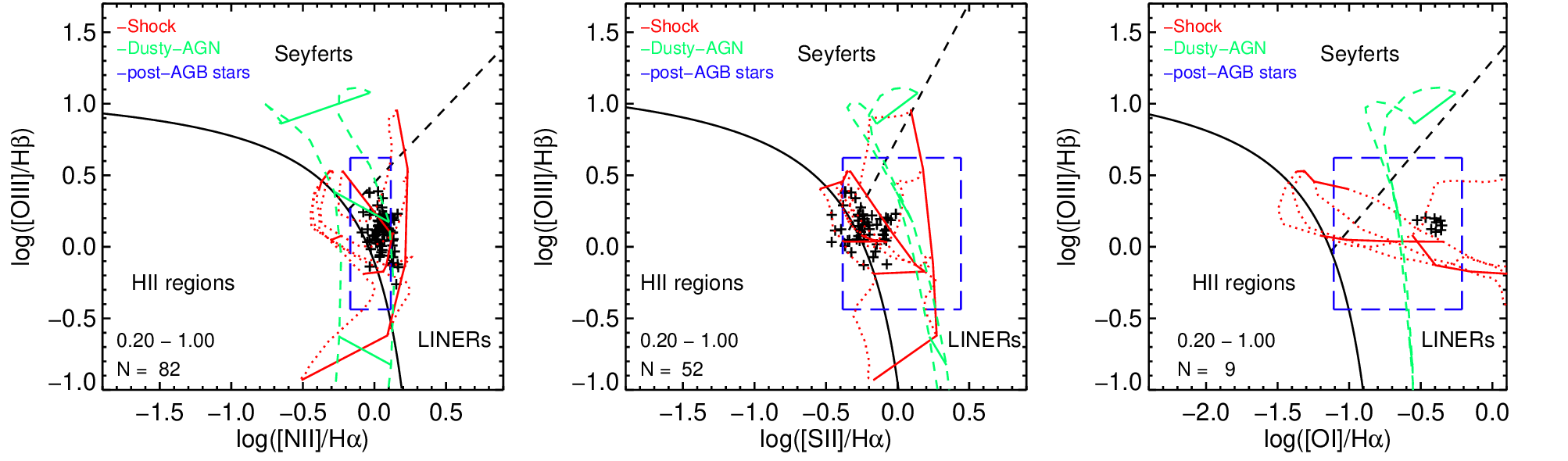}
\caption{Diagnostic diagrams for NGC~5966. Curves as in Fig.~\ref{ngc_bpt}.}
\label{ngc_5966_bpt}
\end{figure*}

The standard diagnostic diagrams \citep[][hereafter BPT]{bal81} are
widely used to probe the dominant ionizing source in
galaxies \citep[e.g.][]{kew06a,keh08,mon10b}. The BPT diagrams, on a
spaxel-by-spaxel basis, for NGC~6762 and NGC~5966 are shown in
Figs.~\ref{ngc_bpt} and \ref{ngc_5966_bpt}, respectively: [O{\sc
iii}]$\lambda$5007/H$\beta$ vs [N{\sc ii}]$\lambda$6584/H$\alpha$
(left panels), [O{\sc iii}]$\lambda$5007/H$\beta$ vs [S{\sc
ii}]$\lambda$$\lambda$6717,6731/H$\alpha$ (middle panels), and [O{\sc
iii}]$\lambda$5007/H$\beta$ vs [O{\sc i}]$\lambda$6300/H$\alpha$
(right panels). The line ratios are not corrected for reddening, but
their reddening dependence is negligible since they are calculated
from lines which are close to each other in wavelength space. As a
guide to the reader, the data points with highest F(H$\alpha$)
[i.e. 0.20 $\lesssim$ F(H$\alpha$)/F(H$\alpha$)$_{peak}$ $\lesssim$
1.00; the high surface brightness zone] are the closest to the
nucleus, within a circular area with radius $\sim$ $4^{\prime\prime}$
(bottom rows of Figs.~\ref{ngc_bpt} and \ref{ngc_5966_bpt}). The
spaxels covering the more external regions of each galaxy [where
F(H$\alpha$)/F(H$\alpha$)$_{peak}$ $\leq$ 0.20] are shown in the top
rows of Figs.~\ref{ngc_bpt} and \ref{ngc_5966_bpt}. For the galaxy
NGC~5966, the [O{\sc i}]$\lambda$6300/H$\alpha$ diagram is presented
only for its inner region owing to the faintness of the [O{\sc
i}]$\lambda$6300 emission (Fig.~\ref{ngc_5966_bpt}). According to the
spectral classification scheme indicated in each figure, our emission
line ratios in the diagnostic diagrams for most positions in both
galaxies fall in the general locus of LINER-like objects. The relative
uncertainty of our line-ratio measurements plotted in
Figs.~\ref{ngc_bpt} and \ref{ngc_5966_bpt} is typically $\la$ 15\%. An
analysis of the ionized gas in the central region of NGC~6762 suggests
that it is dominated by an ionization source other than star
formation \citep[][]{Sanchez2011}.

Three grids of ionization models are overplotted on the BPT diagrams
(Figs.~\ref{ngc_bpt} and \ref{ngc_5966_bpt}). The plotted AGN models
have an electron density, $n_e$ = 100 cm$^{-3}$, metallicities of
solar ($Z$=$Z_\odot$) and twice solar, a range of ionization parameter (-3.6 $<$ log{\it U} $<$ 0.0) and a power-law ionizing spectrum with spectral
index $\alpha$ = -1.4. While the AGN photoionization models
of \citet{gro04} are consistent with most of the spaxels in the [N{\sc
ii}]$\lambda$6584/H$\alpha$ diagrams for NGC~5966, this is not the
case of NGC~6762 where the measurements of [N{\sc
ii}]$\lambda$6584/H$\alpha$ in the brightest area are not reproduced
well by the AGN grids (bottom-left panel of Fig.8). In the [S{\sc
ii}]$\lambda$$\lambda$6717,6731 diagram the data are merely fit by the
AGN models in NGC~6762, and no match between observations and models
is seen for NGC~5966. The measurements of [O{\sc
i}]$\lambda$6300/H$\alpha$ are not reproduced by these models for
either of the two galaxies.  A harder ionizing continuum, with
$\alpha$ = -1.2, will boost [S{\sc ii}]$\lambda$$\lambda$6717,6731 and
[O{\sc i}]$\lambda$6300 relative to H$\alpha$, yielding a better fit
in the [O{\sc i}]$\lambda$6300/H$\alpha$ BPT diagram, while providing
a poorer match with the [S{\sc ii}]$\lambda$$\lambda$6717,6731/H$\alpha$ measurements.


We also compared our results with shock models \citep{all08}. In
Figs.~\ref{ngc_bpt} and \ref{ngc_5966_bpt}, we plot the grids with $Z$=$Z_\odot$, preshock densities
between 0.1 cm$^{-3}$ and 100 cm$^{-3}$, shock velocities from 100 to
1000 km s$^{-1}$, and preshock magnetic field B=1 $\mu$G. Shock models
with a range of magnetic field strengths (e.g. B=5,10 $\mu$G) match
our observations. Interstellar magnetic fields of B $\sim$ 1 - 10
$\mu$G are typical of what is observed in elliptical
galaxies \citep[e.g.][]{mat97}. The low $n_e$ of these models are
consistent with the values derived here (Tables~\ref{apert_ngc6762}
and \ref{apert_ngc5966}). Overall, shock models reproduce the majority
of our data in the three emission-line ratio diagrams. The shock grids
with lower metallicity (e.g. LMC and SMC metallicities) are not
consistent with our measurements.



Furthermore, we compare our observations to the photoionization models
for pAGB stars with $Z$=$Z_\odot$ \citep{bin94}.  These models are consistent with most
of our observations, except for the highest values of [N{\sc
ii}]$\lambda$6584/H$\alpha$ [log([N{\sc ii}]$\lambda$6584/H$\alpha$)
$>$ 0.1]. The model with Z=1/3 Z$_{\odot}$ is shifted towards lower
values on the {\it x}-axis of the three BPTs, and it gives a much
poorer match to the measurements in the [N{\sc
ii}]$\lambda$6584/H$\alpha$ diagram. The pAGB scenario has recently
been revisited by \citet{sta08}, whose extensive grid of
photoionization models (see their Fig. 5) cover most of the regions
occupied by our spatially resolved measurements.

\begin{figure*}
\center
\includegraphics[width=0.30\textwidth,clip]{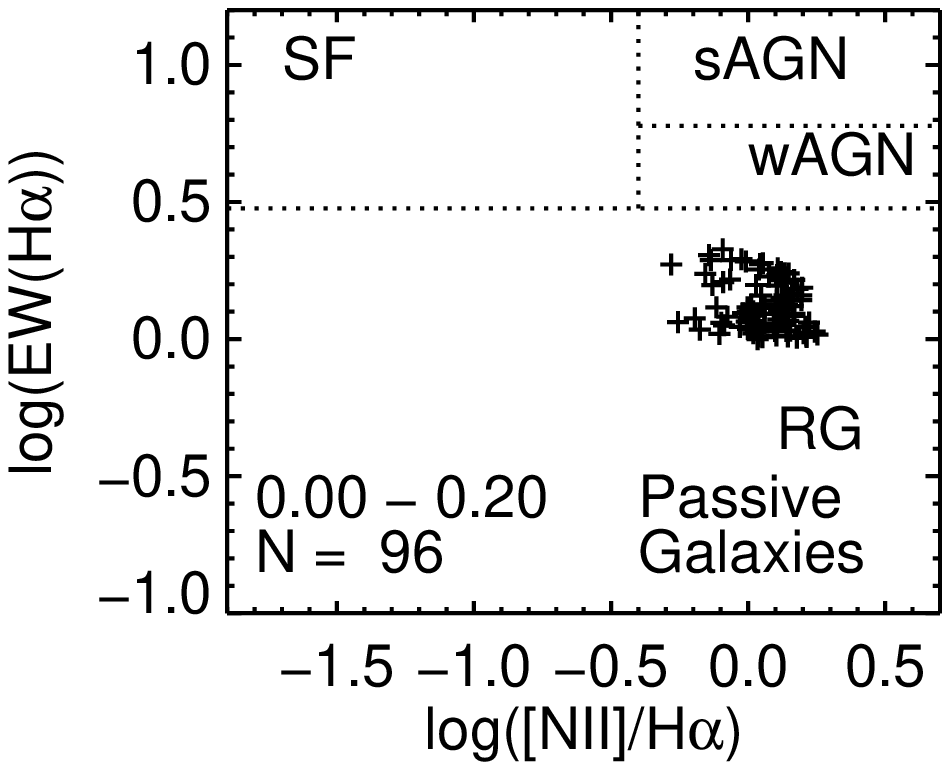}
\includegraphics[width=0.30\textwidth,clip]{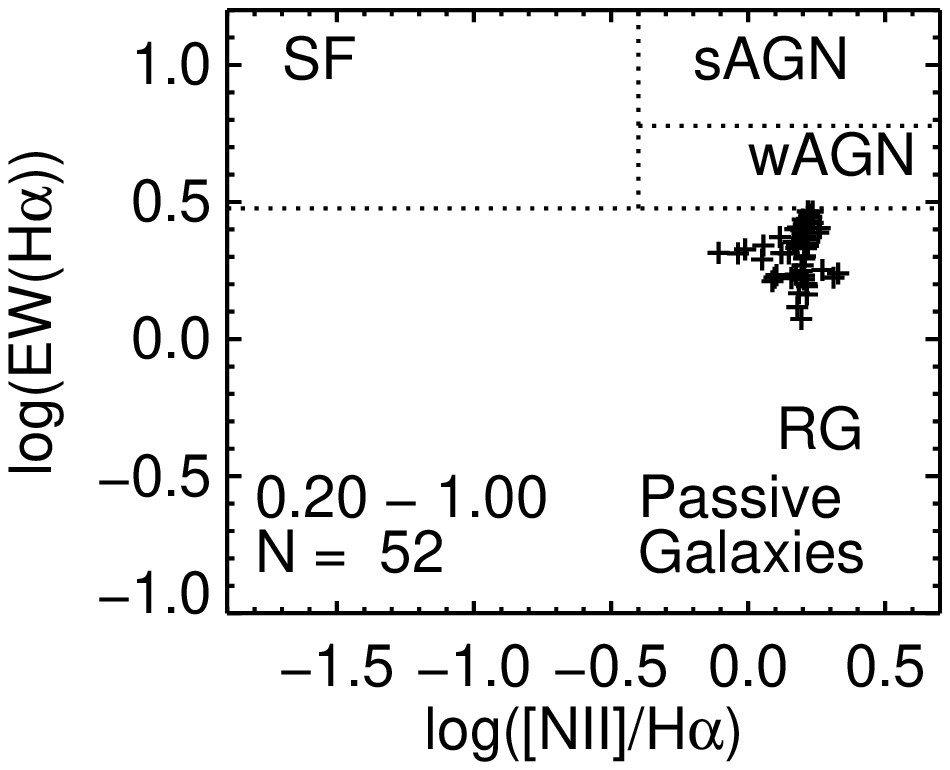}
\caption{EW(H$\alpha$) vs log([N{\sc ii}]$\lambda$6584/H$\alpha$) for NGC~6762. The left and right panels show the spaxels with F(H$\alpha$)/F(H$\alpha$)$_{peak}$ $\leq$ 0.20 and 0.20 $\lesssim$ F(H$\alpha$)/F(H$\alpha$)$_{peak}$ $\lesssim$ 1.00, respectively. SF = pure star-forming galaxies; sAGN = strong AGN; wAGN = weak AGN;  RG = retired galaxies. The dividing lines are transpositions of
the SF/AGN borders from \citet{kew01a,kew06a} and \citet{sta06}, and
the \citet{kew06a} Seyfert/LINER division \citep[see][]{cid10}}
\label{ngc_ew_bpt}
\end{figure*}

\begin{figure*}
\center
\includegraphics[width=0.30\textwidth,clip]{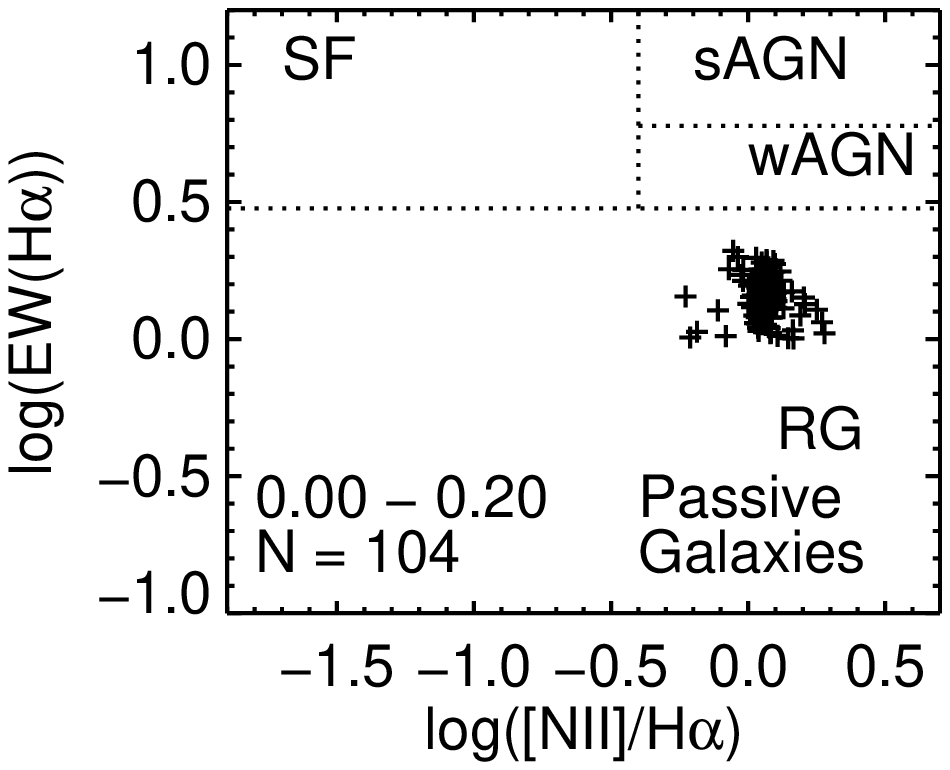}
\includegraphics[width=0.30\textwidth,clip]{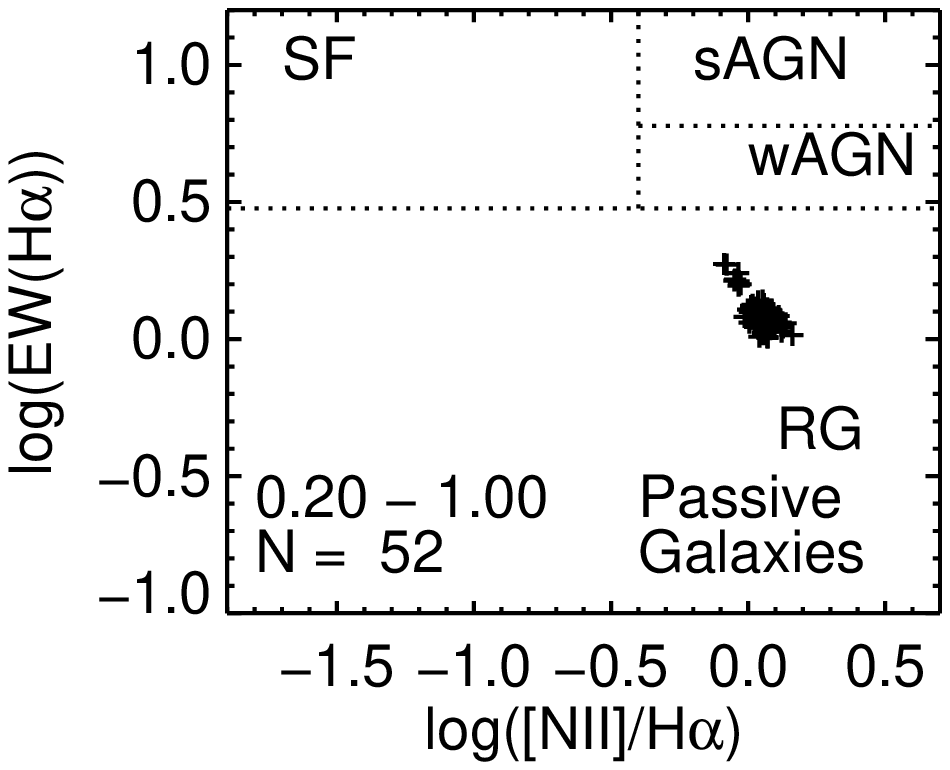}
\caption{EW(H$\alpha$) vs log([N{\sc ii}]$\lambda$6584/H$\alpha$) for NGC~5966. Labels as in Fig.~\ref{ngc_ew_bpt}.}
\label{ngc_5966_ew_bpt}
\end{figure*}

Since all of our emission line ratios appear to fall in the LINER
region of the BPT diagrams (Figs.~\ref{ngc_bpt}
and \ref{ngc_5966_bpt}), we also put our spatially resolved line
measurements in the WHAN diagram (EW(H$\alpha$) vs log([N{\sc
ii}]$\lambda$6584/H$\alpha$); Figs.~\ref{ngc_ew_bpt}
and \ref{ngc_5966_ew_bpt}) introduced by \cite{cid10}. The authors
argue that such a diagram is useful in distinguishing between two
different types of objects that may lead to emission line ratios like
those observed in LINERs and similar to what we observe in NGC~6762 and
NGC~5966. In the WHAN diagram, galaxies with LINER (-like) emission
are thus classified either as objects that present a weak AGN or the
so-called retired galaxies (RG), objects that are not forming stars
anymore and are ionized by their pAGB stars. Just as in the BPT
diagrams, we split our data into two bins: the left and righthand panels of
Figs.~\ref{ngc_ew_bpt} and \ref{ngc_5966_ew_bpt} show the spaxels for
the outer (F(H$\alpha$)/F(H$\alpha$)$_{peak}$ $\leq$ 0.20) and inner
(0.20 $\lesssim$ F(H$\alpha$)/F(H$\alpha$)$_{peak}$ $\lesssim$ 1.00)
regions, respectively. The location of our data in this diagram are,
for most emission line regions, compatible with the gas emission
expected for photoionization by pAGB stars.

\subsection{Kinematics}\label{sec_kinematics}


\begin{figure*}
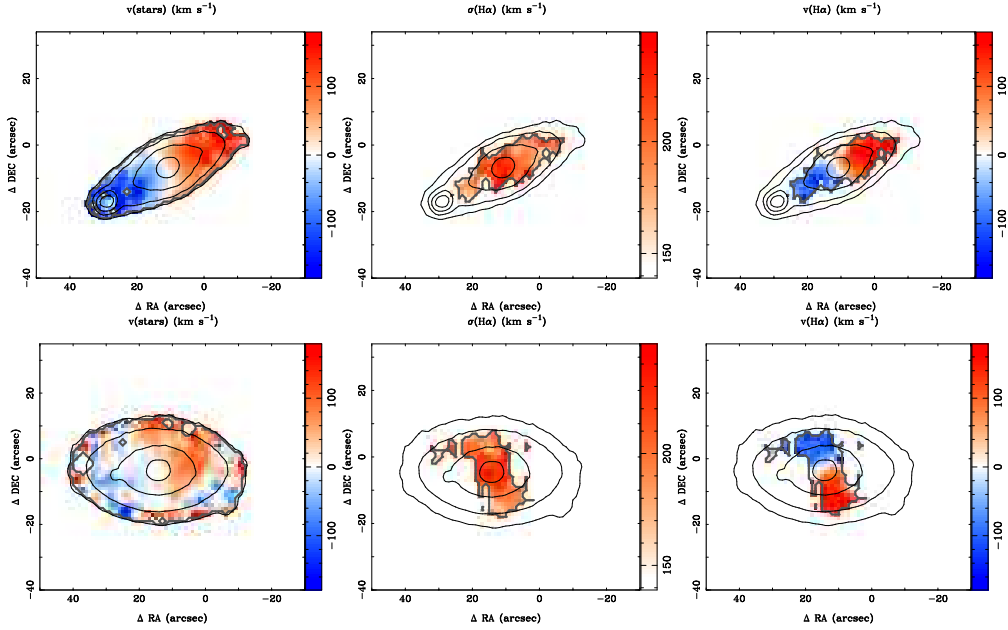

\center
\includegraphics[width=4.1cm,angle=-90]{figures/n6762.l16.ps}
\includegraphics[width=4.1cm,angle=-90]{figures/ngc6762.sigma.ps}
\includegraphics[width=4.1cm,angle=-90]{figures/vel_ha_n6762.ps} \\
\includegraphics[width=4.1cm,angle=-90]{figures/n5966.l16.ps}
\includegraphics[width=4.1cm,angle=-90]{figures/ngc5966.sigma.ps}
\includegraphics[width=4.1cm,angle=-90]{figures/vel_ha_n5966.ps}
\caption{Kinematics maps of NGC~6762 (top row) and NGC~5966 (bottom
row): stellar velocity field (left panels), corrected velocity dispersion maps (middle panels) and radial velocity maps (right panels), as measured from H$\alpha$. The
contours, orientation, and linear scale are the same as in
Figs.~\ref{ngc_fluxes} and \ref{ngc_5966_fluxes}.}
\label{kinematics}
\end{figure*}

Figure~\ref{kinematics} shows, for both galaxies, the stellar velocity
field and the maps of corrected velocity dispersion,
$\sigma(H\alpha)$, and of radial velocity of the ionized gas, derived
from the H$\alpha$ emission line. The values of $\sigma(H\alpha)$ are
corrected for the instrumental profiles as measured from arc
lines. The relative radial velocity of the H$\alpha$ lines ranges from
$\sim$ $-130~{\rm km~s}^{-1}$ to $\sim$ $180~{\rm km~s}^{-1}$ for
NGC~6762, and from $\sim$ $-100~{\rm km~s}^{-1}$ to $\sim$ $160~{\rm
km~s}^{-1}$ for NGC~5966. The typical uncertainty in the velocities
are $ < 15~{\rm km~s}^{-1}$ \citep[see][]{Sanchez2011}. The typical value
of $\sigma(H\alpha)$ is $\sim$ $200~{\rm km~s}^{-1}$ in both NGC~6762
and NGC~5966. We estimated the errors for $\sigma(H\alpha)$ based on a Monte Carlo simulation. The corresponding errors are $\sim$$10-20~{\rm km~s}^{-1}$ for NGC~6762 and $30-40~{\rm km~s}^{-1}$ for NGC~5966. For a sample of $\sim$ 50 ETGs, \citet{sar06}
find ionized gas velocities (estimated using the [O{\sc iii}]$\lambda$5007 line), between $\sim$ $-250~{\rm km~s}^{-1}$ and
$\sim$ $250~{\rm km~s}^{-1}$ and gas velocity dispersions as high as
$250~{\rm km~s}^{-1}$.

NGC~6762 displays an overall smooth rotation pattern along the
SE-NW direction in both the gas and stellar velocity
fields. NGC~5966 shows gas kinematics that are decoupled with respect to that
of the stars. The orientation of the stellar component is roughly
aligned SE to NW, with the stars in the NW part
having a higher recessional velocity. The axis of the ionized gas in
NGC~5966 is roughly orthogonal to that of the stars. Both cases,
i.e. stars-ionized gas kinematically aligned and misaligned, have been
observed in ETGs and help for determining the origin of the ionized gas in
these galaxies \citep[e.g.][references therein]{dav11}.  For
instance, \citet{sar06} conclude that in half of their objects with
gas kinematics decoupled from the stellar kinematics, this decoupling
suggests an external origin for the gas. A deeper analysis of the
complex kinematics in NGC~5966 is beyond the scope of this paper and
will be presented elsewhere.

\section{Spectral classification vs aperture size}

\begin{table*}
\caption{Observed emission line fluxes in units of 10$^{-16}$ erg cm$^{-2}$ s$^{-1}$ and physical properties from different apertures for NGC~6762}
\label{apert_ngc6762}
\centering
%
\begin{tabular}{lcccc}
\hline\hline
Wavelength & NGC6762 & NGC6762 & NGC6762  & NGC6762    \\  
    & Reg 1     & Reg 2  &  Reg 3 & Reg 4  \\ \hline
4861 H$\beta$    & 45.6 $\pm$ 0.7  & 81.4 $\pm$ 1.3  & 123.4 $\pm$ 3.1   & 135.5 $\pm$ 7.8   \\
5007 [O{\sc iii}]   & 155.1 $\pm$ 0.7  & 247.1 $\pm$ 1.3  & 356.5 $\pm$ 3.1   & 402.9 $\pm$ 7.8   \\
6300 [O{\sc i}]       & 46.9 $\pm$ 2.4   & 74.1 $\pm$ 4.8  & 112.3 $\pm$ 8.8  & ---  \\
6563 H$\alpha$  & 161.2 $\pm$ 2.1  & 273.8 $\pm$ 4.5  & 406.4 $\pm$ 7.7  & 451.5 $\pm$ 11.8  \\ 
6584 [N{\sc ii}]     & 244.6 $\pm$ 2.1  & 379.4 $\pm$ 4.5  & 516.3 $\pm$ 7.7  & 558.6 $\pm$ 11.8  \\
6717 [S{\sc ii}]     & 114.3 $\pm$ 0.9  & 180.8 $\pm$ 4.5  & 277.5 $\pm$ 7.7  & 340.2 $\pm$ 11.8   \\
6731 [S{\sc ii}]      & 82.5 $\pm$ 0.9  & 130.4 $\pm$ 4.5  & 169.6 $\pm$ 7.7  & 183.2 $\pm$ 11.8  \\ \hline
log ([O{\sc iii}]$\lambda$5007/H$\beta$)   & 0.53 $\pm$ 0.01 & 0.48 $\pm$ 0.01 & 0.46 $\pm$ 0.01 & 0.47 $\pm$ 0.03   \\
log ([N{\sc ii}]$\lambda$6584/H$\alpha$)  & 0.18 $\pm$ 0.01 & 0.14 $\pm$ 0.01 & 0.10 $\pm$ 0.01 & 0.09 $\pm$ 0.01   \\
log ([S{\sc ii}]$\lambda$6717,6731/H$\alpha$)    & 0.09 $\pm$ 0.01 & 0.06 $\pm$ 0.01 & 0.04 $\pm$ 0.01 & 0.06 $\pm$ 0.02   \\
log ([O{\sc i}]$\lambda$6300/H$\alpha$)  & -0.54 $\pm$ 0.02 & -0.57 $\pm$ 0.03 & -0.56 $\pm$ 0.04 & ---   \\ 
H$\alpha$/H$\beta$  & 3.54 $\pm$ 0.07 & 3.36 $\pm$ 0.08 & 3.29 $\pm$ 0.10 & 3.33 $\pm$ 0.21  \\ \hline 
$c_{H\beta}$     & 0.31 $\pm$ 0.03 & 0.24 $\pm$ 0.03 & 0.21 $\pm$ 0.05 & 0.22 $\pm$ 0.09   \\ 
$EW_{H\alpha}$ (\AA)   & 2.52 $\pm$ 0.04 & 2.01 $\pm$ 0.03 & 1.59 $\pm$ 0.03 & 1.43 $\pm$ 0.04   \\                  
$n_{\rm e}$([S{\sc ii}])($cm^{-3}$)   & $<$ 100 & $<$ 100 & $<$ 100 & $<$ 100  \\ \hline 
\end{tabular}
\tablefoot{Reg 1 $=$ 5$^{\prime\prime}$/diam.; Reg 2 $=$ 10$^{\prime\prime}$/diam.; Reg 3 = spectrum obtained by co-adding all fibers that cover the [N{\sc ii}]-H$\alpha$ emission zone; Reg 4 $=$ 30$^{\prime\prime}$/diam. The quoted uncertainties include measurements errors. }
\end{table*}

\begin{table*}
\caption{Observed emission line fluxes in units of 10$^{-16}$ erg cm$^{-2}$ s$^{-1}$ and physical properties from different apertures for NGC~5966} 
\label{apert_ngc5966}
 \centering 
\begin{minipage}{20.0cm}
\centering
{\tiny
\begin{tabular}{lcccccc}
\hline\hline 
Wavelength & NGC5966 & NGC5966 & NGC5966  & NGC5966  & NGC5966 & NGC5966  \\  
   & Reg 1    & Reg 2  &  Reg 3 & Reg 4   & Region NE  &  Region SW \\ \hline
4861 H$\beta$  & 29.1 $\pm$ 4.5 & 58.7 $\pm$ 8.4   & 88.0  $\pm$ 11.0  & 147.0 $\pm$ 18.2  & 6.9 $\pm$ 0.5  & 11.0 $\pm$ 0.9  \\
5007 [O{\sc iii}]   & 50.2 $\pm$ 4.5 & 92.4 $\pm$ 8.6  & 150.5 $\pm$ 11.0  & 261.6 $\pm$ 11.2  & 11.4 $\pm$ 0.5  & 23.9 $\pm$ 0.9  \\
6300 [O{\sc i}]     & --- & --- & --- & --- & --- & --- \\
6563 H$\alpha$  & 68.8 $\pm$ 2.2  & 139.4 $\pm$ 4.6  & 238.0 $\pm$ 8.0  & --- & 19.7 $\pm$ 0.4  & 37.0 $\pm$ 0.9  \\ 
6584 [N{\sc ii}]     & 91.2 $\pm$ 2.2  & 160.0 $\pm$ 4.6  & 281.8 $\pm$ 8.0  & 330.5 $\pm$ 11.3   & 23.2 $\pm$  0.4  & 38.9 $\pm$ 0.9  \\
6717 [S{\sc ii}]     & 33.4 $\pm$ 0.9  & 62.0 $\pm$ 4.2   & 71.7 $\pm$ 2.9   & 125.8: & 13.0 $\pm$ 0.4  & 11.1 $\pm$ 0.9  \\
6731 [S{\sc ii}]     & 19.2 $\pm$ 0.9  & 38.8 $\pm$ 4.2  & 41.7 $\pm$ 2.9   & 74.2: & --- & 7.0 $\pm$ 0.9  \\ \hline
log ([O{\sc iii}]$\lambda$5007/H$\beta$)  & 0.24 $\pm$ 0.08  & 0.20 $\pm$ 0.07 & 0.23 $\pm$ 0.06  & 0.25 $\pm$ 0.06 & 0.22 $\pm$ 0.04 & 0.34 $\pm$ 0.04 \\
log ([N{\sc ii}]$\lambda$6584/H$\alpha$)  & 0.12 $\pm$ 0.02  & 0.06 $\pm$ 0.02 &  0.07 $\pm$ 0.02  & ---  & 0.07 $\pm$ 0.01  & 0.02 $\pm$ 0.01  \\
log ([S{\sc ii}]$\lambda$6717,6731/H$\alpha$)  & -0.12 $\pm$ 0.02 & -0.14 $\pm$ 0.03 & -0.32 $\pm$ 0.02 & --- & --- & -0.31 $\pm$ 0.03  \\
log ([O{\sc i}]$\lambda$6300/H$\alpha$)   & --- & --- & --- & --- & ---  & ---   \\ 
H$\alpha$/H$\beta$  &  2.37 $\pm$ 0.37  & 2.37 $\pm$ 0.35  & 2.70 $\pm$ 0.35  & ---   & 2.88 $\pm$ 0.22  & 3.37 $\pm$ 0.27  \\ \hline
$c_{H\beta}$          & 0.00 & 0.00 & 0.00 & --- & 0.01 $\pm$ 0.11  & 0.24 $\pm$ 0.12  \\ 
$EW_{H\alpha}$ (\AA)   & 0.97 $\pm$ 0.03 & 0.93 $\pm$ 0.03 & 0.86 $\pm$ 0.03 & --- & 1.64 $\pm$ 0.04  & 1.48 $\pm$ 0.04 \\                  
$n_{\rm e}$([S II])($cm^{-3}$)  & $<$ 100 &  $<$ 100 & $<$ 100 & $<$ 100  & $<$ 100 & $<$ 100 \\ \hline
\end{tabular}} 
\end{minipage}
\tablefoot{Reg 1 $=$ 5$^{\prime\prime}$/diam.; Reg 2 $=$ 10$^{\prime\prime}$/diam.; Reg 3 = spectrum
obtained by co-adding all fibers that cover the [N{\sc ii}]-H$\alpha$ emission zone; Reg 4 $=$ 30$^{\prime\prime}$/diam. The last two columns correspond to the lower (towards SW) and
upper (towards NE) regions of the elongated structure as observed in
the emission-line and ionized gas velocity maps (Figs.~\ref{ngc_5966_fluxes} and ~\ref{kinematics}). The
quoted uncertainties include measurements errors. The colon (:) indicates uncertain value.}
\end{table*}

In this section we present the analysis of 1D spectra extracted within
circular apertures of increasing diameter (5\arcsec, 10\arcsec\ and
30\arcsec) centered on the intensity maximum of the red stellar
continuum. These apertures correspond to radii of 0.35, 0.7, and 2.1
$r_{\rm eff}$ for NGC~6762 and to 0.24, 0.47 and 1.4 $r_{\rm eff}$ for
NGC~5966. We also extracted the integral spectrum by summing the
emission from each spaxel within the [N{\sc ii}]-H$\alpha$ emitting
area of each ETG, covering $\sim$ 300 arcsec$^{2}$ for both galaxies
($\sim$ 14 kpc$^{2}$ $\sim$ 5 $r_{\rm eff}^{2}$ and 33
kpc$^{2}$ $\sim$ 3 $r_{\rm eff}^{2}$ for NGC~6762 and
NGC~5966, respectively). Two additional 1D spectra were extracted for
NGC~5966, corresponding to the SW and NE regions
of the elongated gas structure observed in the emission-line and
velocity maps (Figs.~\ref{ngc_5966_fluxes} and
~\ref{kinematics}). Emission-line fluxes were measured in the 1D
spectra as described in Sect.~\ref{spatial_line_ratios} and given in
Tables~\ref{apert_ngc6762} and \ref{apert_ngc5966}, together with the
spectroscopic properties. Line fluxes quoted in the tables are not
corrected for internal reddening.

We obtained the $n_{e}$ from the [S{\sc
ii}]$\lambda$6717/$\lambda$6731 line ratio using the IRAF nebular
package \citep{sha95}. The derived estimates for $n_{e}$ place all of
the {\it n}-diameter aperture zones in the low-density regime of the
[S{\sc ii}] doublet ($n_{e}$ $<$ 100 cm$^{-3}$).

The logarithmic reddening, $c(H\beta$), was computed from the ratio of
the measured-to-theoretical H$\alpha$/H$\beta$, assuming the Galactic
reddening law of \citet{car89} and an intrisic value of
H$\alpha$/H$\beta$ = 2.86 (Case B, electron temperature $T_{e}$= 10$^{4}$ K, $n_{e}$= 100
cm$^{-3}$). At relatively low densities, the theoretical
H$\alpha$/H$\beta$ values vary from $\sim$ 3.00 ($T_{e}$ = 5 $\times$
10$^{3}$ K) to $\sim$ 2.75 ($T_{e}$ = 20 $\times$ 10$^{3}$ K). Since
we are not able to estimate the $T_{e}$ of the warm ISM,
we decided to adopt the intermediate value of 2.86 in this work. For
some zones in NGC~5966, we adopt $c(H\beta$)=0.0 since the
corresponding H$\alpha$/H$\beta$ values are consistent with no
reddening within the errors. A different assumption from
2.86 would not change our results since most of quantities treated in
this paper [e.g. [O{\sc iii}]$\lambda$5007/H$\beta$, [N{\sc
ii}]$\lambda$6584/H$\alpha$, EW(H$\alpha$)] slightly depend on
reddening corrections.

From the integrated spectra (covering all of the H$\alpha$ emitting
region from each galaxy), we estimate the values of the
reddening-corrected H$\alpha$ luminosities, $L(H\alpha$): 6.0 $\pm$
0.5 $\times$ 10$^{39}$ $erg\,s^{-1}$ (NGC~6762) and 5.8 $\pm$ 0.2
$\times$ 10$^{39}$ $erg\,s^{-1}$ (NGC~5966). These values are within
the H$\alpha$ luminosity range measured for luminous ETGs
generally \citep{mac96}. For NGC~6762, we checked how much
$L(H\alpha$) would change by adopting various values of
H$\alpha$/H$\beta$ (2.75 to 3.00), and found that the variations are
within the quoted uncertainties.

Spectra constructed from apertures with a range of diameters allow us
to evaluate how aperture effects may affect the spectral
classification of the ETGs under study. The observed flux ratios
([O{\sc iii}]$\lambda$5007/H$\beta$, [N{\sc
ii}]$\lambda$6584/H$\alpha$, [S{\sc
ii}]$\lambda\lambda$6717,6731/H$\alpha$, and [O{\sc
i}]$\lambda$6300/H$\alpha$), for both galaxies, are consistent with
LINER-type emission based on the BPT diagrams (Figs.~\ref{ngc_bpt}
and \ref{ngc_5966_bpt}), independent of the aperture size
(Tables~\ref{apert_ngc6762} and \ref{apert_ngc5966}). Splitting the
elongated gas emission in NGC~5966 into two separate regions on either
side of the nucleus suggests that both regions have the same spectral
classification in the diagnostic BPT diagrams
(Table~\ref{apert_ngc5966}). This indicates that the properties of the
ionized gas do not vary significantly across the PPAK FOV for both
galaxies. This is consistent with the results of an analysis of the
radial profiles (Sect. 5.3.1). However, one should be cautious when interpreting the nebular spectra from integrated apertures. The
presence of different ionization sources and the way they are spatially distributed
might play roles in the spectral classification in the various apertures. For
instance, from a sample of luminous infrared galaxies,
\citet{alo09} find  that the nuclear and integrated line-ratios
give different spectral classifications in the BPT diagrams for some of their objects. In this case, this was interpreted as the
result of an increased contribution of extra-nuclear high surface-brightness
HII regions to the integrated emission of these galaxies.

\section{Discussion}\label{discussion} 

Based on our EW and diagnostic line ratios measurements, both NGC~6762
and NGC~5966 are weak emission-line galaxies in which most of spaxels
can be classified as LINER-like
(Figs.~\ref{ngc_bpt}, \ref{ngc_5966_bpt}, \ref{ngc_ew_bpt},
and \ref{ngc_5966_ew_bpt}). Different ionizing mechanisms (e.g. pAGB stars,
AGN, shocks, massive stars) have been proposed to explain LINER-like
excitation in ETGs, but it is still a matter of debate
(Sect.~\ref{intro}). In the previous sections we presented diagnostics
that help distinguish between these different ionizing sources,
including the overall morphology of the emission line gas. In the
following we discuss which one(s) might be more likely to power the
emission lines in these two galaxies based on our analysis.

According to the WHAN diagrams (Figs.~\ref{ngc_ew_bpt}
and \ref{ngc_5966_ew_bpt}), NGC~6762 and NGC~5966 are classified as
RG, objects that have stopped forming stars and are ionized by 
the pAGB stars contained in them \citep{cid10}. A central black hole may be
present in RGs, but it is not expected to dominate the
ionization budget \citep{cid11}. Solar-metallicity photoionization models for pAGB
stars are able to reproduce the majority of our spatially resolved
data \citep[Figs.~\ref{ngc_bpt} and \ref{ngc_5966_bpt};][]{bin94},
except for the high [N{\sc ii}]$\lambda$6584/H$\alpha$ ratios in
the central region of NGC~6762, which are not reproduced even by a
high metallicity model (Z=3 Z$_{\odot}$).  However, one should bear
in mind that in these photoionization models, the metallicity, Z,
is a scaling factor of relative abundances of every element with
respect to H, although it is known that the N/O ratio in galaxies
is not (necessarily) constant \citep[e.g.][]{mol06}. 
The models by \citet{sta08}, which are essentially updated versions of the \citet{bin94} ones, do not consider N/O to be a constant and do extend towards larger [N{\sc ii}]$\lambda$6584/H$\alpha$. Photoionization models by pAGB stars are therefore consistent with the
line ratios in both ETGs studied here.   

To investigate whether pAGB stars can account for the
ionizing flux in NGC~6762 and NGC~5966, we also computed the rate of
the Lyman continuum photons expected from the surface density and age
distribution of pAGB stars that we derived in each spaxel and used it
to predict the Balmer emission fluxes, assuming case B, $T_{\rm e}$ of
$10^4$ K and low densities ($n_{\rm e}\ll 10^4$ cm$^{-3}$). We also
allow for calculating the Balmer emission line fluxes under the
assumption that the warm ISM has the same foreground extinction as the
stars. Since the stellar extinction derived is generally low ($A_V\la
0.3$ mag), this assumption has only a minor influence on the final
results. The predicted Balmer H$\beta$ intensity was computed for
different annuli (see the Appendix). In Fig.~\ref{zones} we plot  the H$\beta$ predicted-to-observed flux ratio (H$\beta_{pred.}$/H$\beta_{obs.}$) for
both galaxies as a function of \rr. Typical
values of H$\beta_{pred.}$/H$\beta_{obs.}$ are close to 1 for
NGC~6762, suggesting that pAGB stars can produce enough
ionizing photons to explain the Balmer line fluxes. In the case of
NGC~6762, the morphology of gas and stellar emission following each
other is an additional and strong argument supporting the hypothesis
that pAGB stars are responsible for ionizing the gas. In NGC~5966, the
H$\beta_{pred.}$/H$\beta_{obs.}$ values are slightly greater than 1 on
average for \rr $\lesssim$ 8 arcsec, and decrease to $<$ 1 for larger
radii where the elongated gas emission dominates, indicating that
another excitation source is likely needed. We discuss this further at
the end of this section.

To probe the ionization by an AGN, we compare our observed
line ratios ([O{\sc iii}]$\lambda$5007/H$\beta$, [N{\sc
ii}]$\lambda$6584/H$\alpha$, [S{\sc
ii}]$\lambda\lambda$6717,6731/H$\alpha$, [O{\sc
i}]$\lambda$6300/H$\alpha$) with the AGN models of \citet{gro04}. In Figs.~\ref{ngc_bpt} and \ref{ngc_5966_bpt} we
represent the AGN-grids that better reproduce the majority of the
spaxels in both galaxies (see Sect. 5.3.2 for details). However, for
both galaxies, none of the AGN photoionization models are able to
reproduce our spatially-resolved line ratios in all of the three BPT
diagrams simultaneously. The relative radial constancy of the [O{\sc
iii}]$\lambda$5007/H$\beta$ and [N{\sc ii}]$\lambda$6584/H$\alpha$
ratios (Fig.~\ref{zones}) also argue against an AGN as the dominant
ionization source in both NGC~6762 and NGC~5966 unless the ratio of
ionizing photon intensity to the gas density is approximately
constant.
  
Examples of extended LINER-like excitation from shocks can be found in
the literature \citep[e.g.][]{mon06,far10,ric11}. Here, the detection
of [O{\sc i}]$\lambda$6300 suggests the presence of
shocks \citep[e.g.][]{dop76}. To assess the role of
excitation by shocks, the observations in the BPTs are compared to the
fast shock models from \citet{all08} (Figs.~\ref{ngc_bpt}
and \ref{ngc_5966_bpt}). A large fraction of the spaxels in both
galaxies are reproduced by shock grids with velocities higher than
$\sim$ 200 km s$^{-1}$. Outflows and accretion into a central
black hole and collisions between gas clouds have been suggested as possible sources of mechanical energy that are able
to account for such high velocity shocks \citep[e.g.][and references
therein]{dop95,dop97,ani10}. From our data, the estimated average gas
velocity dispersion $\sigma$, as measured from the H$\alpha$ line, is
$\sim$ 200 km s$^{-1}$ for both objects (Sect. 5.4), suggesting that
fast shock-ionization cannot be ruled out in our galaxies. X-ray data
would help in constraining the role of shocks in ionizing
the gas at such velocities. In any case, if
shock excitation was the dominant ionizing source in the two galaxies,
we would expect to see a more extended [O{\sc i}]$\lambda$6300
emission, following the morphology of the strong emission
lines \citep[e.g.][]{far10}.

NGC~5966 is especially interesting because it has similar ionization
characteristics to NGC~6762 but presents a gas emission
morphology that is completely different from its underlying stellar
population (again, unlike NGC~6762). NGC~5966 exhibits an elongated
emitting gas component spanning $\sim$ 6 kpc that is roughly orthogonal
to the stellar emission (Sect.~\ref{maps}). This component could represent
a decoupled rotating disk resulting from a merger event, as found previously in other ETGs galaxies \citep[e.g.][]{ser06}. The
outflow scenario appears as an alternative interpretation for the gas with
biconical gas emission. Similar
features, called ``ionization cones'' have been seen in starburst,
Seyfert, and LINER-(like) galaxies with strong line emission for the
gas \citep[e.g.][]{marquez98,mor98,arr01,mon06,sharp10}. As far as we
know, this would be the first time that such a large gaseous bicone has been discovered in a weak emission-line galaxy.  
In the outflow scenario, what could power the gas motion? Our
analysis shows that star formation cannot be this source. The
STARLIGHT fits do indicate a small contribution (0.1\% by mass) of young stars in the nucleus, but these are almost
certainly old blue stars that are not well accounted for by current
evolutionary synthesis models \citep[][]{kol08,cid10a}. The weakness of H$\alpha$ reinforces the
conclusion that these are "fake bursts" as described by \citet{ocv10},
otherwise EW($H\alpha$) should be much stronger. As discussed above, our
energetic balance approach indicates that an additional excitation
mechanism other than pABG stars might operate at large radii (Fig.7). Morphologically, a nuclear power source appears the most
likely alternative to form a symmetrical biconical
feature like the one we observe \citep[e.g.][]{tad89}. The radio source associated with the
nucleus in NGC~5966 argue in favor of the presence of an AGN, and in
fact this galaxy has been classified as an AGN based on its FIR/radio
flux ratio \citep[e.g.][and references therein]{con02}.
High-resolution spectroscopic analysis is needed to determine the
presence of an outflow.  Clearly, NGC~5966 deserves closer
investigation to unveil the origin of its elongated ionized gas which
will be presented in a future paper. The case of NGC~5966
suggests the exciting possibility that the CALIFA survey may reveal
more of them, allowing us to investigate the nature and ionizing source
of such biconical emission in ETGs generally.

\section{Summary}\label{fim} 

In this work we present the first optical IFS study of the warm ionized ISM in
the two ETGs, NGC~6762 and NGC~5966, which are part of the CALIFA
survey. Using the STARLIGHT spectral synthesis code, we modeled and subtracted the
stellar component from the observed spectra at each spaxel of the PPak data
cubes. The pure nebular emission-line spectra were then used to probe the
nature of the ionized gas. In the following we list the main results derived from this
work. \\

\begin{itemize}

\item The warm ionized gas was probed through the
use of the optical emission lines (H$\beta$,
[O{\sc iii}]$\lambda$5007, [O{\sc i}]$\lambda$6300, [N{\sc ii}]$\lambda$6584, H$\alpha$,
[S{\sc ii}]$\lambda$6717,6731) in both NGC~6762 and NGC~5966. The two galaxies are
very faint emission-line objects relative to their stellar continua with EW(H$\alpha$) values $\lesssim$ 3.00 \AA. While in NGC~6762, the gas
and stellar morphology are strikingly similar, this is not the case for
NGC~5966. This galaxy shows an elongated ionized gas structure, oriented
approximately orthogonally to the major axis of the stellar ellipsoid.
Differences are also reflected in their kinematics where the stellar
and gas kinematics are aligned in NGC~6762 and misaligned in NGC~5966.   \\

\item The radial profiles of diagnostic line ratios, [O{\sc iii}]$\lambda$5007/H$\beta$ and
[N{\sc ii}]$\lambda$6584/H$\alpha$, show that they are roughly constant with radius for both galaxies. This indicates that the dominant ionizing source is not confined
to the nuclear region in the two objects and that the ionized gas properties are homogeneous in the emission line
regions across each galaxy. \\


\item We showed that the spectral classification of both ETGs does not depend on the aperture size. This result might have implications for interpreting the nebular spectra of more distant ETGs where spatially-resolved data
are more difficult to obtain.\\ 

\item From an analysis of the BPT diagrams, [N{\sc ii}]$\lambda$6584/H$\alpha$, [S{\sc ii}]$\lambda$6717,6731/H$\alpha$,
and [O{\sc i}]$\lambda$6300/H$\alpha$, for the spatially-resolved
emission lines, both galaxies contain a LINER nucleus and extended
LINER-like emission across the PPak FOV. According to the WHAN diagram
[EW(H$\alpha$) vs log([N{\sc ii}]$\lambda$6584/H$\alpha$)], both objects
are located in the region which suggests that the emission line gas is
ionized by pAGB stars.  \\ 
  
\item In NGC~6762, different lines of evidence (e.g. H$\beta_{pred.}$/H$\beta_{obs.}$
$\sim$ 1, gas and stellar emission showing the same morphology) argue in favor of
pAGB stars being the dominant ionization source. In the case of
NGC~5966, the differing gas and stellar morphologies, and the energetic balance indicate that an additional ionization source other
than pAGB stars is needed. The existence of a nuclear radio source in
NGC~5966 suggests that an AGN
might be present in this galaxy, and may be responsible for the extended
($\sim$ 6 kpc) elongated ionized gas emission. Shock-ionization cannot be ruled out in either galaxy. \\ 

\item An ionization cone is a possible interpretation for the elongated
gas feature in NGC~5966, which would be the first ionization cone
associated with a weak emission-line galaxy. A decoupled rotating disk
appears as an alternative way to explain the morphology of the ionized gas. At present we are unable to make a definitive statement
about the origin of this elongated gas structure. A deeper study of
this object will be presented in a subsequent paper. \\

\end{itemize}

The CALIFA survey will ultimately provide a sample of $\sim$100 ETGs.
We will extend the analysis presented in this work to the remaining
CALIFA ETGs and address many of the issues discussed
here statistically. Further, multiwavelength data will be helpful for better
understanding which of the mechanisms for photoionizing the gas is
dominant. For example, MIR and X-ray high spatial resolution data
would help in detecting a central unresolved source and checking for any
spatial correlation with the optical spectral maps.

\bigskip 
 
\noindent \textbf{Acknowledgements}\\ 

This paper is based on data of the Calar Alto Legacy Integral Field
Area Survey, CALIFA\footnote{http://www.caha.es/CALIFA/}, funded by
the Spanish Ministery of Science under grant ICTS-2009-10, and the
Centro Astron\'omico Hispano-Alem\'an. We wish to thank the anonymous referee for his/her useful comments and suggestions. This work has been partially
funded by research projects AYA2007-67965-C03-02 and
AYA2010-21887-C04-01 from the Spanish PNAYA and CSD2006-00070 1st
Science with GTC of the MICINN. CK, as a Humboldt Fellow, acknowledges
support from the Alexander von Humboldt Foundation, Germany. PP is
supported by a Ciencia 2008 contract, funded by FCT/MCTES (Portugal)
and POPH/FSE (EC).  AM-I is grateful for the hospitality of the 3D
Spectroscopy group at the Leibniz-Institut f\"{u}r Astrophysik Potsdam
where part of this work was performed. JMG is supported by a
post-doctoral grant, funded by FCT/MCTES (Portugal) and POPH/FSE
(EC). JBH is supported by a Federation Fellowship from the Australian
Research Council. RCF is supported by grant 5760-10-0 from CAPES
(Brazil). RAM is funded by the spanish programme of International Campus of Excellence (CEI). Financial support from the Spanish grant AYA2010-15169 and
from the Junta de Andalucia through TIC-114 and the Excellence Project
P08-TIC-03531 is acknowledged.  

The STARLIGHT project is supported by the Brazilian agencies CNPq,
CAPES, and FAPESP. This paper uses the plotting package jmaplot
developed by Jes\'us Ma\'{\i}z-Apell\'aniz (available at
http://dae45.iaa.csic.es:8080$\sim$jmaiz/software). This research 
made use of the NASA/IPAC Extragalactic Database (NED) which is
operated by the Jet Propulsion Laboratory, California Institute of
Technology, under contract with the National Aeronautics and Space
Administration.

This paper makes use of the Sloan Digital Sky Survey data.
Funding for SDSS and SDSS-II has been provided by the Alfred P. Sloan
Foundation, the Participating Institutions, the National Science
Foundation, the US Department of Energy, the National Aeronautics and
Space Administration, the Japanese Monbukagakusho, the Max Planck
Society, and the Higher Education Funding Council for England. The
SDSS web site is http://www.sdss.org/.

SDSS is managed by the Astrophysical Research Consortium for the
Participating Institutions. The Participating Institutions are the
American Museum of Natural History, Astrophysical Institute Potsdam,
University of Basel, University of Cambridge, Case Western Reserve
University, University of Chicago, Drexel University, Fermilab, the
Institute for Advanced Study, the Japan Participation Group, Johns
Hopkins University, the Joint Institute for Nuclear Astrophysics, the
Kavli Institute for Particle Astrophysics and Cosmology, the Korean
Scientist Group, the Chinese Academy of Sciences (LAMOST), Los Alamos
National Laboratory, the Max-Planck-Institute for Astronomy (MPIA),
the Max-Planck-Institute for Astrophysics (MPA), New Mexico State
University, Ohio State University, University of Pittsburgh,
University of Portsmouth, Princeton University, the United States
Naval Observatory, and the University of Washington.
 
 
\bibliography{mybib}
\bibliographystyle{./aa}


\appendix

\section{Derivation of irregular annuli \label{profiles}}
\begin{figure*}[hb]
\center
\includegraphics[width=5.0cm,angle=0]{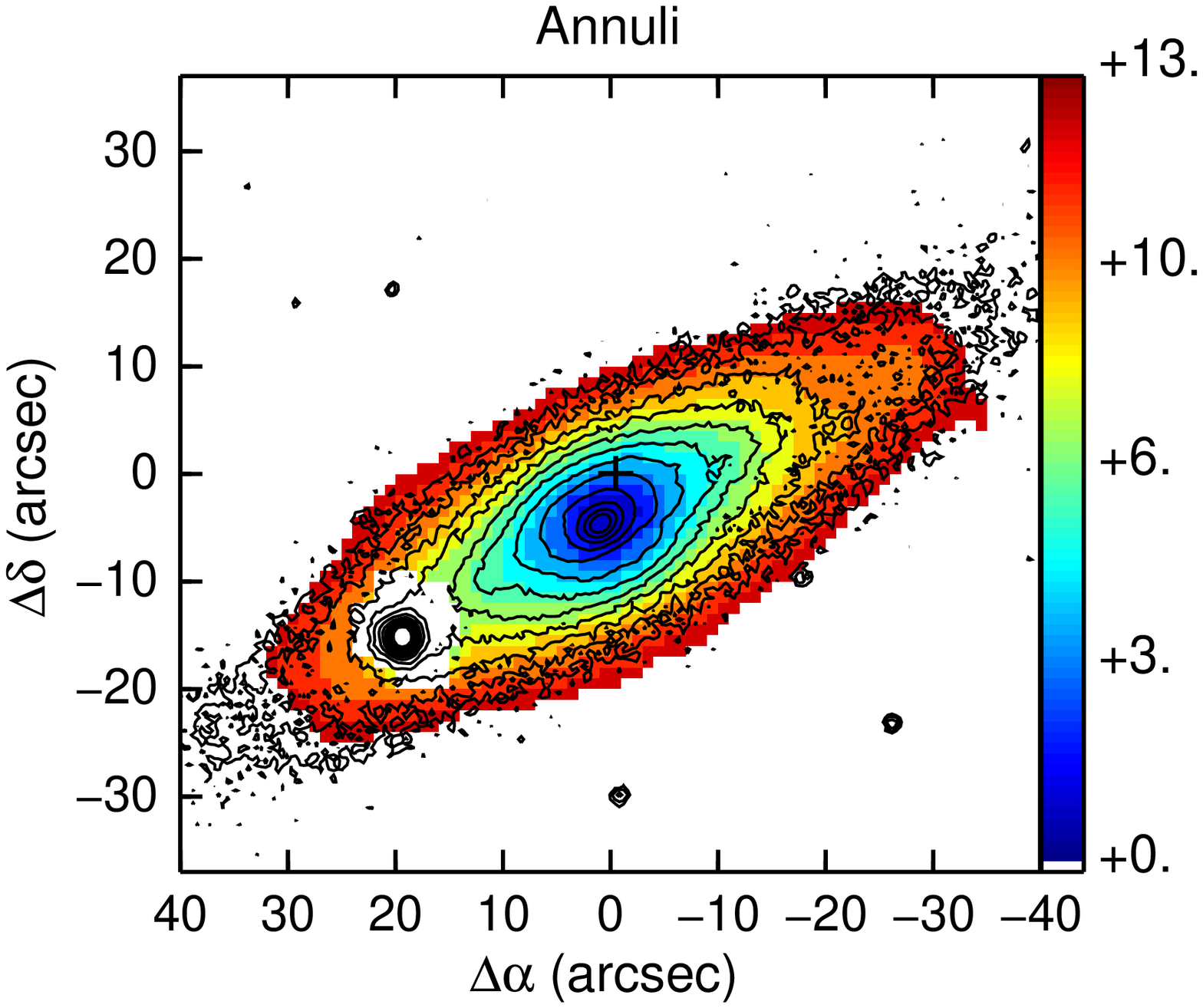}
\includegraphics[width=5.0cm,angle=0]{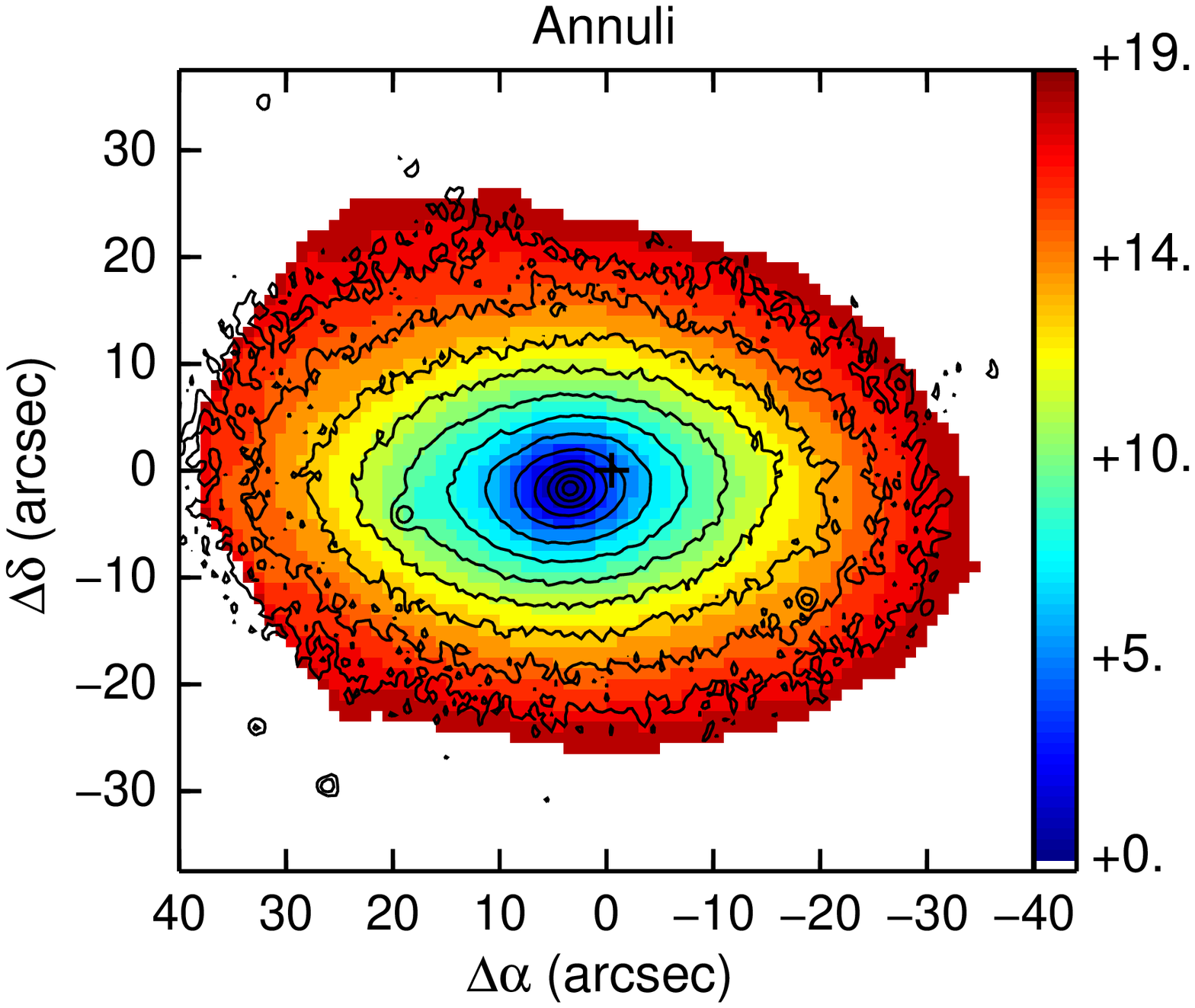}
\caption{NGC~6762 (left panel) and NGC~5966 (right panel): the vertical color bar indicates (from the center outwards) the irregular annuli used for deriving radial profiles 
in Fig.~\ref{zones}. The overlaid contours are computed from SDSS $g$ band images and go from 18 to 23.5 \sbb in steps of 0.5 mag.}
\label{annuli}
\end{figure*}

In this work, the characteristics of the emission line gas are investigated within different annuli
in order to better evaluate their dependence on the
galactocentric radius (Sects.~\ref{spatial_line_ratios} and \ref{discussion}) and reduce the scatter relative to the modeling and
post-processing of individual spaxel spectra. 
The annuli were computed with a slightly modified version of 
the surface photometry method {\vs iv} \citep{Papaderos02-IZw18}. 
This method permits a simultaneous processing of coaligned images of a galaxy 
in several bands and does not require a (generally subjective) choice of a 
galaxy center, nor does it implicitly assume that the galaxy can be
approximated by the superposition of axis-symmetric luminosity components.
A key feature of this method lies in the computation of photon statistics within 
automatically generated irregular annuli that are adapted to the galaxy
morphology in one or several reference passbands (Fig.~\ref{annuli}).
This concept distinguishes method {\vs iv} from conventional surface photometry techniques
that compute surface brightness profiles essentially through ellipse-fitting to isophotes or photon
statistics within elliptical annuli (e.g. method {\vs I}
of \cite{Papaderos96a}, the IRAF task ELLIPSE, or FIT/ELL3 in MIDAS) 
or approximate a galaxy by a single or several 2D axis-symmetric components, 
such as GIM2D \citep{Simard98} and GALFIT \citep{Peng02}. 
The photometric radius \rr\ of the annulus mapping the
surface brightness interval between $\mu$ and $\mu+\delta\mu$ in the reference
frame is given as 
[($A_{\mu}$+$A_{\mu+\Delta\mu}$)/2\,$\pi$]$^{0.5}$ where $A$ is the area 
($\sq$\arcsec) subtended by the galaxy's isophote at a given $\mu$. 
As method {\vs iv} allows adjusting both the reference frame(s) used 
for generating the morphologically adapted annuli and their number, 
it offers a handy tool for analyzing radial trends in galaxies.
Here, we used the stellar continuum emission, extracted between 6390 and 6490 \AA~
from the 3D$_{\rm obs}$ cubes as reference frame for the generation of morphologically adapted 
annuli. We coadded the spaxels within each annulus in
order to create 1D spectra to be used in investigating radial trends.

\end{document}